\pdfoutput=1
\documentclass[
    aps,pre,twocolumn,
    reprint,
    superscriptaddress,
    footinbib,
    floatfix,
    amssymb,
    longbibliography
]{revtex4-2}

\usepackage[utf8]{inputenc}
\DeclareUnicodeCharacter{2212}{−}

\usepackage{amsmath}
\usepackage{amssymb}
\usepackage{siunitx}
\usepackage{bm}
\usepackage{dsfont}

\usepackage{graphicx}
\usepackage{nicefrac}

\usepackage[table,dvipsnames]{xcolor}
\usepackage{multirow}
\usepackage{array}
\usepackage{booktabs}

\usepackage{rotating}

\definecolor{linkColor}{rgb}{0,0.3,0.7}
\usepackage{hyperref}
\hypersetup{
    colorlinks=true,
    allcolors=linkColor,
    pdfborder={0 0 0},
    pdfencoding=auto
}

\usepackage{etoolbox}
\usepackage[normalem]{ulem}  

\definecolor{myGreen}{RGB}{19,132,23}
\definecolor{highlightColor}{RGB}{200,0,0}
\definecolor{lightgray}{rgb}{0.75, 0.75, 0.75}

\definecolor{colA}{HTML}{E8E0F0}
\definecolor{colAtxt}{HTML}{3C3489}
\definecolor{colM}{HTML}{D4F0E4}
\definecolor{colMtxt}{HTML}{085041}
\definecolor{rowlbl}{HTML}{EEEEEE}
\definecolor{secbar}{HTML}{DCDCDC}
\definecolor{secbarII}{HTML}{E0E8EF}  %
\definecolor{critbg}{HTML}{F5F5F5}
\definecolor{chk}{HTML}{0F6E56}
\definecolor{crs}{HTML}{A32D2D}
\definecolor{bbctxt}{HTML}{888888}

\usepackage{pgfplots}
\pgfplotsset{width=7cm,compat=newest}
\pgfmathdeclarefunction{gauss}{2}{%
  \pgfmathparse{1/(#2*sqrt(2*pi))*exp(-((x-#1)^2)/(2*#2^2))}%
}
\usepgfplotslibrary{groupplots,dateplot}
\usepgfmodule{plot}

\usetikzlibrary{
    arrows,
    arrows.meta,
    calc,
    decorations.markings,
    decorations.pathreplacing,
    fit,
    patterns,
    positioning,
    shapes.arrows
}

\tikzset{
 big arrow/.style={
   decoration={markings,mark=at position 1 with {\arrow[scale=2.5,#1]{>}}},
   postaction={decorate},
   shorten >=0.4pt},
 big arrow/.default=black}

\tikzset{
   set arrow inside/.code={\pgfqkeys{/tikz/arrow inside}{#1}},
   set arrow inside={end/.initial=>, opt/.initial=},
   /pgf/decoration/Mark/.style={
       mark/.expanded=at position #1 with
       {
           \noexpand\arrow[\pgfkeysvalueof{/tikz/arrow inside/opt}]{\pgfkeysvalueof{/tikz/arrow inside/end}}
       }
   },
   arrow inside/.style 2 args={
       set arrow inside={#1},
       postaction={
           decorate,decoration={
               markings,Mark/.list={#2}
           }
       }
   },
}

\begin{document}

\title{Classification of Intracellular Protein Patterns from Reactive Equilibria}

\author{Henrik Weyer}
\thanks{H.W.\ and C.Y.L.\ contributed equally to this work.}
\affiliation{Arnold Sommerfeld Center for Theoretical Physics and Center for NanoScience, Department of Physics, Ludwig-Maximilians-Universit\"at M\"unchen, Theresienstra\ss e 37, D-80333 M\"unchen, Germany}
\affiliation{Present address: Kavli Institute for Theoretical Physics, University of California Santa Barbara,
Santa Barbara, CA~93106, USA}
\author{Ching Yee Leung}
\thanks{H.W.\ and C.Y.L.\ contributed equally to this work.}
\affiliation{Arnold Sommerfeld Center for Theoretical Physics and Center for NanoScience, Department of Physics, Ludwig-Maximilians-Universit\"at M\"unchen, Theresienstra\ss e 37, D-80333 M\"unchen, Germany}
\author{Erwin Frey}
\email{frey@lmu.de}
\affiliation{Arnold Sommerfeld Center for Theoretical Physics and Center for NanoScience, Department of Physics, Ludwig-Maximilians-Universit\"at M\"unchen, Theresienstra\ss e 37, D-80333 M\"unchen, Germany}
\affiliation{Max Planck School Matter to Life, Hofgartenstra\ss e 8, D-80539 Munich, Germany}

\date{\today}
	
\begin{abstract}
Self-organized spatial patterns are central to nonequilibrium physics and cell biology, yet locating instabilities in multi-component, reaction-diffusion networks remains challenging because standard eigenvalue analyses scale with the number of biochemical states and rely on reaction kinetics often poorly constrained by experiments.
Exploiting the common mass-conserving structure of protein reaction kinetics and the fact that nonlinear feedback is typically confined to membrane reactions while lateral membrane diffusion is negligible,
we develop a geometric classification that predicts stationary, and approximately also oscillatory, pattern-forming instabilities from reactive equilibria.
The developed criteria reduce the stability analysis from the full component space to the space of conserved species.
On this reduced space, slope matrices---describing the change of equilibrium cytosolic densities with respect to total species densities---govern onset.
Based on densities in chemical equilibrium, this approach eliminates the requirement of comprehensive kinetic knowledge frequently lacking in experimental systems.
Thus, a broad range of instabilities can be understood as \emph{mass-redistribution instabilities}---self-amplifying mass redistribution caused by shifting local equilibria.
We apply these criteria to models for the \textit{Escherichia coli} Min system and the \textit{Caenorhabditis elegans} polarity system and show that
the reduction extends to the mixed-dimensional dynamics in systems coupling bulk cytosolic dynamics with membrane dynamics on the boundary.
Together, these results provide an interpretable, broadly applicable, and experimentally accessible framework for diagnosing and designing pattern formation in multicomponent nonequilibrium systems on the basis of conservation laws.
\end{abstract}

\maketitle

\section{Introduction}
Determining when a homogeneous state first becomes unstable is a central problem across nonequilibrium physics, materials science, and the physics of living systems. 
The loss of stability against non-uniform perturbations marks the onset of pattern formation, spontaneous flows, and phase--separation--like demixing in settings ranging from equilibrium matter to active matter and reaction--diffusion networks.
In equilibrium, such thresholds can be inferred from a thermodynamic potential, but nonequilibrium steady states do not obey detailed balance and are in general not characterized by a free-energy landscape whose local geometry decides stability.
Instead, couplings generated by advection, active stresses, or reaction fluxes are generically non-variational and may involve multiple slow fields that interact through non-symmetric mobility and reaction matrices.
Accordingly, locating the threshold of instability is a \emph{dynamical question of linear response}: one must determine whether infinitesimal perturbations about a homogeneous steady state grow or decay in time. 

Performing a \emph{linear stability analysis} (LSA) requires a fully specified dynamic model for the fields ${\mathbf u(\mathbf x,t)}$, for instance, an active matter or reaction–diffusion field theory.
Linearizing around a homogeneous steady state leads, in an $N$-component system, to an eigenvalue problem for a ${N\times N}$-matrix, which is typically resolved case-by-case and thus in a highly model-specific manner, offering little structural insight beyond the particular model at hand.

This challenge is particularly acute for intracellular protein networks, not only because the number of components in biological networks can be very high, but also because the details of the interactions may not be fully known.
Predicting which \emph{functions} protein interaction networks perform is central not only to understanding cell biology, but also to engineering bottom-up biochemical networks with life-like capabilities (e.g., synthetic cells). 
A particularly intriguing function is spatiotemporal self-organization driven by NTPase cycles: reaction-diffusion mechanisms convert chemical free energy into intracellular patterns that orchestrate downstream processes and can, in bottom-up reconstitutions,  generate shape transformations~\cite{Litschel.etal2018,Reverte-Lopez.etal2024} and autonomous motility~\citep{Fu.etal2023} of synthetic compartments.
However, even in view of these advances, we still lack broadly applicable criteria that predict \emph{when} patterns arise in realistic multi-component reaction--diffusion networks and identify which interactions control and drive protein pattern formation.

In thermal equilibrium systems, the onset of instability and symmetry breaking can be inferred from an analysis of the free energy landscape alone without reference to the dynamics. 
For example, in multi-component mixtures, the onset of phase separation (loss of local stability) can be determined from the loss of convexity of the free energy density $f(\boldsymbol{\phi})$ in composition space $\boldsymbol{\phi}$.
Phase separation occurs if the Hessian
\begin{align}
H_{ij}(\boldsymbol{\phi}) \;=\; \frac{\partial^2 f(\boldsymbol{\phi})}{\partial \phi_i\,\partial \phi_j}
\end{align}
is no longer positive definite~\citep{Chaikin.Lubensky1995}; see also Ref.~\citep{Mao.etal2019} for a recent discussion of the resulting high-dimensional phase behavior.
Beyond diagnosing instability, the geometry of the free energy landscape also determines coexistence—binodals, tie-lines, and critical points—via the common-tangent (hyperplane) construction, independent of kinetics.
No comparably universal, geometry-based classification exists yet for non-equilibrium systems. 

Existing approaches for reaction-diffusion systems try to connect the onset of lateral instability to specific properties of the linearized dynamics or to the topology of the reaction networks. 
Algebraic tests based on the principal minors of the reaction Jacobian provide conditions for stationary Turing instabilities~\citep{Satnoianu.etal2000,Villar-Sepulveda.Champneys2023}, and sparsity results quantify how many nonzero Jacobian entries are required for Turing onset~\citep{Hambric.etal2022}. 
Graph-theoretic analyses identify ``destabilizing subgraphs'' in the reaction Jacobian as the structural signatures of stationary Turing instabilities~\citep{Mincheva.Roussel2006,Diego.etal2018}.
Moreover, it has been shown how the diffusion matrix must be structured to induce lateral instabilities given a reaction Jacobian, and vice versa, how the Jacobian must be structured given a diffusion matrix \cite{Villar-Sepulveda.etal2025}.
Together, these works show that multi-component reaction–diffusion systems destabilize via a destabilizing (activator-like) module coupled to a stabilizing (inhibitor- or substrate-like) module whose effective diffusion is faster; thus activator/inhibitor/substrate labels attach to modules rather than to single components, unlike in classical two-component models~\citep{Gierer.Meinhardt1972,Murray2003,Landge.etal2020}. 
In practice, the principal-minor and graph-theoretic routes grow rapidly in complexity with component number, and whether a given parameter set is unstable hinges on entries of the reaction Jacobian (linearized reaction rates), which are highly nonlinear functions of the underlying kinetics. 
A useful simplification for \emph{stationary} Turing instabilities reduces the analysis to the diffusive species—eliminating non-diffusers and shrinking the eigenvalue problem significantly when they are numerous~\citep{Smith.Dalchau2018}. 
Moreover, complementary studies try to relate instability to network topology by numerically screening multi-species model classes~\citep{Marcon.etal2016,Zheng.etal2016,Scholes.etal2019, Haas.Goldstein2021} and by developing efficient parameter-screening algorithms~\citep{Solomatina.etal2022}; by construction, these adopt particular reaction forms (e.g., linear or Hill-type reactions). 

Experimentally, comprehensive interaction networks are frequently not known, and
to complicate matters, intracellular kinetic rates are difficult to measure.
In contrast, equilibrium concentrations as functions of the conserved total densities (steady-state titrations) are often more accessible.
Hence, criteria for \emph{dynamic} instability purely in terms of \emph{steady-state} properties of the interaction networks---akin to the curvature criterion for equilibrium phase separation---would be a critical advance toward an experimentally applicable classification of biological pattern formation.

Here, we show that mass conservation of the different protein species---together with nonlinear reactions and slow diffusion of membrane-bound components---allows the stability problem to be recast in geometric terms: stability of the homogeneous steady states relates to the changes of the chemical equilibrium densities with respect to the total species densities; in other words, the slopes of the reactive-equilibrium (nullcline) manifold.
Concretely, we analyze multi-component mass-conserving (McRD) models describing intracellular pattern formation~\citep{Jilkine.Edelstein-Keshet2011,Trong.etal2014,Halatek.etal2018,Frey.Brauns2022,Burkart.etal2022a, Frey.Weyer2026}, for which such a geometric criterion has previously been shown for two-component systems \cite{Brauns.etal2020} and to naturally hold in the long-wavelength limit due to quasi-stationary dynamics \cite{Brauns.etal2021}.
Here, we derive a geometric formulation for arbitrary stationary instabilities by considering characteristic properties of the underlying protein interaction networks.
If cytosolic densities change mildly compared to membrane densities, our approach also leads to approximate criteria for weakly oscillatory pattern-forming instabilities.
Thus, our theory shows that pattern formation in a broad range of McRD systems is driven by mass-redistribution instabilities, that is, positive feedback in mass redistribution driven by shifting local equilibria, and independent of local reaction dynamics.
We show that when some species have multiple cytosolic states, the homogeneous state can first lose stability at a finite wavelength related to the diffusion length set by the rate of conversion between the cytosolic components. 
As a corollary, if membrane diffusion is negligible and each species has a single cytosolic state, stationary onset always occurs at zero wavenumber.
Our formulation collapses the analysis from all biochemical states to the space of \emph{conserved species} (conservation laws). 
As a result, the criteria do not rely on a full specification of the interaction network, the reaction rate laws, or the reaction Jacobian. 
Rather, we tie the dynamic instability directly to properties of the reactive equilibria, making the criteria experimentally accessible from measurements in steady state.
Taken together, the derived criteria mirror the curvature criterion for spinodal decomposition in multi-component equilibrium mixtures, but here diagnose the far-from-equilibrium, NTPase-driven patterning of reaction--diffusion systems. 

To isolate the physics of mass redistribution, our framework builds on two generic features of intracellular protein pattern formation~\citep{Frey.Brauns2022,Burkart.etal2022a, Frey.Weyer2026}: 
(i) patterns arise from NTPase-driven attachment--detachment cycles at membranes, where membrane-bound states diffuse much more slowly than cytosolic states; and
(ii) nonlinear feedback is localized to the membrane (where protein densities are high), whereas the cytosol mainly harbors conversion reactions that convert proteins into binding-competent states (possibly nonlinear, e.g., dimerization) but do not themselves provide feedback.

The remainder of this paper is organized as follows. 
We begin by illustrating the mass-conserving reaction–diffusion (McRD) framework for intracellular protein networks, defining conserved total densities, reactive-equilibria (nullcline) manifolds, and mass-redistribution potentials (Sec.~\ref{sec:RDS}). We then revisit the long-wavelength (mass-redistribution) instability: starting from a minimal example and generalizing to multiple species, we recover the LQSS nullcline-slope criterion and its specialization to one cytosolic component per species (Sec.~\ref{sec:long-wavelength-limit}). Section \ref{sec:3c} extends the analysis to stationary finite-wavelength onset by constructing marginal modes at ${q>0}$ and deriving a solvability condition that links short-wave instability to the slopes of local reactive equilibria. The full multi-species, multi-component generalization, including compact matrix formulations and the effects of explicit bulk-boundary coupling with diffusive filtering, is developed in Appendix \ref{app:multi-cyt-components} and Appendix \ref{app:bbc}. Building on these results, Section \ref{sec:classification} presents a species-level classification for multi-species McRD systems, stating sufficient tests and exact conditions in terms of slope matrices, together with the assumptions under which they apply. Building on the finite-$q$ criteria. Section~\ref{sec:approx-dispRel-oscillatory} develops approximate dispersion relations, treating oscillatory onsets, and quantifying the impact of membrane diffusion. These results set the stage for Section \ref{sec:application}, which applies the classification to the \textit{Escherichia coli} Min and the \textit{Caenorhabditis elegans} polarity (PAR) networks. We thereby show how these two biochemical systems lead to negative slope eigenvalues through distinct routes. Section \ref{sec:conclusion} then summarizes the framework and outlines experimental applications and theoretical extensions. Technical details are collected in the Appendices: Appendix \ref{app:model} specifies the MinDE and PAR models used for quantitative comparison with simulations. Appendix \ref{app:multi-cyt-components} gives the general multi-species/multi-component finite-$q$ criteria (including matrix formulations), and Appendix \ref{app:bbc} treats explicit bulk–boundary coupling.

\section{Mass-conserving reaction--diffusion systems}
\label{sec:RDS}

\begin{figure}[t]
\centering
\includegraphics[width=\columnwidth]{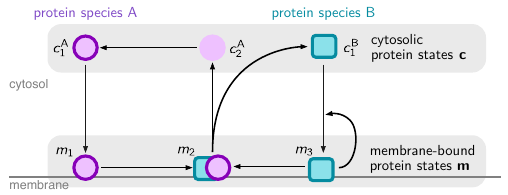}
\caption{
Schematic of a two-species membrane--cytosol reaction network exemplifying the multi-component systems considered in this work.
Species~A (left, magenta circles) and species~B (right, cyan squares) each comprise membrane-bound states~$\mathbf{m}$ (top, shaded region) and cytosolic states~$\mathbf{c}$ (bottom).
Vertical arrows indicate attachment to and detachment from the membrane; horizontal arrows indicate interconversion among membrane-bound states (${m_1 \leftrightarrow m_2 \leftrightarrow m_3}$) and among cytosolic states (e.g., ${c_2^{A} \to c_1^{A}}$).
For each species, a single cytosolic state ($c_1^{A}$, $c_1^{B}$) is binding-competent, i.e., capable of attaching to the membrane; all other cytosolic states are binding-incompetent.
The illustrated network contains ${N_{\mathrm{species}}=2}$ protein species but ${N_{\mathrm{comp}}=6}$ components: ${\mathbf{m}=(m_1,m_2,m_3)}$ and ${\mathbf{c}=(c_1^{A},c_2^{A},c_1^{B})}$, where $c_i^{\alpha}$ denotes the $i$th cytosolic state of species~$\alpha$.
}
\label{fig:cartoonNetwork}
\end{figure}

To analyze intracellular protein pattern formation driven by the interplay of diffusion and mass-conserving biochemical reactions, we consider the spatio-temporal dynamics of the density fields $\mathbf{u}(\mathbf{x},t)$.
Each component of this vector field represents a distinct biochemical state of a protein species or complex [Fig.~\ref{fig:cartoonNetwork}].
In a cell biology context, the components of $\mathbf{u}$ naturally separate into membrane-bound states $\mathbf{m}$ and cytosolic states $\mathbf{c}$, such that
\begin{align}
    \mathbf{u} (\mathbf{x},t)
    = 
    \big(
    \mathbf{m}(\mathbf{x},t), \mathbf{c}(\mathbf{x},t)
    \big)^{\!\top}
    \, .
\end{align}
Because proteins can adopt multiple such states, the total number of components $N_\mathrm{comp}$ is generally much larger than the number of underlying protein species $N_\mathrm{species}$, which we label by a Greek index~${\alpha=\mathrm{A}, \mathrm{B}, \dots}$.
 
In intracellular pattern-forming systems, proteins cycle between the cell membrane—a two-dimensional surface—and the cytosol—a three-dimensional volume~\citep{Frey.Weyer2026}. Because the membrane is a surface, the relevant protein densities on it are areal (number per area), whereas cytosolic concentrations are volumetric (number per volume). For a given total protein copy number, this difference in dimensionality implies that effective densities on the membrane can exceed those in the cytosol by a factor on the order of the bulk-to-surface ratio. The resulting high densities for bulk heights above a few micrometers favor bimolecular and higher-order reactions among membrane-bound species, while making such encounters comparatively rare in the cytosol.

Cytosolic components are therefore simply different conformational or oligomeric states of individual species.
A typical example are NTPases that undergo nucleotide exchange between an inactive NDP-bound and an active NTP-bound state.
For each species~$\alpha$, we denote the $n_\alpha$ cytosolic state concentrations by $c_1^\alpha, \dots, c_{n_\alpha}^\alpha$.
Because each cytosolic component belongs to exactly one species, $\mathbf{c}$ decomposes into species-specific blocks,
\begin{align}
    \mathbf{c} = (c^\mathrm{A}_1,\dots,c^\mathrm{A}_{n_\mathrm{A}},c^\mathrm{B}_1,\dots)
    \, .
\end{align}
In contrast, membrane-bound proteins can form heteromeric complexes involving multiple species---for example, the MinDE complex in which MinE stimulates the ATPase activity of MinD~\citep{Frey.Weyer2026}.
Membrane-bound components are therefore indexed by a single label~$m_j$ without a species superscript; the stoichiometric factors introduced below [cf.\ Eq.~\eqref{eq:def-tot-dens}] track the species content of each membrane component.
 
Throughout, we assume that for each species exactly one cytosolic state is \emph{binding-competent}, i.e., capable of attaching to the membrane; all other cytosolic states are binding-incompetent.
We collect the binding-competent concentrations in the vector ${\mathbf{c}_1 = (c_1^\mathrm{A},c_1^\mathrm{B},\dots)}$, so that the membrane reaction term depends only on the membrane-bound densities and the binding-competent cytosolic components, ${\mathbf R_m(\mathbf{u}) = \mathbf R_m(\mathbf{m},\mathbf{c}_1)}$.
We demonstrate in Sec.~\ref{sec:classification} that this assumption captures the common structure of a broad range of models for intracellular protein patterning.
 
As a concrete instance, in the skeleton model of the \textit{Escherichia coli} Min protein system~\citep{Huang.etal2003, Fange.Elf2006, Halatek.Frey2012}, the binding-competent states are ATP-bound MinD and cytosolic MinE, ${\mathbf{c}_1 = (c_{\text{D}^{\mathrm{ATP}}},\,c_\text{E})^\top}$, and the membrane states are ${\mathbf{m}=(m_\text{D},\,m_\text{DE})^\top}$, describing membrane-bound MinD and MinDE complexes.
We will exemplify our theory using the Min system as well as a second biological example in Sec.~\ref{sec:application} and describe the models in more detail in App.~\ref{app:model}.
 
The components of ${\mathbf{u}}$ are coupled through nonlinear reaction terms that account for conformational switching, binding and unbinding interactions, and membrane attachment or detachment (Fig.~\ref{fig:cartoonNetwork}).
Together with diffusion, these mass-conserving reactions form the reaction--diffusion network that underlies spatial protein pattern formation.
Assuming that membrane diffusion is negligible compared to cytosolic diffusion, the dynamics reduce to
\begin{subequations}
\label{eq:RDS}
\begin{align}
    \partial_t \mathbf{m} (\mathbf{x},t) &= {\mathbf R}_m(\mathbf{u})
    \,,
    \label{eq:RDS-mem}\\
    \partial_t \mathbf{c} (\mathbf{x},t) &= \mathbf{D}_c \boldsymbol{\nabla}^2 \mathbf{c} + {\mathbf R}_c (\mathbf{u})
    \ ,
    \label{eq:RDS-cyt}
\end{align}
\end{subequations}
defined on the $d$-dimensional domain $\Omega\subseteq\mathbb{R}^d$ with no-flux or periodic boundary conditions.
The diagonal matrix
\begin{align}
    \mathbf{D}_c \equiv \mathrm{diag}\big(D_{c,1}^{\mathrm A},\dots,D_{c,n_{\mathrm A}}^{\mathrm A},\ D_{c,1}^{\mathrm B},\dots\big)
\end{align}
contains the cytosolic diffusion coefficients, where $D_{c,i}^\alpha$ is the diffusion coefficient of the $i$th cytosolic state of species~$\alpha$.
The vector ${{\mathbf R} = ({\mathbf R}_m,{\mathbf R}_c)^{\!\top}}$ describes the local reaction kinetics of all individual components.
For simplicity, we consider a thin layer approximation and neglect density variations in the direction perpendicular to the membrane~\citep{Halatek.Frey2018}.
In Sec.~\ref{sec:classification} and App.~\ref{app:bbc}, we show how this simplified framework generalizes to systems with explicit bulk-boundary coupling, where cytosolic dynamics in the volume perpendicular to the membrane are fully accounted for.

\subsection{Mass conservation and mass-redistribution potential}
\label{sec:mass-conservation}
 
\smallskip
 
\emph{Stoichiometric factors.---} To specify how many molecules of each species $\alpha$ are distributed across the components ${\mathbf{u} = (\mathbf{m},\mathbf{c})^{\! \top}}$, we introduce stoichiometric vectors~\footnote{
As an illustration, consider a system with two protein species A and B with component vector ${\mathbf{u}=(m_\text{AA}, \,  \, m_\text{AB}, \, c_\text{A}, \, c_\text{B})^{\! \top}}$: a cytosolic monomer $c_\text{A}$, a membrane dimer $m_{AA}$, a cytosolic monomer $c_\text{B}$, and a heteromeric membrane complex $m_\text{AB}$ containing one species A and one species B. The stoichiometric vectors are ${\mathbf{s}_\text{A}=(2,1,1,0)^{\! \top}}$ and ${\mathbf{s}_\text{B}=(0,1,0,1)^{\! \top}}$, giving the total species densities ${\rho_\text{A}=2m_\text{AA}+m_\text{AB}+c_\text{A}}$ and ${\rho_\text{B}=m_\text{AB}+c_\text{B}}$. This example shows how $\mathbf{s}_\alpha$ accounts for both higher-order states and heteromeric complexes.
} 
${\mathbf{s}_\alpha = (\mathbf{s}_m^\alpha, \mathbf{s}_c^\alpha)}^{\! \top}$, whose entries count how many molecules of species $\alpha$ are contained in the respective membrane-bound or cytosolic component~\footnote{The species-specific block structure of $\mathbf{c}$ introduced above can be expressed as an orthogonality condition on the cytosolic stoichiometric vectors: $\mathbf{s}_c^\alpha\cdot \mathbf{s}_c^\beta=0$ for all $\alpha \neq \beta$.}.
The total density $\rho_\alpha$ of species $\alpha$ is then given by
\begin{equation}\label{eq:def-tot-dens}
    \rho_\alpha 
    = 
    \mathbf{s}_\alpha \cdot \mathbf{u}
    = \mathbf{s}_m^\alpha \cdot \mathbf{m}
    + \mathbf{s}_c^\alpha \cdot \mathbf{c}
    \, ,
\end{equation}
where $\mathbf{s}_m^\alpha$ and $\mathbf{s}_c^\alpha$ denote the stoichiometric factors for the membrane and cytosolic components, respectively.
 
\smallskip
 
\emph{Continuity equations.---} The dynamics of the total densities ${\rho_\alpha\equiv\mathbf s_\alpha\cdot\mathbf u}$ of each species follow directly from taking the inner product of the stoichiometric vector $\mathbf{s}_\alpha$ with the evolution equations, Eq.~\eqref{eq:RDS},
\begin{equation}
    \partial_t \rho_\alpha (\mathbf{x},t) = \mathbf{s}^\alpha_c \cdot \mathbf{D}_c \boldsymbol{\nabla}^2 \mathbf{c} (\mathbf{x},t)
    \, .
\label{eq:continuity_general}    
\end{equation}
Here we use the fact that, to conserve the total number of molecules of each species ${\alpha}$, the reaction terms must satisfy molecular balance,
\begin{equation}\label{eq:reacTerm-conservationConstraint}
    \mathbf{s}_\alpha \cdot {\mathbf R} 
    = 0 \,,
    \quad \text{for all } \alpha \, .
\end{equation}
Under no-flux or periodic boundary conditions this implies conservation of the spatial averages
\begin{align}  
    \bar{\rho}_\alpha 
    \equiv 
    \frac{1}{|\Omega|} \int_\Omega \mathrm{d}^d x\, \rho_\alpha \, .
\end{align}

\paragraph*{Mass-redistribution potentials.---} The continuity equations, Eq.~\eqref{eq:continuity_general}, introduce the mass-redistribution current $\mathbf j_\alpha(\mathbf x,t)$ for species $\alpha$ by
\begin{equation}
    \partial_t \rho_\alpha(\mathbf x,t)
    = -\,\boldsymbol{\nabla}\!\cdot \mathbf j_\alpha(\mathbf x,t)\,.
\end{equation}
Consequently, the current is
(in the absence of membrane diffusion) a linear combination of gradients in the cytosolic densities
\begin{equation}
    \mathbf j_\alpha(\mathbf x,t)
    = - \sum_{i=1}^{n_\alpha} s_{\mathrm c,i}^\alpha\, D_{\mathrm c,i}^\alpha \,\boldsymbol{\nabla} c_i^\alpha(\mathbf x,t)\,,
\end{equation}
where the sum runs over the cytosolic states $i$ of species $\alpha$; $D_{\mathrm c,i}^\alpha$ are cytosolic diffusion coefficients; and ${s_{\mathrm c,i}^\alpha}$ are stoichiometric factors counting how many molecules of species $\alpha$ are present in cytosolic component $i$, as introduced above [cf.\ Eq.~\eqref{eq:def-tot-dens}].
 
To express the current compactly, we define the \emph{mass-redistribution potentials}~\footnote{
If membrane diffusion is finite, the mass-redistribution potentials include additional contributions from the membrane components, weighted by their respective diffusion coefficients; see Sec.~\ref{sec:approx-dispRel-oscillatory}.}
\begin{align}
    \eta_\alpha(\mathbf x,t)
    \equiv
    \sum_{i=1}^{n_\alpha} s_{c,i}^\alpha\,\frac{D_{c,i}^\alpha}{D_{c,1}^\alpha}\,c_i^\alpha(\mathbf x,t)\,,
\label{eq:eta_general}
\end{align}
where the diffusion coefficient $D_{c,1}^\alpha$ of the binding-competent state serves as reference, so that $\eta_\alpha$ has units of concentration.
Physically, $\eta_\alpha$ is the diffusion-weighted cytosolic density of species~$\alpha$: each cytosolic state contributes in proportion to its diffusive mobility, making $\eta_\alpha$ the effective concentration available for lateral mass transport.
In terms of $\eta_\alpha$, the current becomes
\begin{equation}
    \mathbf j_\alpha(\mathbf x,t) = -\,D_{c,1}^\alpha\,\boldsymbol{\nabla}\eta_\alpha(\mathbf x,t),
\end{equation}
and the continuity equation takes the form~\citep{Otsuji.etal2007,Ishihara.etal2007,Halatek.Frey2018,Brauns.etal2020,Brauns.etal2021,Brauns.etal2021b}
\begin{equation}
\label{eq:cont-eqs}
    \partial_t \rho_\alpha(\mathbf x,t) = D_{c,1}^\alpha\,\boldsymbol{\nabla}^2 \eta_\alpha(\mathbf x,t) \, .
\end{equation}
Thus, gradients in the scalar potentials $\eta_\alpha$ fully encode how cytosolic diffusion redistributes the conserved total densities $\rho_\alpha$, and they will play a central role in the analysis of the spatial redistribution instabilities below.

\subsection{ Reactive equilibria (nullcline)  manifolds}
\label{sec:reactive_equilibria}

Because reactions conserve the total density of each species [cf.\ Eq.~\eqref{eq:reacTerm-conservationConstraint}], the total densities $\bm{\rho}$ are fixed by the initial conditions---by how many proteins of each species one mixes together---and remain unchanged by the local reaction kinetics.
This constraint implies that
the reactive dynamics evolves within the \emph{mass--conservation subspaces}:
\begin{equation}
    \mathcal M(\bm\rho)
    \equiv \big\{\, \mathbf u=(\mathbf m,\mathbf c)^\top \;\big|\; \mathbf{S}\,\mathbf u=\bm\rho \,\big\},
\end{equation}
where $\mathbf{S}$ is the constant stoichiometric matrix whose rows are the species stoichiometric vectors ${\mathbf s_\alpha^\top}$ and ${\bm{\rho}= (\rho_\mathrm{A},\rho_\mathrm{B}, \dots)^{\top}}$.
Thus, ${\mathcal M(\bm\rho)}$ are the subspaces of the phase space in which the total densities of all species remain constant at values $\bm\rho$.

The reactive flow within each mass-conservation subspace will be characterized by a set of reactive equilibria $\mathcal E^*(\bm\rho)$ at which the flow vanishes, that is, it holds ${{\mathbf R}(\mathbf u)=0}$ at the equilibria.
The densities $\mathbf{u}$ of these reactive equilibria will smoothly change as the total densities $\bm{\rho}$, that is, the mass-conservation subspace, is varied.
Collecting this family of equilibria, we define the \emph{reactive equilibrium (nullcline) manifold} 
\begin{equation}
    \mathcal N
    \equiv \big\{\, \mathbf u \;\big|\; {\mathbf R}(\mathbf u)=0 \,\big\}.
\end{equation}
The set of reactive equilibria at total densities ${\bm\rho}$ is the intersection (see Sec.~\ref{sec:heuristics_2cmcrd} for an example)
\begin{equation}
    \mathcal E^*(\bm\rho)
    = \mathcal M(\bm\rho)\cap \mathcal N.
\end{equation}
When this intersection is a singleton, we write the \emph{nullcline manifold} as
\begin{equation}
\mathbf u^*(\bm\rho)=(\mathbf m^*(\bm\rho),\mathbf c^*(\bm\rho))^{\!\top} \, ;
\end{equation}
otherwise we index the different branches of this manifold by ${k}$ as ${\mathbf u_k^*(\bm\rho)}$.

The reactive equilibria ${\mathbf u_k^*(\bm\rho)}$ are the \emph{homogeneous steady states} of the spatially extended reaction-diffusion system given initially fixed total densities $\bm{\rho}$.
To understand their pattern-forming instabilities in the spatially extended setting, we take the homogeneously stable equilibria as the base states and ask whether they are unstable against spatially non-uniform perturbations inducing diffusion-mediated mass redistribution.
If multiple equilibria exist for given ${\bm\rho}$ (multistability), the linear stability of each candidate must be assessed separately.
Away from bifurcation points, each single equilibrium lies on one of $k$ smooth solution branches ${\mathbf u_k^*(\bm\rho)}$ that vary continuously with the conserved total densities.
Therefore, we will drop the index $k$ for simplicity and denote by $\mathbf u^*(\bm\rho)$ the nullcline branch that contains the homogeneous steady state under consideration.

\subsection{Linear stability analysis}
\label{sec:lsa}

We fix the total densities $\bm\rho=\bm\rho_\mathrm{hss}$ and select as base state one homogeneous steady state (hss) to assess its linear stability
\begin{align}
    \mathbf u_{\mathrm{hss}}=\mathbf{u}^*(\bm{\rho}_\mathrm{hss}) 
    \in
    \mathcal E^*(\bm\rho_\mathrm{hss})=\mathcal M(\bm\rho_\mathrm{hss})\cap\mathcal N \, .
\end{align}
Assuming that the state is stable against uniform perturbations, we ask whether it is destabilized by spatially varying perturbations.
Linearizing the reaction--diffusion dynamics Eq.~\eqref{eq:RDS} about $\mathbf u_{\mathrm{hss}}$ gives
\begin{equation}
    \partial_t \,
    \delta\mathbf u
    = 
    \big(\mathbf{J} - \mathbf{D}\,\nabla^2\big)\,\delta\mathbf u
    \, ,
\end{equation}
with the Jacobian of the reaction term~\footnote{%
We define the Jacobian matrix $\partial_{\mathbf{y}}\mathbf{w}$ of a vector-valued function $\mathbf{w}$ with respect to the vector argument $\mathbf{y}$ componentwise by
${\big[ \partial_{\mathbf{y}}\mathbf{w} \big]_{\alpha\beta} \equiv \frac{\partial w_\alpha}{\partial y_\beta} \equiv \partial_{y_\beta} w_\alpha}$.} 
\begin{equation}
   \mathbf{J} \equiv \partial_{\mathbf u}{\mathbf R}\big|_\mathrm{hss} \, . 
\end{equation}
Here, $\mathbf{D}$ is the diagonal diffusion matrix
\begin{equation}
    \mathbf{D} \;\equiv\;
    \begin{pmatrix}
        \mathbf 0 & \mathbf 0 \\
        \mathbf 0 & \mathbf{D}_c
    \end{pmatrix} \, ,
\end{equation}
where $\mathbf{D}_c$ is the diagonal matrix of cytosolic diffusion coefficients introduced above, i.e., no Laplacian acts on membrane components, while cytosolic components ${c_i^\alpha}$ diffuse with coefficients ${D_{c,i}^\alpha}$.

\smallskip

\emph{Eigenmodes and growth rates.---}
With the eigenmode ansatz ${\delta\mathbf u =  \delta\mathbf u_p \, \mathrm{e}^{\sigma_p(q)\,t + i \mathbf q\cdot\mathbf x}}$, where ${\mathbf q\in\mathbb R^d}$ is the wavevector and ${q\equiv\|\mathbf q\|}$ the wavenumber, we obtain the eigenvalue problem
\begin{equation}
 \label{eq:LSA}   \sigma_p(q)\,\delta\mathbf u_p
    = \big(\mathbf{J} - \mathbf{D}\,q^2\big)\,\delta\mathbf u_p,
\end{equation}
where ${\sigma_p (q)\in\mathbb C}$  denotes the growth rates corresponding to the $p^\text{th}$ eigenvector $\delta \mathbf{u}_p(q)$; $p \in \{ 1, \ldots \, , N_\text{comp} \}$.
Lateral instability is signaled by ${\max_{q>0}\mathrm{Re}\,\sigma_p(q)>0}$, while stability to uniform perturbations in the well-mixed limit requires ${\mathrm{Re}\,\sigma_p(0)\leq}0$.
Because only the wavenumber $q$ appears in the eigenvalue problem, the linear stability analysis is independent of the spatial dimension $d$ of the domain.

Importantly, if the eigenvalue problem, that is, the Jacobian is high-dimensional, it is in general difficult to relate the sign of its eigenvalues to simple properties of the Jacobian, and thus of the reaction term ${\mathbf R}$~\citep{Villar-Sepulveda.Champneys2023,Villar-Sepulveda.etal2025,Mincheva.Roussel2006,Diego.etal2018}. 
In the following, we will show that mass conservation provides physical constraints that allow exact progress for stationary (see Secs.~\ref{sec:long-wavelength-limit},~\ref{sec:3c},~\ref{sec:classification}) and approximate progress for oscillatory instabilities (see Sec.~\ref{sec:approx-dispRel-oscillatory}).
Specifically, we analyze the zero crossings of the real part of the dispersion relation, building on and extending Ref.~\citep{Smith.Dalchau2018}.
We show that for mass-conserving systems, the zero crossings, and approximately also the dispersion relation, are determined by strongly simplified criteria based on geometric properties of the nullcline manifold, that is, the reactive equilibria.

\section{Mass-redistribution instability in the long-wavelength limit}
\label{sec:long-wavelength-limit}

To understand the mechanism of pattern formation in McRD systems, we first analyze long-wavelength instabilities.
At these scales, local reactions rapidly restore local reactive balance compared to the timescale of diffusive redistribution.
So, at each point of the spatially extended system, the component densities remain near the reactive equilibrium values set by the conserved total densities.
In this \emph{local quasi–steady-state} (LQSS) regime, the total densities obey a nonlinear diffusion law with an \emph{effective diffusion matrix} set by the \emph{slopes of the nullcline manifold}; instability occurs when this matrix acquires an eigenvalue with negative real part (anti-diffusion).

In the following, we first illustrate these concepts for a minimal setting: a single-species, two-component McRD system [Fig.~\ref{fig:2c-mass-redistribution}] \cite{Brauns.etal2020,Halatek.etal2018}. 
We then generalize the LQSS slope criterion to multi-species systems. 
Finally, we show that for models with a single cytosolic component per species, any non-oscillatory onset occurs at zero wavenumber, so the LQSS slope criterion is exact in such models.

\subsection{One-species system with two components}
\label{sec:heuristics_2cmcrd}

The simplest McRD model consists of a single protein species with one cytosolic and one membrane-bound component, described by the dynamic equations:
\begin{subequations}
\label{eq:2cMcRD}
\begin{align}
    \partial_t m 
    &= R_m(m,c)
    \,, \\
    \partial_t c 
    &= D_c \boldsymbol{\nabla}^2 c - R_m(m,c)
    \, ,
    \label{eq:2cMcRD-cytosolic}
\end{align}
\end{subequations}
with a single reaction term $R_m(m,c)$. 
Despite its simplicity, this model can exhibit pattern formation  and has been widely studied as a minimal model for cell polarity~\citep{Otsuji.etal2007,Altschuler.etal2008,Goryachev.Pokhilko2008,Mori.etal2008,Edelstein-Keshet.etal2013,Brauns.etal2020}.

\begin{figure}[!t]
\centering
\includegraphics[width=\columnwidth]{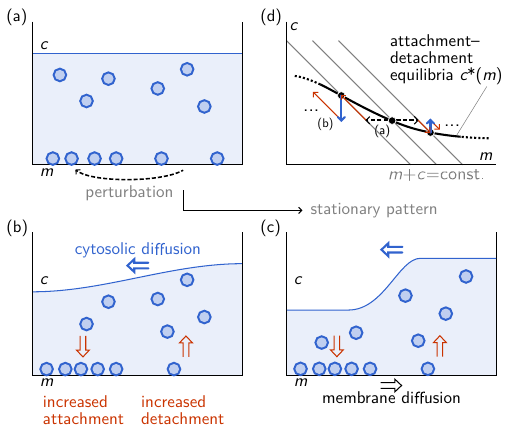}
\caption{
Illustration of the mass-redistribution instability in two-component mass-conserving reaction-diffusion systems; redrawn from Ref.~\citep{Halatek.etal2018}.
(a) The homogeneous steady state has uniform cytosolic density $c(\mathbf{x},t)$ and membrane density $m(\mathbf{x},t)$.
We consider a small perturbation that increases the membrane density on the left side of the system and reduces it on the right side.
(b) If the reactive equilibrium between attachment and detachment shifts in favor of attachment with an increased total density of the proteins, the cytosolic density decreases on the left and increases on the right (red vertical arrows).
The resulting gradient in the cytosolic density redistributes further proteins to the left (blue horizontal arrow), amplifying the initial perturbation. 
(c) The steady-state pattern is reached when weak membrane diffusion (black arrow) balances the cytosolic protein flux.
(d) The mass-redistribution instability occurs when the reaction nullcline, viewed as the curve
$c^*(m)$ in the $(m,c)$ plane (black solid curve), possesses a negative slope
$(\partial_m c^* < 0)$ at the steady state.
The initial perturbation (black dashed arrows) triggers reactive relaxation (red thin arrows) along the gray diagonal lines of constant local total density,
$m+c=\rho$, toward the reactive equilibria (black dots) corresponding to the changed total density in the left and right parts of the system [cf.\ panel (b)].
These reactive processes create cytosolic gradients that further amplify the initial density difference (blue thick arrows) [cf.\ panel (b)].
}
\label{fig:2c-mass-redistribution}
\end{figure}

Specializing the framework of Sec.~\ref{sec:RDS} to this two-component system, the reaction nullcline $\mathcal N$ is the curve $R_m(m,c)=0$ in the $(m,c)$ plane and the mass-conservation subspace $\mathcal M(\rho_\mathrm{hss})$ is the line $m+c=\rho_\mathrm{hss}$.
The homogeneous steady state(s) for a given total density is their intersection $\mathcal E^*(\rho_\mathrm{hss})$ [Fig.~\ref{fig:2c-mass-redistribution}(d)].
We work on the smooth nullcline branch $c^*(\rho)$ which contains the homogeneously stable hss that we seek to analyze in the following.

\smallskip 

\emph{Heuristic instability mechanism.---} The mechanism of pattern formation in this system can be illustrated as follows~\citep{Halatek.etal2018}.
Consider, for simplicity, a one-dimensional system where a small perturbation of the homogeneous steady state increases the membrane density on the left and decreases it on the right [Fig.~\ref{fig:2c-mass-redistribution}(a)].
If the increased membrane density shifts the reactive equilibrium between membrane attachment and detachment toward further attachment, then the cytosolic protein concentration decreases on the left while it increases on the right [Fig.~\ref{fig:2c-mass-redistribution}(b)].
Cytosolic diffusion then will redistribute even more proteins to the left.
As a result, more proteins attach to the membrane on the left and detach on the right, further enhancing the gradients in both cytosolic and membrane densities.
This self-enhancing feedback—mediated by the redistribution of mass through cytosolic diffusion—amplifies the initial perturbation and destabilizes the homogeneous steady state.
The final stationary pattern will be attained when the diffusive flux due to weak membrane diffusion (neglected in our analysis) balances the cytosolic diffusive flux [Fig.~\ref{fig:2c-mass-redistribution}(c)].
In the idealized limit of vanishing membrane diffusion, the cytosolic flux must vanish at steady state, so the cytosolic density becomes spatially uniform; the pattern then resides entirely in the membrane density, with bulk regions sitting at different equilibrium values of $m$ at the same shared cytosolic density. This is possible because the nonlinearity of the nullcline (\textsf{N}-shape) admits multiple membrane-equilibrium values at a single cytosolic density~\citep{Brauns.etal2020,Mori.etal2008}.
In summary, the condition for such a \textit{mass-redistribution instability} is that the cytosolic equilibrium density $c^*(\rho)$ decreases with the total density
\begin{align}
    \partial_\rho c^*(\rho)|_\mathrm{hss} 
    < 0 \, ,
\end{align} 
known as the \emph{nullcline-slope criterion} [Fig.~\ref{fig:2c-mass-redistribution}(d)] \citep{Brauns.etal2020}.

\smallskip 

\emph{Anti-diffusion.---} Analytically, the criterion for instability can also be derived by examining the dynamics of the total density ${\rho = m + c}$.
Summing the two-component equations yields the continuity equation:
\begin{equation}
    \partial_t \rho = D_c \boldsymbol{\nabla}^2 \eta,\quad\text{with}\quad \eta = c\,.
\end{equation}
In the regime of a long-wavelength instability, reactions occur much faster than diffusion. Under this adiabatic assumption---referred to in the following as the \textit{local quasi-steady-state (LQSS) approximation}---one replaces $\eta$ by the local reactive equilibrium ${\eta^*(\rho)= c^*(\rho)}$, defined implicitly by the nullcline condition ${R_m(m^*(\rho), c^*(\rho))=0}$. Thus, the dynamics of the total density reduce to a closed nonlinear diffusion equation:
\begin{equation}
    \partial_t \rho = D_c \boldsymbol{\nabla}^2 \eta^*(\rho) 
    = D_c \boldsymbol{\nabla} \cdot \left[\partial_\rho \eta^*(\rho) \boldsymbol{\nabla} \rho\right].
\end{equation}
This nonlinear diffusion process has an effective density-dependent diffusion coefficient:
\begin{equation}
    D_c^\mathrm{eff}(\rho) = D_c\,\partial_\rho \eta^*(\rho)\,.
\end{equation}
An instability arises precisely when this effective diffusion coefficient becomes negative (\textit{anti-diffusion}), i.e., when the slope of the equilibrium curve satisfies:
\begin{equation}
    \partial_\rho \eta^*(\rho) 
    = 
    \partial_\rho c^*(\rho)
    < 
    0 
    \,,
\end{equation}
in agreement with the previously derived slope criterion.

\subsection{Multi-species systems and LQSS nullcline slope criterion}
\label{sec:qss}

The local quasi-steady-state (LQSS) approximation introduced above for the two-component system can be generalized to arbitrary McRD systems involving multiple protein species and biochemical states~\citep{Brauns.etal2021}. 
As for the two-component system, this generalization builds on the idea that, in the long-wavelength limit, diffusive transport becomes slow relative to local reaction kinetics.
Under these conditions, the densities of all cytosolic and membrane-bound components rapidly equilibrate to their local reactive equilibria $\mathbf{u}^*(\bm{\rho})$. 
Therefore, we can approximate the mass-redistribution potentials by their values at the reactive equilibrium [cf.\ Eq.~\eqref{eq:eta_general}] 
\begin{align}
    \eta_\alpha^* (\bm{\rho}) 
    :=
    \sum_{i=1}^{n_\alpha} 
    s_{c,i}^\alpha\,
    \frac{D_{c,i}^\alpha}{D_{c,1}^\alpha}\,
    c_i^{*\alpha} (\bm{\rho}) 
    \,,
\end{align}
Then, the continuity equations for the mass densities [Eq.~\eqref{eq:cont-eqs}] become closed equations for the total protein densities $\rho_\alpha$:
\begin{equation}
    \partial_t \rho_\alpha (\mathbf{x},t)
    =
    \sum\nolimits_\beta \boldsymbol{\nabla} \cdot
    \big[ 
    D_{c,1}^\alpha \,  
    \partial_{\rho_\beta} \eta^*_\alpha (\mathbf{\bm\rho})\, 
    \boldsymbol{\nabla} \rho_\beta (\mathbf{x},t)\big]
\end{equation}
This is a \emph{nonlinear diffusion equation} 
\begin{equation}
    \partial_t \bm\rho 
    = \boldsymbol{\nabla}^2 
    \big[\,\mathbf{D}_{c,1}\,\boldsymbol\eta^*(\bm\rho)\,\big]
    = \boldsymbol{\nabla} \cdot
    \big[\,
    \mathbf{D}^{\mathrm{eff}}(\bm\rho)\,\bm\nabla \bm\rho
    \,\big]
    \, ,
\end{equation}
with the effective diffusion matrix
\begin{equation}
    D_{\alpha \beta}^\text{eff} (\bm{\rho})
    \;:=\;
    D_{c,1}^\alpha \,  
    \partial_{\rho_\beta} \eta^*_\alpha (\bm{\rho}) =
    \mathbf{D}_{c,1} \, \big[ \partial_{\bm\rho}\boldsymbol\eta^{*}(\bm\rho) \big]_{\alpha \beta}\,.
    \label{eq:Deff-def}
\end{equation}
The off-diagonal elements of this matrix, which is in general not symmetric, encode cross-couplings between different protein species: a gradient in the total density of species $\beta$ can drive a redistribution current of species $\alpha$ by changing the local reactive equilibrium densities of $\alpha$.

The stability of the homogeneous steady state can be analyzed by examining the eigenvalues of the effective diffusion matrix $\mathbf{D}^{\mathrm{eff}}$. 
Positive eigenvalues correspond to stabilizing diffusive behavior. 
In contrast, negative eigenvalues indicate anti-diffusive (negative diffusion) behavior, where perturbations amplify and mass spontaneously accumulates, signaling an instability. 
Thus, the presence of at least one eigenvalue of $\mathbf{D}^{\mathrm{eff}}$ with a negative real part provides a direct criterion for long-wavelength pattern-forming instabilities in McRD systems.
The negative-eigenvalue condition for the $N_\mathrm{species}\times N_\mathrm{species}$ matrix thus generalizes the scalar slope criterion ${\partial_\rho c^*<0}$ of the single-species system to multi-species systems.
It defines the \emph{LQSS nullcline-slope criterion} for the existence of a long-wavelength instability---classified as a type-II instability in the classification scheme of Cross and Hohenberg~\citep{Cross.Hohenberg1993}.
When $D_{c,1}^{\alpha}=D$ for all $\alpha$,
\begin{subequations}\label{eq:equal-D}
\begin{align}
\mathbf{D}^{\mathrm{eff}}(\bm\rho) &= D\, \partial_{\bm\rho}\boldsymbol\eta^{*}(\bm\rho)
\end{align}
\end{subequations}
so all onset criteria can be checked directly on the Jacobian ${\mathbf{J}_\eta =  \partial_{\bm\rho}\boldsymbol\eta^{*}(\bm\rho)}$.

Mathematically, the above eigenvalue condition for the long-wavelength limit can be derived using degenerate perturbation theory in the limit ${q \to 0}$ ~\citep{Brauns.etal2021}.
By extending the definition of the mass-redistribution potential to incorporate membrane diffusion, the LQSS criterion for long-wavelength instabilities applies more broadly to general McRD systems ~\citep{Brauns.etal2021}.
In particular, this criterion is independent of the number of binding-competent cytosolic components per species.
In the next section, we show that for systems with a single binding-competent state per species the LQSS criterion is exact for stationary instabilities.

\subsection{One cytosolic component per protein species}
\label{sec:single-cyt-comp}

We now consider McRD systems in which each protein species has exactly one cytosolic component. 
This includes, for instance, the minimal two-component model discussed in the heuristic analysis in Sec.~\ref{sec:heuristics_2cmcrd} as well as the MinDE skeleton model with instantaneous nucleotide exchange (cf.\ Appendix~\ref{app:model}).
As we will show next, in this class of systems, stationary instabilities can only arise in the long-wavelength limit. 
This implies that the heuristic mechanism of mass redistribution and the LQSS nullcline-slope criterion fully account for all stationary pattern-forming instabilities. 
Other mechanisms for stationary lateral instability via short-wavelength (type-I) instabilities, in contrast, are not possible in this setting.

This can be seen by analyzing the zero crossings of the dispersion relation.
Assuming a stationary instability, we demand that the imaginary part of the eigenvalue is zero at a zero crossing of the real part, implying ${\sigma =0}$ at the zero crossings of the dispersion relation.
Then, since we assume that membrane components do not diffuse, the membrane block of Eq.~\eqref{eq:LSA} at $\sigma=0$ reduces to the requirement that the membrane reactions vanish exactly:
\begin{equation}
\label{eq:membrane-zero-eq}
    0 
    = 
   {\mathbf R}_m 
    \left(
    \mathbf{u}_\mathrm{hss} + \delta\mathbf{u}
    \right)
    \, ,
\end{equation}
Linearizing around $\mathbf{u}_\mathrm{hss}$, this condition requires the eigenmode to lie in the null space of the membrane-reaction Jacobian:
\begin{equation}
\label{eq:kernel_mem_react_Jacob}
    \partial_\mathbf{u}{\mathbf R}_m |_\mathrm{hss} \, \delta \mathbf{u} = 0 .
\end{equation}
Geometrically, this means the marginal mode must be tangent to the manifold of membrane equilibria (the membrane nullcline).
The mass-conservation constraints, ${0 = \mathbf{s}_\alpha \cdot {\mathbf R} = R_\mathrm{c,1}^\alpha + \mathbf{s}_m^\alpha \cdot{\mathbf R}_m}$~\footnote{Without loss of generality, we chose the normalization of the stoichiometric vectors by $s_{c,1}^\alpha=1$.}, then imply that the cytosolic reaction terms must also vanish at a zero of the dispersion relation: ${0=R_\mathrm{c,1}^\alpha (\mathbf{u}_\mathrm{hss} + \delta\mathbf{u})}$. 
Taken together, the full reaction Jacobian acting on the eigenmode vanishes, ${\partial_\mathbf{u} {\mathbf R} |_\mathrm{hss} \, \delta \mathbf{u} = 0}$, and Eq.~\eqref{eq:LSA} reduces to ${0 = -q^2 \mathbf{D}_c \, \delta\mathbf{c}}$.
Since $\mathbf{D}_c$ is diagonal with strictly positive entries, this condition implies that the eigenmode $\delta\mathbf{c}$ must vanish unless ${q = 0}$. 
Therefore, the dispersion relation can only cross zero at ${q = 0}$.
It follows that any stationary lateral instability in such systems must start at ${q = 0}$, corresponding to a long-wavelength (type-II) instability~\footnote{Finite membrane diffusion ensures that the dispersion relation attains a negative real part at large wavenumbers $q$.
In the approximation of zero membrane diffusion the dispersion relation remains positive in the limit ${q \to \infty}$ (see also Ref.~\cite{Villar-Sepulveda.etal2025}).}.
Accordingly, the stability of the system is fully determined by its behavior in the long-wavelength limit.

In summary, the LQSS nullcline-slope criterion---meaning the existence of an eigenvalue with negative real part for the effective diffusion matrix---is exact for stationary instabilities in systems with only one cytosolic component per species.  
It fully characterizes the onset of stationary pattern formation and relates it directly to properties of the reactive equilibria, independent of the specific reaction rates.

\section{Generalizing slope criteria to stationary short-wavelength instabilities}
\label{sec:3c}

The local quasi-steady-state (LQSS) approximation provides a useful framework for analyzing long-wavelength instabilities. 
However, the LQSS approximation cannot capture \emph{finite-wavenumber} (Turing-type) instabilities, in which the homogeneous state—stable to uniform perturbations—first loses stability at a nonzero wavenumber~\footnote{
They correspond to the so-called type-I instabilities in the classification of Cross and Hohenberg~\citep{Cross.Hohenberg1993}.
Because of the conservation law(s) ensuring zero eigenvalues at ${q=0}$, the short-wavelength instabilities arising in McRD systems are more specifically referred to as ``conserved Turing instabilities''~\citep{Frohoff-Hulsmann.Thiele2023a}.}.
Such conserved Turing instabilities can arise in systems with more than one cytosolic component per species, as we will see in the following.
We ask: Can these (stationary) conserved Turing instabilities also be predicted and understood using similar ``nullcline-slope criteria'' that relate the instability conditions to properties of the reactive equilibria?

To show that such kind of criteria exist, we first analyze a minimal single-species system comprising three components. We will show, both by heuristic arguments and formal mathematical analysis, that this system can exhibit a type-I (stationary finite-wavelength) instability that is not captured by the LQSS approximation, but is still captured by an analogous slope criterion.

\subsection{Heuristic analysis of cytosolic mass-redistribution} 
\label{sec:heuristics_3cmcrd}

\begin{figure}[!t]
	\includegraphics[width=1\columnwidth]{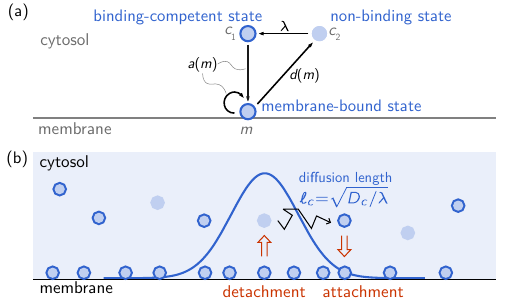}
	\caption{
Illustration of a minimal McRD system with one species and two cytosolic components.
(a) We consider a single protein species with two cytosolic states: a binding-competent state $c_1$ (blue with outline) and a non-binding (purely cytosolic) state $c_2$ (blue without outline). 
(b) After detaching (upward arrow), proteins enter the non-binding cytosolic state and diffuse (black, zigzag arrow) until they regain binding competence at rate $\lambda$; the typical distance covered before reactivation is ${\ell_c=\sqrt{D_{c_2}/\lambda}}$, after which they can reattach to the membrane (downward arrow).}
\label{fig:3c-diffusion}
\end{figure}

As illustrated in Fig.~\ref{fig:3c-diffusion}(a), we consider a minimal single-species setting with one membrane-bound state $m$ and two cytosolic states: a \emph{binding-competent} state $c_1$ (dark blue) and a \emph{non-binding} state $c_2$ (light blue) \cite{Gai.etal2020a,Chiou.etal2021,Toffenetti.etal2026}.
Membrane-bound proteins detach into the non-binding state and subsequently regain binding competence by conversion ${c_2 \to c_1}$ at rate $\lambda$ (e.g., nucleotide exchange).
Because detachment and reactivation are separated in time, proteins explore the cytosol in the non-binding state before they can reattach.
The corresponding \emph{cytosolic diffusion length}
\begin{equation}
    \ell_c \equiv \sqrt{D_{c_2}/\lambda}
\end{equation}
is the typical distance travelled in the non-binding state before conversion [Fig.~\ref{fig:3c-diffusion}(b)]; for nonlinear conversion reactions $\lambda(c_2)\,c_2$, the screening length is defined using the linearized conversion rate $\lambda^\mathrm{lin} = \partial_{c_2}[\lambda(c_2)\,c_2]|_\mathrm{hss}$ (see Appendix~\ref{app:multi-cyt-components}).

\begin{figure*}[!htb]
	\includegraphics[width=\linewidth]{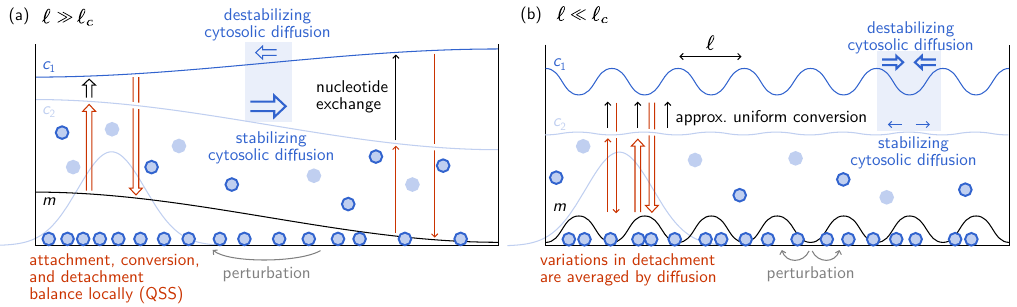}
\caption{
Long- and short-wavelength instabilities of the three-component McRD model introduced in Fig.~\ref{fig:3c-diffusion}.
In both panels, the lower black curve shows the membrane profile $m$, and the upper two curves show the cytosolic profiles $c_1$ (binding-competent, dark blue) and $c_2$ (non-binding, light blue); arrows indicate attachment (upward), detachment (downward), and conversion ($c_2 \to c_1$, black thin).
The length scale of diffusive averaging in the $c_2$ component is indicated by the light blue diffusion kernel shown in the background of the attachment and detachment arrows.
(a) In the long-wavelength limit ($q\ell_c \ll 1$), the spatial separation between detachment and attachment is negligible. 
Attachment, conversion, and detachment locally balance (LQSS).
A lateral perturbation of $\rho$ shifts the local reactive equilibrium $\mathbf u^*(\rho)$, producing cytosolic profiles that drive mass redistribution. With autocatalytic membrane recruitment, the $c_1$ and $c_2$ profiles may have opposite signs: ${\partial_\rho c_1^*<0}$ (destabilizing) and ${\partial_\rho c_2^*>0}$ (stabilizing).
The sign of the net effective diffusivity ${D_\mathrm{eff} = D_{c_1}\partial_\rho c_1^* + D_{c_2}\partial_\rho c_2^*}$ is set by which contribution dominates (blue horizontal arrows).
(b) For short-wavelength perturbations ($q\ell_c \gg 1$), diffusion in the non-binding state $c_2$ before conversion into $c_1$ rapidly suppresses spatial variations, rendering the ${c_2 \to c_1}$ conversion effectively uniform. 
In this regime, gradients of the binding-competent state $c_1$ generated by autocatalytic membrane recruitment are no longer counteracted by gradients in $c_2$ and can dominate (compare blue horizontal arrows in both panels), driving a loss of stability of the homogeneous steady state at sufficiently large $q$ even if the system is stable in the long-wavelength limit.
}
\label{fig:3c-mass-redistribution}
\end{figure*}

\smallskip

\emph{Long-wavelength limit.---} Consider spatial perturbations with a characteristic length $\ell \equiv 1/q$ much larger than the cytosolic diffusion length $\ell_c$ (${q\,\ell_c\ll 1}$) [Fig.~\ref{fig:3c-mass-redistribution}(a)]. 
On such scales, the spatial offset $\sim\ell_c$ between detachment and subsequent reattachment is small compared to the wavelength $\ell$ over which densities vary, so attachment and detachment fluxes can be treated as locally balanced to leading order in $q\ell_c$.
Accordingly, the LQSS approximation applies on such long length scales: the local state follows the reactive-equilibrium manifold,
 ${\mathbf{u}(\mathbf{x},t)\approx \mathbf{u}^*(\rho(\mathbf{x},t))}$, parameterized by the conserved total density $\rho=m+c_1+c_2$.

A non-uniform $\rho(\mathbf x)$ then produces concomitant profiles in all components $\mathbf u^*(\rho(\mathbf x))$ along the domain [Fig.~\ref{fig:3c-mass-redistribution}(a)]. 
Cytosolic gradients in $c_1$ and $c_2$ drive a diffusive mass-redistribution flux determined by spatial gradients of the mass-redistribution potential

\begin{equation}
    \eta \equiv c_1 + \frac{D_{c_2}}{D_{c_1}}\,c_2\,.
\end{equation}
In LQSS, ${\nabla c_i \simeq \partial_\rho c_i^\ast\,\nabla\rho}$ and thus ${\nabla\eta \simeq \partial_\rho \eta^\ast\,\nabla\rho}$, with
\begin{equation}
    \partial_\rho\eta^\ast
    =
    \partial_\rho c_1^\ast
    +
    \frac{D_{c_2}}{D_{c_1}}\,\partial_\rho c_2^\ast
    \, .
\end{equation}
The effective diffusivity governing mass redistribution at $q\to 0$ is therefore [cf.\ Eq.~\eqref{eq:Deff-def}]
\begin{equation}
    D_{\mathrm{eff}}(\rho)=D_{c_1}\,\partial_\rho\eta^\ast
    =
    D_{c_1}\,\partial_\rho c_1^\ast
    +
    D_{c_2}\,\partial_\rho c_2^\ast
    \, ,
\label{eq:DefR_two-component}
\end{equation}
and the homogeneous steady state is unstable at long wavelength if ${D_{\mathrm{eff}}(\rho_\mathrm{hss})<0}$, the LQSS slope criterion (cf.\ Sec.~\ref{sec:long-wavelength-limit}).

The two cytosolic components contribute to $D_\mathrm{eff}$ with potentially opposite signs.
With autocatalytic membrane recruitment, the attachment rate rises in regions of larger membrane density, inducing faster depletion of the binding-competent cytosolic component, and thus, results in a reduced steady-state cytosolic density and ${\partial_\rho c_1^*<0}$ (destabilizing nullcline slope) if the autocatalytic behavior is sufficiently strong (App.~\ref{app:self-recruitment}). 
In contrast, detachment typically increases with the membrane density, resulting in an increased steady-state density of the non-binding pool $c_2$ in regions of high membrane density and ${\partial_\rho c_2^*>0}$ (stabilizing).
Such counteracting contributions are depicted in  Fig.~\ref{fig:3c-mass-redistribution}(a).
Long-wavelength stability is then set by which contribution dominates in ${D_\mathrm{eff}=D_{c_1}\partial_\rho c_1^* + D_{c_2}\partial_\rho c_2^*}$: 
when the stabilizing $c_2$ term outweighs the destabilizing $c_1$ term, ${D_\mathrm{eff}>0}$ (equivalently, ${\partial_\rho\eta^*>0}$) and the homogeneous steady state is stable at long wavelength.
This regime---long-wavelength stable despite $\partial_\rho c_1^*<0$---is precisely the one that admits a short-wavelength instability, as we show below.

\smallskip

\emph{Short-wavelength limit ($q\ell_c\gg 1$).---}
On lateral scales ${\ell \ll \ell_c}$, the diffusive mixing of proteins detached into the non-binding state $c_2$ over distances $\sim\ell_c$ before reactivation averages out spatial variations in the detachment. Consequently, $c_2$ becomes nearly uniform ($\nabla c_2 \simeq 0$).
The conversion $c_2 \to c_1$ then acts as an approximately spatially uniform source of the binding-competent pool $c_1$ [Fig.~\ref{fig:3c-mass-redistribution}(b)].
The stabilizing counter-flux associated with $c_2$ is strongly attenuated, and mass redistribution is governed predominantly by gradients in $c_1$.

Because the ${c_2\!\to\!c_1}$ source is effectively uniform, lateral gradients in $c_1$ arise primarily from spatial variations in
membrane attachment.
For autocatalytic attachment, a region with slightly elevated membrane density attaches $c_1$ faster and thereby
locally depletes cytosolic $c_1$.
Hence $\nabla c_1$ points away from the high-membrane-density region, and the resulting diffusive flux $\mathbf{j}_1=-D_{c_1}\nabla c_1$ supplies additional proteins to the region of already elevated membrane density.
The arriving proteins are attached at the elevated rate, increasing the local membrane density further and reinforcing the initial perturbation.
This closes a positive feedback loop that drives a pattern-forming instability at length scales ${q^{-1}\gtrsim \ell_c}$.

\smallskip

\emph{Instability criterion.---} 
Therefore, at short wavelengths, the destabilizing feedback is controlled by $c_1$ alone.
The heuristic picture thus identifies ${\partial_\rho c_1^*|_\mathrm{hss} < 0}$ as the condition for short-wavelength instability if ${\partial_\rho \eta^*|_\mathrm{hss} > 0}$ causes long-wavelength instability.
The next section confirms this by a formal marginal-mode analysis and derives the onset wavenumber $q_\mathrm{min}$.

\subsection{Finite-wavelength stability criterion from marginal modes}
\label{}

Building on the heuristic analysis in the previous subsection, we now formally derive this criterion from the linear stability analysis of the one-species, three-component McRD model with state vector ${\bm{u}=(m,c_1,c_2)^{\! \top}}$.
Only the cytosolic species diffuse laterally, encoded by the diagonal diffusion matrix ${\mathbf{D}=\mathrm{diag}(0,D_{c_1},D_{c_2})}$, resulting in the dynamics: 
\begin{align}
    \partial_t \bm{u} = \mathbf{D}\,\boldsymbol{\nabla}^2 \bm{u} + {\mathbf R}(\bm{u}), 
\label{eq:3cMcRD-vector}
\end{align}
with the reaction vector
\begin{align}
    {\mathbf R}(\bm{u}) =
    \begin{pmatrix}
    \phantom{+} \;\; a(m)\,c_1 - d(m) \, m \;\;\\
    -\,a(m)\,c_1 + \lambda\,c_2\\
    d(m) \, m - \lambda\,c_2
\end{pmatrix}.
\label{eq:3cMcRD-f}
\end{align}
Here, $a(m)$ and $d(m)$ represent the attachment rate of the binding-competent component and the detachment rate of the membrane-bound protein. For simplicity, we consider a linear conversion rate $\lambda$.
Linearizing Eq.~\eqref{eq:3cMcRD-vector} around $\bm{u}_\mathrm{hss}$ yields the (right) eigenproblem (cf.\ Sec.~\ref{sec:lsa})
\begin{align}
    \sigma(q) \, \delta\bm{u} 
    =
    \big[
    \mathbf{J}-\mathbf{D} \, q^2 
    \big] \, 
    \delta\bm{u}
    \equiv \mathbf{M}(q)\,\delta\bm{u}
    \, .
\label{eq:3cMcRD-eigenproblem}
\end{align}
We seek a condition that decides whether the homogeneous steady state undergoes a stationary lateral instability, i.e., we again assume that the imaginary part of the growth rate vanishes (when the real part vanishes).
Then, a short-wavelength instability corresponds to the existence of a finite wavenumber ${q_\mathrm{min}>0}$ at which the growth rate vanishes (cf.\ Sec.~\ref{sec:single-cyt-comp}).
Thus, a stationary short-wavelength instability arises if there exist nontrivial solutions of ${\mathbf{M}(q)\,\delta\mathbf{u}=\mathbf{0}}$ for positive wavenumbers ${q>0}$.
Equivalently, we have to determine whether solutions $q>0$ exist to ${\det\mathbf{M}(q)=0}$.

Rather than solving for eigenvalue and eigenvector simultaneously, we work right at marginality (${\sigma =0}$) and first construct the \emph{marginal} eigenvector using the equations for the membrane and non-binding cytosolic component, exploiting the local geometry of the reactive equilibria in phase space.
Substituting the marginal eigenvector into the linearized continuity equation and imposing marginality---which, for dynamics at finite $q$, is equivalent to a uniform mass-redistribution potential, ${\delta\eta=0}$---collapses the original ${3 \times 3}$ problem to a single scalar onset condition for $q$ (self-consistency condition).
A positive root establishes the finite-wavelength instability and the corresponding lower edge ${ q_{\min} }$ of the band of unstable modes.

\begin{figure}[!t]
\centering
\includegraphics[width=\columnwidth]{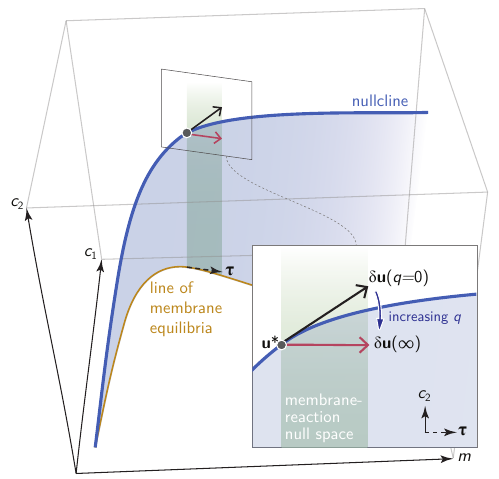}
\caption{%
Geometric interpretation of marginal stability in the $(m,c_1,c_2)$ phase space.
The nullcline $\mathbf{u}^*(\rho)$ (thick solid curve, blue) projects onto the $(m,c_1)$--plane as the line of membrane equilibria ${R_m(m,c_1) = 0}$ (blue-shaded manifold and thin solid curve, orange), with projected tangent $\boldsymbol{\tau}$ (dashed arrow).
The null space of the membrane-reaction Jacobian is shown as the shaded plane (dark green shading).
\textit{Inset:} vicinity of the homogeneous steady state $\mathbf{u}^*$ (gray dot).
For ${q \to 0}$, the marginal eigenvector ${\delta\mathbf{u} (q\!=\!0)}$ is tangent to the nullcline (black arrow); for ${q \to \infty}$, the $c_2$--component is suppressed and ${\delta\mathbf{u} (q\!=\!\infty)}$ lies in a plane parallel to the $(m,c_1)$--plane (red horizontal arrow).
}
\label{fig:marginal-stability-geometry}
\end{figure}

\smallskip

\emph{Marginal membrane dynamics.---}
We start with constructing the marginal eigenvector.
Thus, we seek to determine the \emph{direction} of $\delta\mathbf{u}(q)$ in component space (its ratios across $m,c_1,c_2$) up to an arbitrary overall normalization.
Since the membrane equation has no diffusive contribution, marginality imposes the constraint:
\begin{equation}
    \label{eq:3cMcRD-membraneEq}
    \partial_m R_m \, \delta m 
    + 
    \partial_{c_1} R_m \, \delta c_1 
    = 
    0 \, .
\end{equation} 
Geometrically, this constraint requires that the \emph{marginal} perturbation $(\delta m,\delta c_1)$ must be tangent to the line of membrane equilibria ${R_m (m,c_1) = 0}$ in the $(m,c_1)$--plane [Fig.~\ref{fig:marginal-stability-geometry}]; that is, it lies in the null space of the membrane-reaction Jacobian row $(\partial_m R_m, \partial_{c_1} R_m)$ . 
To obtain an explicit vector spanning this null space, we note that the equilibrium condition for the membrane reactions, $R_m(m^*(\rho),c_1^*(\rho))=0$, holds as a function of the total protein density along the nullcline $\mathbf{u}^*(\rho)$ of the full reaction term. Differentiating with respect to $\rho$ yields
\begin{align}
    \partial_m R_m \,\partial_\rho m^* \;+\; \partial_{c_1} R_m \,\partial_\rho c_1^* \;=\; 0 .
\end{align}
This shows that the projected tangent $(\partial_\rho m^*,\partial_\rho c_1^*)^{\! \top}$ of the nullcline  satisfies the same constraint as the marginal perturbation, Eq.~\eqref{eq:3cMcRD-membraneEq}. 
Since the nullspace of $(\partial_m R_m, \partial_{c_1} R_m)$ is one-dimensional, the two vectors must be parallel, and we can use the projected nullcline tangent to parameterize the perturbation vector ${(\delta m, \delta c_1)^{\! \top}}$.
Choosing an appropriate proportionality factor $\widetilde{\delta\rho}$, we set 
\begin{align}
    \begin{pmatrix}\delta m\\[2pt]\delta c_1\end{pmatrix}
    \;=\;
    \begin{pmatrix}\partial_\rho m^*\\[2pt]\partial_\rho c_1^*\end{pmatrix}\,
    \widetilde{\delta\rho} 
    \, .
\label{eq:3cMcRD-chemEq-parameterization}
\end{align}
This parameterization fixes the 
admissible $(m,c_1)$ direction uniquely, independently of~$q$. Importantly, $\widetilde{\delta\rho}$ is a parameterization variable 
measuring displacement along the nullcline; it should not be confused 
with the physical total-density perturbation 
$\delta\rho = \delta m + \delta c_1 + \delta c_2$. 
The two coincide in the long-wavelength limit, where the full marginal 
mode is tangent to the nullcline.
However, they differ at finite $q$ because gradients in the 
$c_2$--component are suppressed, as we derive next
(see App.~\ref{app:perturbation_rho_detail}).

\smallskip

\emph{Non-binding cytosolic component.---}
With $(\delta m,\delta c_1)$ fixed by the marginal membrane constraint, the amplitude of the non-binding cytosolic component $\delta c_2$ follows from linearizing the $c_2$-component of Eq.~\eqref{eq:3cMcRD-vector} at $\sigma = 0$:
\begin{equation}
\label{eq:def-eq-c2-mode}
    \left(
    \lambda+D_{c_2} \, q^2
    \right) 
    \delta c_{2}
    = 
    \partial_m \big[d(m) \, m \big]_\mathrm{hss} \, \delta m 
    \, .
\end{equation}
Because the non-binding cytosolic pool $c_2$ 
couples to the membrane only via the detachment 
flux $d(m)\,m$, its marginal response is 
driven by the membrane perturbation $\delta m$.

Along the nullcline manifold the steady-state balance 
$d(m^*(\rho))\,m^*(\rho) = \lambda\,c_2^*(\rho)$ 
[cf.\ Eq.~\eqref{eq:3cMcRD-f}] 
links $c_2$ to $m$; differentiating with respect to 
$\rho$ gives
\begin{equation}
\label{eq:nullcline-c2-rho}
    \lambda\, \partial_\rho c_2^* 
    = 
    \partial_m \big[d(m)\,m \big]_\mathrm{hss} \, \partial_\rho m^*
    \, ,
\end{equation}
which shows that the $\rho$--dependence of $c_2^*$ is inherited from 
the membrane equilibrium $m^*(\rho)$ through the detachment coupling. 
Inserting $\delta m = \partial_\rho m^*\,\widetilde{\delta\rho}$ 
[Eq.~\eqref{eq:3cMcRD-chemEq-parameterization}] into 
Eq.~\eqref{eq:def-eq-c2-mode} and using 
Eq.~\eqref{eq:nullcline-c2-rho} to replace 
$\partial_m[d(m)\,m]\,\partial_\rho m^*$ by 
$\lambda\,\partial_\rho c_2^*$ gives
\begin{equation}
\label{eq:suppression-inactive-cytosolic-3c}
    \delta c_{2} = \left(1+\frac{D_{c_2}}{\lambda} \, q^2\right)^{-1} 
    \partial_\rho c_2^* \, \widetilde{\delta\rho}
    \, .
\end{equation}
Collecting components then yields the \emph{marginal perturbation 
vector}
\begin{align}
    \delta\mathbf u
    =
    \begin{pmatrix}
        \partial_\rho m^*\\[2pt]
        \partial_\rho c_1^*\\[2pt]
        \bigl[1+(q\ell_c)^2\bigr]^{-1}\partial_\rho c_2^*
    \end{pmatrix}
    \widetilde{\delta\rho} \, ,
\label{eq:marginal_perturbaton_vector}
\end{align}
with the screening length 
${\ell_c = \sqrt{D_{c_2}/\lambda}}$, as defined above.
The prefactor 
${\bigl[1+(q\ell_c)^2\bigr]^{-1}}$ acting on the non-binding 
component is a \emph{low-pass filter} with the diffusion length $\ell_c$ acting as the screening length: 
long-wavelength modulations (${q \ell_c \ll 1}$) pass essentially unchanged, while short-wavelength modulations (${q \ell_c \gg 1}$) are 
smoothed by diffusion before conversion into the binding-competent form, in agreement with the heuristic argument.
Consequently, the marginal eigenvector 
transitions monotonically between two 
limits: as ${q\to0}$, $\delta\mathbf u$ is tangent to the full 
nullcline, whereas for ${q\to\infty}$ the $c_2$--component is 
suppressed (thick dark vs.\ thin 
light arrows in Fig.~\ref{fig:marginal-stability-geometry}).

\smallskip

\emph{Flux balance and short-wavelength instability condition.---}
Equation~\eqref{eq:marginal_perturbaton_vector} shows that the only wavevector dependence of the marginal mode enters through the non-binding cytosolic component: the $(m,c_1)$-projection is fixed by marginal membrane kinetics, while the $c_2$-amplitude is attenuated by the screening factor $\bigl[1+(q\ell_c)^2\bigr]^{-1}$.
Consequently, the perturbation of the mass-redistribution potential along the marginal direction,
\begin{align}
\delta\eta(q)
=
\delta c_1 + \frac{D_{c_2}}{D_{c_1}}\,\delta c_2,
\end{align}
varies monotonically with $q$.
Using Eq.~\eqref{eq:marginal_perturbaton_vector} one finds
\begin{align}
\delta\eta(q)
&=
\left[
\partial_\rho c_1^\ast
+
\frac{D_{c_2}}{D_{c_1}}\,
\frac{1}{1+(q\ell_c)^2}\,
\partial_\rho c_2^\ast
\right]\widetilde{\delta\rho}
\nonumber \\
&=
\frac{1}{1+(q\ell_c)^2}\,
\Bigl[(q\ell_c)^2\,\partial_\rho c_1^\ast+\partial_\rho\eta^\ast\Bigr]\,
\widetilde{\delta\rho} .
\label{eq:eta-interpolation}
\end{align}
At long wavelengths ($q\ell_c\ll 1$), this expression 
$\delta\eta(q)\to 
(\partial_\rho\eta^\ast)\,\widetilde{\delta\rho}$ recovers
the full nullcline slope (LQSS regime); whereas at short wavelength 
($q\ell_c\gg 1$), one has
$\delta\eta(q)\to 
(\partial_\rho c_1^\ast)\,\widetilde{\delta\rho}$ reflecting the 
suppression of the non-binding pool.

\smallskip

\emph{Variations in the mass-redistribution potential drive diffusive fluxes.---}

Summing over all three component equations in Eq.~\eqref{eq:3cMcRD-eigenproblem}, one obtains the linearized continuity equation for the total density ${\rho=m+c_1+c_2}$ [cf.\ Eq.~\eqref{eq:cont-eqs}]:
\begin{align}
    \sigma\,\delta\rho 
    = -\,D_{c_1}q^2\,\delta c_1 - D_{c_2}q^2\,\delta c_2
    = -\,D_{c_1}q^2\,\delta\eta,
\label{eq:cont}
\end{align}
where $\delta \eta$ is the shift in the mass-redistribution potential defined above.
At marginality $(\sigma=0)$ and finite wavenumber $q>0$, Eq.~\eqref{eq:cont} imposes the \emph{diffusive flux–balance constraint}
\begin{equation}
    \delta\eta = 0 \, .
\label{eq:continuity_constraint}
\end{equation}
Consequently, a marginal finite-$q$ mode must leave $\eta$ invariant: 
any nonzero $\delta\eta$ would drive a redistribution flux 
${\propto \nabla\eta}$, giving $\sigma \neq 0$.
Evaluating Eq.~\eqref{eq:continuity_constraint} using the marginal 
eigenvector, Eq.~\eqref{eq:marginal_perturbaton_vector}, gives
\begin{equation}            
    (q\ell_c)^2\,   
    \partial_\rho c_1^*+\partial_\rho \eta^*
    =0
    \, ,
\label{eq:solvability}
\end{equation}
which has a nontrivial root
\begin{equation}\label{eq:3c-qmin}
    q_{\min}^2
    =
    \frac{\partial_\rho \eta^*}{-\partial_\rho c_1^*}\,\ell_c^{-2}
    \,,
\end{equation}
provided $\partial_\rho\eta^*$ and $\partial_\rho c_1^*$ have 
opposite signs.
This is precisely the scenario identified above: the gradients in 
$c_1$ and $c_2$ counteract each other, and $q_\mathrm{min}$ marks 
the wavenumber where their effects exactly compensate.

Taken together, the conditions for a short-wavelength instability, also ensuring that the system is laterally stable at long wavelengths, are:
\begin{equation}
\label{eq:short-wave-criterion}
    \partial_\rho \eta^* > 0
    \qquad\text{and}\qquad
    \partial_\rho c_1^* < 0 \, .
\end{equation}
When both conditions are satisfied, a band of unstable modes arises at wavenumbers $q > q_\mathrm{min}$.
While ${\partial_\rho \eta^*>0}$ forbids an instability for ${q<q_c}$, for ${q>q_c}$, the destabilizing gradients in $c_1$, caused by ${\partial_\rho c_1^* < 0}$ dominate.

Note that we find only a single zero crossing, so the band formally extends to $q\to\infty$; in practice, weak membrane diffusion suppresses the instability at short wavelengths and introduces a finite upper cutoff of the band of unstable modes. The practical importance of this nullcline-slope criterion is that this condition can be checked directly from the reactive equilibria, without performing the full linear stability analysis and having access to all rate constants of all reactions.
We thus find steady-state conditions for finite-wavelength dynamic instabilities far from equilibrium.

\smallskip

In Appendix~\ref{app:multi-cyt-components}, we generalize this 
derivation to systems with multiple protein species.
The scalar solvability condition, Eq.~\eqref{eq:solvability}, is then 
replaced by a matrix condition: the bracketed expression in 
Eq.~\eqref{eq:eta-interpolation} generalizes to a matrix whose 
determinant must vanish at some $q_\mathrm{min} > 0$ for a 
short-wavelength instability.
The underlying mechanism remains the same---suppression of gradients in 
non-binding components---but in multi-species systems the required 
negative nullcline slopes can arise from the interplay between 
different species.
The resulting criteria, still formulated in terms of nullcline slope 
matrices, are summarized in Sec.~\ref{sec:classification}.

\section{Classification of stationary pattern-forming instabilities in intracellular protein systems}
\label{sec:classification}

\begin{figure*}[t]
\centering

\begin{minipage}[t]{0.32\textwidth}
  \centering
  \textbf{(a)} PAR polarity system\\[4pt]
  \includegraphics[width=\linewidth]{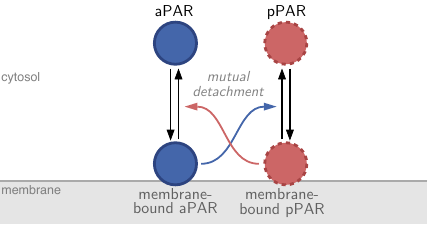}
\end{minipage}%
\hfill
\begin{minipage}[t]{0.32\textwidth}
  \centering
  \textbf{(b)} MinDE skeleton model\\[4pt]
  \includegraphics[width=\linewidth]{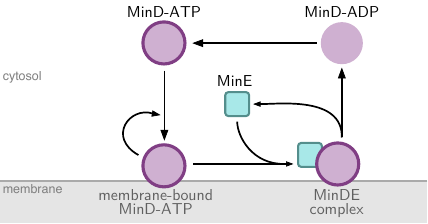}
\end{minipage}%
\hfill
\begin{minipage}[t]{0.32\textwidth}
  \centering
  \textbf{(c)} MinDE with explicit dimerization\\[4pt]
   \includegraphics[width=\linewidth]{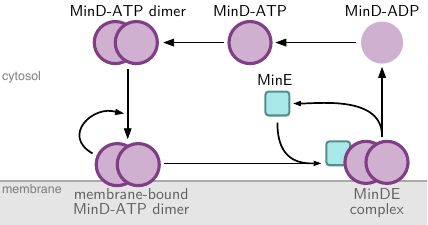}
\end{minipage}

\caption{%
    Reaction networks of three representative intracellular 
    protein systems illustrating the cytosolic-component 
    classes defined in this work.
    (a)~PAR polarity system (\textit{C.\ elegans}):
    two antagonistic species (aPAR, pPAR), each cycling 
    between a single cytosolic and a membrane-bound form.
    Mutual detachment of the membrane-bound species 
    (colored arrows) provides the nonlinear cross-antagonism 
    that drives symmetry breaking.
    (b)~MinDE skeleton model (\textit{E.\ coli}):
    MinD exists in two cytosolic states---a 
    binding-competent form (MinD-ATP) and an inactive form 
    (MinD-ADP) connected by nucleotide exchange---while 
    MinE has a single cytosolic state.
    Cooperative self-recruitment of MinD to the membrane 
    (curved arrow) and MinE-stimulated detachment with 
    hydrolysis close the reaction cycle.
    (c)~MinDE model with explicit dimerization:
    an additional cytosolic conversion step 
    (MinD-ATP monomer to MinD-ATP dimer) extends the MinD 
    pathway to three components, of which only the dimer is binding-competent (cf.\ Appendix~\ref{app:cytSeries}).
    The three systems thus have one~(a), two~(b), and 
    three~(c) cytosolic components per species, 
    corresponding to the classes introduced in 
    Fig.~\ref{fig:classification}.}
\label{fig:model-systems}
\end{figure*}

Intracellular pattern formation is realized by a diverse set of 
protein systems that share the mass-conserving reaction--diffusion 
structure analyzed in the preceding sections.
Three representative examples are shown in 
Fig.~\ref{fig:model-systems}: the PAR polarity system in 
\textit{C.\ elegans}, the MinDE oscillation system in 
\textit{E.\ coli}, and a variant of the latter with explicit cytosolic MinD 
dimerization.
From the perspective of our analysis, the key distinction is the
\emph{maximum number of cytosolic components per protein species}.
In the PAR system [Fig.~\ref{fig:model-systems}(a)], each 
species---aPAR and pPAR---has a single cytosolic state and cycles 
to the membrane, where mutual detachment provides the nonlinear 
feedback for polarization.
In the MinDE skeleton model [Fig.~\ref{fig:model-systems}(b)], 
MinE has one cytosolic state, but MinD has two: a binding-competent 
form (MinD-ATP) and an inactive form (MinD-ADP) produced upon 
MinE-stimulated hydrolysis, with nucleotide exchange at 
rate~$\lambda$ closing the cycle.
Adding explicit dimerization [Fig.~\ref{fig:model-systems}(c)] 
introduces a third cytosolic component for MinD, as only the 
dimerized form is assumed to be binding-competent.

\begin{figure*}[!t]
\setlength{\tabcolsep}{5pt}
\renewcommand{\arraystretch}{1.3}
\centering
\begin{tabular}{@{} 
  >{\raggedright\arraybackslash}p{1.5cm} 
  >{\raggedright\arraybackslash}p{4.05cm} 
  >{\raggedright\arraybackslash}p{4.55cm} 
  >{\raggedright\arraybackslash}p{5.55cm} 
  >{\raggedright\arraybackslash}p{0.05cm} 
@{}}
\toprule
\multicolumn{2}{@{}c}{\cellcolor{colA}\textcolor{colAtxt}{\textbf{One cytosolic component}}}
&
\multicolumn{3}{>{\columncolor{colM}}c}{\textcolor{colMtxt}{\textbf{Multiple cytosolic components}}}
\\
\midrule
& (a) one component 
& (b) $\leq$ two components 
& (c) cascade  
\\[4pt]
Example
& PAR system [Fig.~\ref{fig:model-systems}(a)]
& Min skeleton [Fig.~\ref{fig:model-systems}(b)]
& \multicolumn{1}{>{\raggedright\arraybackslash}p{7.3cm}}{%
    Min skeleton + dimerization [Fig.~\ref{fig:model-systems}(c)]}
\\[6pt]
Network
&
\raisebox{-0.2\height}{\includegraphics[width=3.5cm]{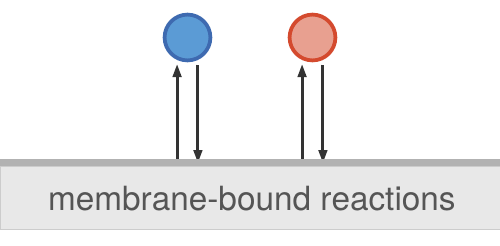}}
&
\raisebox{-0.2\height}{\includegraphics[width=3.5cm]{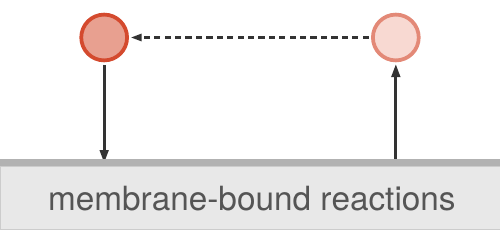}}
&
\raisebox{-0.2\height}{\includegraphics[width=3.5cm]{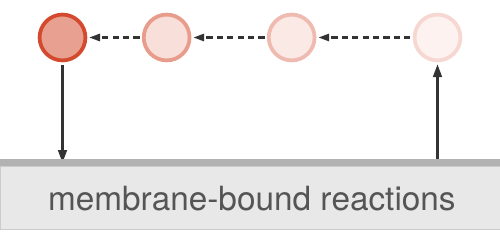}}
\\
\\
\multicolumn{5}{@{}l@{}}{%
  \colorbox{secbar}{\makebox[\dimexpr\textwidth-2\tabcolsep][l]{%
    \hspace{4pt}%
    \textbf{Long-wavelength (type-II) instability}
    \textemdash\ onset at $q = 0$, all classes%
  }}}\\[6pt]
Instability condition
& \multicolumn{4}{@{}l}{%
  $\mathbf{D}_{c,1}\,
   \partial_{\boldsymbol{\rho}}\boldsymbol{\eta}^*
   \big|_\mathrm{hss}$
  has eigenvalue with negative real part%
}\\
Sufficient
& \multicolumn{4}{@{}l}{%
  $\partial_{\boldsymbol{\rho}}\boldsymbol{\eta}^*
   \big|_\mathrm{hss}$
  has an odd number of negative eigenvalues%
}\\
\\[-5pt]
\multicolumn{5}{@{}l@{}}{%
  \colorbox{secbarII}{\makebox[\dimexpr\textwidth-2\tabcolsep][l]{%
    \hspace{4pt}%
    \textbf{Short-wavelength (type-I; conserved Turing) instability}%
    \qquad
    $s(q) \equiv 
     \det\!\big(\mathbf{Q}(q)\,
     \partial_{\boldsymbol{\eta}}\mathbf{c}_1^* 
     + \mathbf{I}\big)$%
  }}}\\[6pt]
Instability condition
& \textcolor{crs}{not possible}
& $\mathbf{Q} = q^2\,
   \mathrm{diag}\bigl((\ell_c^\alpha)^2\bigr)$\newline
  $s(q_\mathrm{min}) = 0$
& $\mathbf{Q}(q)$: App.~\ref{app:cytSeries}\newline
  $s(q_\mathrm{min}) = 0$
\\
\cmidrule(lr){2-2}\cmidrule(lr){3-5}
Sufficient
& \textcolor{crs}{not possible}
& \multicolumn{3}{@{}l}{%
  Classes (b)\textendash(c):\;
  $\partial_{\boldsymbol{\rho}}\mathbf{c}_1^*
   \big|_\mathrm{hss}$
  has an odd number of negative eigenvalues
}
\\
\bottomrule
\end{tabular}
\caption{%
    Classification of stationary mass-redistribution instabilities
    in systems with negligible membrane diffusion ($D_m = 0$).
    Columns correspond to the maximum number of cytosolic
    components per conserved species~$\alpha$;
    Fig.~\ref{fig:model-systems} shows example reaction networks
    for classes~(a)\textendash(c).
    For each species, $c_1^\alpha$ is the binding-competent
    cytosolic component; remaining components
    $c_i^\alpha$ ($i \geq 2$) are binding-incompetent and undergo
    conversion toward $c_1^\alpha$ by possibly nonlinear reactions
    with terms~$\lambda_\alpha(c_i^\alpha)\,c_i^\alpha$.
    \emph{Type-II instability} (onset at $q=0$):
    the onset criterion and sufficient condition
    [Eqs.~\eqref{eq:equal-D};
    Sec.~\ref{sec:single-cyt-comp}] involve the
    nullcline-slope matrix
    $\partial_{\boldsymbol{\rho}}\boldsymbol{\eta}^*$
    and are identical across all classes.
    \emph{Stationary Type-I (conserved Turing) instability}:
    not possible in class~(a)
    (Sec.~\ref{sec:single-cyt-comp});
    present in classes~(b,c), where finite
    conversion rates suppress short-wavelength gradients
    in non-binding pools (Sec.~\ref{sec:3c}).
    Instability requires the solvability condition
    $s(q_\mathrm{min})=0$
    [Eq.~\eqref{eq:s-of-q}]
    at some $q_\mathrm{min}>0$,
    with the filtering matrix~$\mathbf{Q}(q)$ encoding
    low-pass suppression via screening lengths
    $\ell_{c,i}^\alpha
    = \sqrt{D_i^\alpha/\lambda_i^\alpha}$
    (Appendix~\ref{app:multi-cyt-components}).
   We expect these results to extend to the case of explicit 
    bulk-boundary coupling, with the short-wavelength criterion modified by a generalized filtering matrix $\mathbf{Q}_\mathrm{bbc}(q)$, as derived in Appendix~\ref{app:bbc}.
}
\label{fig:classification}
\end{figure*}

In this section, we use this cytosolic count as the organizing axis for a classification of stationary lateral instabilities in multi-species McRD systems (Fig.~\ref{fig:classification}).
Building on the long-wavelength slope criterion (Sec.~\ref{sec:long-wavelength-limit}) and the finite-wavelength marginal-mode analysis (Sec.~\ref{sec:3c}), we state concise instability criteria that can be read off from \emph{nullcline-slope matrices} at the homogeneous steady state.
Stationary instabilities can occur both at long and short wavelengths, corresponding to type-II and type-I instabilities in the  classification of Ref.~\citep{Cross.Hohenberg1993}.

\smallskip 

\textit{Single cytosolic component per species\;---}
The simplest case is that of models with only a single cytosolic component per species 
[Fig.~\ref{fig:classification}(a); see 
Fig.~\ref{fig:model-systems}(a) for the PAR system as a concrete 
example].
As derived in Sec.~\ref{sec:single-cyt-comp}, these systems 
can only become unstable by a stationary instability at long wavelength (type-II onset), with 
the instability predicted by the LQSS criterion: the matrix 
${\mathbf{D}_c^1\partial_{\bm{\rho}}\bm{\eta}^*}$ must have at 
least one eigenvalue with negative real part.
This generalizes the scalar condition 
$\partial_\rho\eta^*(\rho) < 0$ of the single-species two-component 
system to multi-species systems.

This class covers both minimal and more elaborate McRD models.
At the minimal end are two-component models for small-GTPase patterning~\citep{Goryachev.Pokhilko2008,Edelstein-Keshet.etal2013}.
More detailed realizations include models for eukaryotic cell motility~\citep{Maree.etal2006}, Cdc42 polarization in budding yeast~\citep{Klunder.etal2013}, and the reconstituted Rab5 system~\footnote{
    The models for the Rab5 system have only single cytosolic components if the guanosine nucleotide dissociation inhibitor (GDI) is not modeled explicitly.
    If the GDI is modeled as done in Ref.~~\citep{Solomatina.etal2022}, it has two cytosolic components (no membrane-bound states), of which one is the complex with Rab5-ADP.
    Because the complex with Rab5-ADP is the only cytosolic component of Rab5, the arguments from Sec.~\ref{sec:single-cyt-comp} apply to this cytosolic complex.
    Mass conservation of the GDI then implies that they also extend to its own individual cytosolic component.
    Therefore, also this extended model, although not strictly fulfilling the condition of one cytosolic component per species, does only show stationary long-wavelength instabilities.
    }
~\citep{Cezanne.etal2020,Solomatina.etal2022}.
Moreover, models for the PAR polarity system in \textit{C.\ elegans}~\citep{Goehring.etal2011,Trong.etal2014} and a reconstituted lipid kinase--phosphatase system~\citep{Hansen.etal2019} contain only a single cytosolic component per species.

\smallskip

\textit{Systems with multiple cytosolic components per species\;---} 
In contrast, if at least one protein species has two or more cytosolic components, the system may additionally show short-wavelength (type-I, conserved Turing) instabilities [Fig.~\ref{fig:classification}(b,c)].
These instabilities arise from the mechanism identified in 
Sec.~\ref{sec:3c}: finite conversion rates suppress gradients in 
the non-binding components, allowing the binding-competent 
components to drive a short-wavelength instability even when the 
system is stable at long wavelength.

Within this multi-component class, we distinguish two main cytosolic
topologies [Fig.~\ref{fig:classification}(b,c)].
In class~(b), each species has at most two cytosolic
components: an inactive form~$c_2^\alpha$, produced upon detachment
from the membrane, is converted into the binding-competent
form~$c_1^\alpha$ at rate~$\lambda_\alpha$
[Fig.~\ref{fig:model-systems}(b)].
Class~(c) generalizes this to a cascade of conversion steps
$c_n^\alpha \!\to\! \cdots \!\to\! c_1^\alpha$, where detachment
feeds exclusively into the terminal inactive
component~$c_n^\alpha$; the MinDE model with explicit cytosolic
dimerization [Fig.~\ref{fig:model-systems}(c)] is a concrete
realization.
A more general version of the cascade topology, in which detachment can
feed into several non-binding cytosolic components rather than only the
terminal one, is treated in Appendix~\ref{app:cytSeries}.

The exact instability criteria for the multi-component classes shown
in Fig.~\ref{fig:classification}(b,c),
derived in App.~\ref{app:multi-cyt-components}, all involve a filtering matrix $\mathbf{Q}(q)$ that encodes the increased relative contribution of the binding-competent state due to the low-pass suppression of gradients in non-binding components.
It therefore depends on
the ratios $q/\ell_{c,i}^\alpha$ with the screening lengths
$\ell_{c,i}^\alpha = \sqrt{D_i^\alpha/\lambda_i^\alpha}$ for the cytosolic components ${i>1}$ of species $\alpha$.
Physically, at short wavelengths (large $q$), proteins in non-binding states diffuse across spatial variations so rapidly that they effectively ``average out'' or homogenize their own concentration gradients before they are converted back into a binding-competent form.

On length scales much larger than these screening lengths ($q\ell_{c,i}^\alpha \ll 1$), $\mathbf{Q}(q)\to 0$, meaning all cytosolic components contribute fully to the mass-redistribution potential (LQSS regime).
However, on scales smaller than $\ell_{c,i}^\alpha$ ($q\ell_{c,i}^\alpha \gg 1$), $\mathbf{Q}(q)$ grows, progressively amplifying the effect of the binding-competent $c_1$ component. For the two multi-component classes shown in Fig.~\ref{fig:classification}(b,c), a short-wavelength instability requires that the
determinant
\begin{align}
    s(q) = \det\!\big(\mathbf{Q}(q)\,
    \partial_{\boldsymbol{\eta}}\mathbf{c}_1^* + \mathbf{I}\big)
\label{eq:s-of-q}
\end{align}
changes sign at some finite $q_\mathrm{min} > 0$, generalizing the
scalar condition Eq.~\eqref{eq:solvability} for the exemplary three-component system.
If detachment feeds into several non-binding components, the same
marginal-mode construction gives the generalized solvability condition
${\det\!\big(\sum_{i\geq2} \mathbf{Q}_i(q_\mathrm{min})\,
\partial_{\boldsymbol{\eta}}\mathbf{c}_i^* + \mathbf{I}\big)=0}$,
as derived in Appendix~\ref{app:cytSeries}.
The explicit form of $\mathbf{Q}$ depends on the network topology:
for systems with at most two cytosolic components per species
[Fig.~\ref{fig:classification}(b)], it simplifies to
${\mathbf{Q} = q^2\,\mathbf{L}^2}$ with
${\mathbf{L} = \mathrm{diag}\bigl(\ell_c^\alpha\bigr)}$,
so that the criterion reduces to the product
$q_{\mathrm{min}}^2\mathbf{L}^2\,\partial_{\boldsymbol{\eta}}\mathbf{c}_1^*$
having eigenvalue~$-1$;
for a series of cytosolic components with detachment into the last
level only [Fig.~\ref{fig:classification}(c)], the structure of
$\mathbf{Q}(q)$ is more involved but it also increases monotonously in $q$.
The full expressions for $\mathbf{Q}$ in each case are derived in
Appendix~\ref{app:multi-cyt-components}.
While the exact onset condition depends on the conversion
rates~$\lambda_\alpha$ through the screening
lengths~$\ell_{c,i}^\alpha$, a sufficient criterion for the multi-component classes shown in Fig.~\ref{fig:classification}(b,c) is that the slope matrix
$\partial_{\bm{\rho}}\mathbf{c}_1^*$ of the binding-competent
components has an odd number of negative eigenvalues
(App.~\ref{app:multi-cyt-components}).
This sufficient condition involves only the reactive equilibria and
is independent of the rates $\lambda_\alpha$.
The same sufficient condition also applies to the more general
detachment topology discussed in Appendix~\ref{app:cytSeries}.

The ``skeleton model'' of the \textit{E.\ coli} Min system 
[Fig.~\ref{fig:model-systems}(b)] 
falls into class~(b)
~\citep{Huang.etal2003,Halatek.Frey2012} (see also 
Appendix~\ref{app:model}), as well as a model of the PAR system 
including finite dephosphorylation (reactivation) rates in the 
cytosol~\citep{Gessele.etal2020}.
It also applies to the ``MinE-switch'' model of the Min 
system~\citep{Denk.etal2018,Ren.etal2025} if the reactive MinE 
state is included via a local quasi-steady-state 
approximation~\citep{Meindlhumer.etal2023}.
An example of a model with a series of cytosolic components 
[Fig.~\ref{fig:model-systems}(c)] is 
obtained by modeling MinD dimerization as a cytosolic 
process, placing it in class~(c) of 
Fig.~\ref{fig:classification}~\citep{Szeto.etal2002,Hu.Lutkenhaus2003,Szeto.etal2003,Mileykovskaya.etal2003,Taghbalout.etal2006,Ramm.etal2019}(see discussion in Appendix~\ref{app:cytSeries}).

\smallskip

\textit{Bulk--boundary coupling\;---}
Because protein pattern formation is driven by membrane attachment 
and detachment, a careful description must consider the resulting bulk--boundary coupling of 
three-dimensional dynamics in the cytosol and two-dimensional 
membrane dynamics. The preceding analysis neglected the extent of the cytosol 
perpendicular to the membrane; here we summarize how our framework 
generalizes when this coupling is included explicitly.
As shown in Appendix~\ref{app:bbc}, the instability criteria retain 
the same structure---they are formulated in terms of nullcline-slope 
matrices---but 
with a modified filtering matrix $\mathbf{Q}_\mathrm{bbc}(q)$ that 
accounts for the diffusive spreading of the non-binding components in a three- instead of two-dimensional cytosolic volume.
For long-wavelength instabilities, the onset and sufficient 
conditions coincide with those obtained without bulk--boundary 
coupling.
We have derived the short-wavelength criteria explicitly for systems 
with at most two cytosolic components per species; we expect that 
the treatment extends to the cytosolic-series cases~(c), and to the
more general detachment topologies discussed in
Appendix~\ref{app:cytSeries}, in close analogy to the derivation
without bulk--boundary coupling.

\smallskip

\textit{Summary.\;---}
The main cases discussed above are summarized in
Fig.~\ref{fig:classification}, which organizes the instability
criteria by the cytosolic topology of the reaction network.
Although the representative networks shown in the table focus on
classes~(a)\textendash(c), the criteria highlight the common structure of the
classification: long-wavelength instabilities are governed by the
unfiltered nullcline-slope matrix, whereas short-wavelength
instabilities are governed by filtered slope matrices.

Instead of the simple scalar 
conditions for single-species systems, the instability criteria for 
multi-species systems are formulated in terms of the eigenvalues of 
(filtered) slope matrices.
These matrix conditions capture the coupling between the different protein 
species.
Importantly, each criterion 
is formulated in the space of total species densities 
rather than the space of all components, 
reducing the dimensionality to the number of conservation laws.
This formulation in terms of conserved densities alone underlines 
that the instability is driven by self-amplifying mass 
redistribution induced by shifting local equilibria.
The relations for eigenvalues of the slope matrices are analogous to the Hessian criterion for spinodal decomposition in equilibrium phase separation:
The Hessian of the local free energy is the slope matrix of the local chemical potentials.
 
The criteria derived above predict 
\emph{stationary} instabilities; however, some of the 
above-mentioned systems of intracellular pattern formation show 
oscillatory instabilities.
In the next section, we extend our arguments perturbatively to 
oscillatory instabilities and weak membrane diffusion. 

\section{Dispersion relation, oscillatory instabilities, and membrane diffusion}
\label{sec:approx-dispRel-oscillatory}

The instability criteria derived in the preceding sections are exact for stationary instabilities, where the growth rate is purely real at the stability boundary (${\sigma = 0}$).
However, the growth rate may have a nonzero imaginary part at the stability boundary
 (${\mathrm{Re}\,\sigma = 0}$, ${\mathrm{Im}\,\sigma \neq 0}$).
We denote such instabilities as oscillatory instabilities.
These cannot be located by the criteria derived from the condition ${\sigma=0}$ in the previous sections.

Locating these oscillatory instabilities exactly, is more complicated.
Viewing $q$ as a parameter of the system, the stability boundary corresponds to a Hopf bifurcation (${\mathrm{Re}\,\sigma(q) = 0}$, ${\mathrm{Im}\,\sigma(q) \neq 0}$), and the Routh-Hurwitz conditions would have to be tested to locate it (see, e.g., Ref.~\cite{Wiggins2003}).
These conditions are difficult to interpret physically, complicating criteria for oscillatory lateral instabilities \cite{Smith.Dalchau2018,Diego.etal2018}.
Alternatively, one has to determine the full complex dispersion relation, not merely its zeros.
 
We now show that, under suitable assumptions on the reaction kinetics, the full dispersion relation can be calculated perturbatively from the nullcline-slope matrices that describe the reactive equilibria of membrane attachment and detachment.
The same analysis also allows one to include weak membrane diffusion, which was neglected in the previous sections.
 
\medskip

\textit{Assumptions.}\;---
The mapping onto nullcline slopes of Secs.~\ref{sec:long-wavelength-limit}--\ref{sec:classification} exploited the marginal-mode construction possible for ${\sigma = 0}$.
For finite $\sigma$, such a construction of the eigenmodes becomes approximate; the following three assumptions control the corrections.
 
First, we assume that cytosolic density variations are small relative to membrane density variations during pattern formation~\footnote{
In systems with an extended bulk (bulk--boundary coupling), we compare the changes in membrane density to the changes in the cytosolic densities integrated perpendicular to the membrane over the height of the cytosolic bulk.},
so that the nullcline slopes satisfy ${|\partial_{\rho_\alpha}c_i^{\beta*}| \ll 1}$ and the eigenvector components obey ${|\delta c_i^\alpha/\delta\rho_\alpha|,\, |\delta c_i^\alpha/\delta m_j| \ll 1}$.
We denote correction terms of this order by $\mathcal{O}_c$.
This assumption reflects the fact that pattern formation requires nonlinear feedback in membrane attachment and detachment. If this feedback is rather weak, large membrane-density variations are required to create weak destabilizing gradients in the cytosolic densities.
Moreover, if the pattern-forming proteins have a strong preference for membrane binding, cytosolic densities are small, and high-contrast patterns in the membrane densities that reliably guide downstream processes require large changes in the membrane densities compared to their associated cytosolic densities.

Second, we assume that membrane diffusion is weak compared to cytosolic diffusion.
We denote correction terms suppressed by a ratio of membrane-to-cytosolic diffusion coefficients by $\mathcal{O}_D$.
Together, assumptions~1 and~2 imply ${|\sigma|/(D_{c,i}^{\alpha}q^2)=\mathcal{O}_c\ll 1}$ [cf.\ Eq.~\eqref{eq:cont-eqs}]: the growth rate is small compared to the rate of cytosolic diffusion.
This ensures that $\sigma$ is insignificant in the cytosolic equations, so the non-binding components remain low-pass filtered exactly as in the stationary analysis.
 
Third, we assume, additionally, that the rate of mass redistribution is comparable or smaller than the rates of the local membrane reactions.
For conciseness, we denote the typical membrane reaction rate by $r$, which can be taken to be the magnitude of the least negative real part of the eigenvalues of the Jacobian $\partial_\mathbf{m}{\mathbf R}_m$ of the membrane reaction term, i.e., as the slowest relaxation rate.
We require ${q^2 D_{c_i^\alpha}/r\lesssim 1}$ and describe terms of this order by $\mathcal{O}_r$.
Consequently, this assumption restricts the largest wavenumbers at which our approximation holds. 
This additional assumption is needed for the membrane equation: combined with ${|\sigma|/(D_{c,i}^{\alpha}q^2)\ll 1}$ from above, it yields ${|\sigma|/r\ll 1}$, ensuring that the finite growth rate can be treated as a lower order correction compared to the membrane reaction rates.
The membrane equation then reduces to local reactive balance at leading order, as in the stationary case.

The precise role of each assumption will become apparent in the perturbation analysis below, where the correction terms are tracked explicitly at each step.
While the assumptions have a clear physical interpretation, whether the assumptions hold in a given system must ultimately be checked against the full linear stability analysis; we do so for the skeleton Min model in Sec.~\ref{sec:application}.
 
\medskip

\textit{Approximate dispersion relation.}\;---
With these approximations, we now perform a linear stability analysis to derive the dispersion relation, retracing the marginal-mode construction of Sec.~\ref{sec:3c} and tracking corrections at each step.
For simplicity, we restrict this analysis to systems with at most two cytosolic components per species and without bulk--boundary coupling [see Eq.~\eqref{eq:RDS-2cyt}].
Generalizing this approximation to cytosolic series and explicit bulk--boundary coupling, following the treatments in Appendices~\ref{app:cytSeries} and~\ref{app:bbc}, is an interesting future task.
 
As in the stationary analysis (Sec.~\ref{sec:3c}; App.~\ref{app:2c}), we parametrize the binding-competent components of the perturbation eigenvector by a nullcline displacement $\widetilde{\delta\bm\rho}$,
\begin{equation}\label{eq:sigma-c1}
    \delta\mathbf{c}_1\equiv \partial_{\bm{\rho}}\mathbf{c}_1^{\alpha*}|_\mathrm{hss}\widetilde{\delta\bm{\rho}} \, .
\end{equation}
This is well-defined whenever the square matrix ${\partial_{\bm{\rho}}\mathbf{c}_1^{\alpha*}|_\mathrm{hss}}$ is invertible (i.e., away from bifurcations of the local reactions).
The linearized membrane equations read [cf.\ Eq.~\eqref{eq:3cMcRD-eigenproblem}]
\begin{equation}
    (\sigma + \mathbf{D}_m q^2) \, \delta \mathbf{m} = \partial_{\mathbf{u}}{\mathbf R}_m|_\mathrm{hss} \, \delta\mathbf{u} \, .
\end{equation}
where ${\mathbf{D}_m=\mathrm{diag}(D_{m,1},\dots)}$ collects the membrane diffusion coefficients.
Because $|\sigma|/r \ll 1$ and $D_m \ll D_c$ (assumptions~1,~2, and~3), the left-hand side is a small perturbation, and at leading order the membrane equation reduces to local reactive balance,
\begin{equation}
    0 = \partial_{\mathbf{u}}{\mathbf R}_m|_\mathrm{hss} \, \delta\mathbf{u}\; \big[1+(\mathcal{O}_c+\mathcal{O}_D)\mathcal{O}_r\big] \, ,
\end{equation}
as in the stationary case.
Using Eq.~\eqref{eq:sigma-c1}, the membrane perturbation is therefore again parametrized by the nullcline slopes,
\begin{equation}
\label{eq:sigma-m}
    \delta\mathbf{m} = \partial_{\bm{\rho}}\mathbf{m}^*|_\mathrm{hss} \widetilde{\delta\bm{\rho}} \; \big[ 1+(\mathcal{O}_c+\mathcal{O}_D) \, \mathcal{O}_r \big] \, .
\end{equation}
 
For the non-binding cytosolic components, the same low-pass filtering as in the stationary analysis applies.
Because $|\sigma|/(D_{c,2}^{\alpha} q^2) \ll 1$ (assumptions~1 and~2), the growth rate can be absorbed into a correction term $(1+\mathcal{O}_c)$ of $c_2^\alpha$ in the linearized equation for $c_2^\alpha$ [cf.\ Eq.~\eqref{eq:inactive-species}], and the finite (linearized) conversion rate $\lambda_\alpha^\mathrm{lin}$ suppresses density variations on wavelengths shorter than the diffusion length ${(\ell_c^\alpha)^2 = D_{c,2}^{\alpha}/\lambda_\alpha^\mathrm{lin}}$ [cf.\ Eqs.~\eqref{eq:suppression-inactive-cytosolic-3c},~\eqref{eq:suppression-inactive-cytosolic}].
This yields the low-pass filtered profiles
\begin{equation}\label{eq:sigma-c2}
    \delta \mathbf{c}_2 
    = 
    (1+ \mathbf{L}^2 q^2)^{-1} \partial_{\bm{\rho}}\mathbf{c}_2^*|_\mathrm{hss} \, \widetilde{\delta\bm{\rho}}
    \; \big[ 1+\mathcal{O}_D\mathcal{O}_r+\mathcal{O}_c \big] \, .
\end{equation}
where ${\mathbf{L}=\mathrm{diag}(\ell_c^\mathrm{A},\ell_c^\mathrm{B},\dots)}$ is the diagonal matrix of diffusion lengths.
The correction terms arise from the finite ratio $\sigma/(D_{c,2}^{\alpha} q^2)$ and from the corrections in $\delta\mathbf{m}$ [Eq.~\eqref{eq:sigma-m}].
 
Inserting the approximate eigenvector components [Eqs.~\eqref{eq:sigma-c1},~\eqref{eq:sigma-m},~\eqref{eq:sigma-c2}] into the linearized continuity equation for species $\alpha$ [cf.\ Eq.~\eqref{eq:cont-eqs}] yields
\begin{align}\label{eq:sigma-cont-eq}
    \sigma\,\delta\rho_\alpha 
    &= -q^2 D_{c,1}^{\alpha}
    \biggl[\partial_{\bm{\rho}}c_1^{\alpha*}
    + \frac{s_{c,2}^\alpha}{1+(\ell_c^\alpha q)^2}\,
    \frac{D_{c,2}^{\alpha}}{D_{c,1}^{\alpha}}\,
    \partial_{\bm{\rho}}c_2^{\alpha*} \nonumber\\
    &\qquad
    + \mathbf{s}_m^\alpha\cdot 
    \frac{\mathbf{D}_m}{D_{c,1}^{\alpha}}\,
    \partial_{\bm{\rho}}\mathbf{m}^*\biggr]
    \, \widetilde{\delta\bm{\rho}}\;
    \bigl[ 1+\mathcal{O}_{D}\mathcal{O}_{r}+\mathcal{O}_{c} \bigr] \, .
\end{align}

The right-hand side now includes a membrane-diffusion contribution (the $\mathbf{D}_m$ term), absent in the previous analysis.
This relation almost constitutes a reduced eigenvalue problem for the growth rate $\sigma$. However,
the left-hand side involves the physical total-density perturbation ${\delta\rho_\alpha = \mathbf{s}_\alpha \cdot \delta\mathbf{u}}$, while the right-hand side is expressed in terms of the nullcline displacement $\widetilde{\delta\bm{\rho}}$.
As discussed in App.~\ref{app:perturbation_rho_detail}, the two differ at finite $q$ due to the wavenumber-dependent filtering of the non-binding components.
Nonetheless, under approximation~2, we find
\begin{equation}
    \delta\rho_\alpha 
    = \widetilde{\delta\rho}_\alpha
    - \frac{s_{c,2}^\alpha\,(\ell_c^\alpha q)^2}
    {1+(\ell_c^\alpha q)^2}\,
    (\partial_{\bm{\rho}}c_2^{*\alpha})\,
    \widetilde{\delta\bm{\rho}}
    = \widetilde{\delta\rho}_\alpha\;
    \big[ 1+\mathcal{O}_c \big] \, .
\end{equation}
Thus, replacing $\widetilde{\delta\bm{\rho}}$ by $\delta\bm{\rho}$ on the right-hand side, we arrive at the approximate eigenvalue problem for the complex dispersion relation (${\alpha = 1, \dots, N_\mathrm{species}}$)
\begin{align}\label{eq:sigma-approx}
    \sigma\,\delta\rho_\alpha 
    &\approx -q^2 D_{c,1}^{\alpha}
    \biggl[\partial_{\bm{\rho}}c_1^{\alpha*}
    + \frac{s_{c,2}^\alpha}{1+(\ell_c^\alpha q)^2}\,
    \frac{D_{c,2}^{\alpha}}{D_{c,1}^{\alpha}}\,
    \partial_{\bm{\rho}}c_2^{\alpha*} \nonumber\\
    &\qquad
    + \mathbf{s}_m^\alpha\cdot 
    \frac{\mathbf{D}_m}{D_{c,1}^{\alpha}}\,
    \partial_{\bm{\rho}}\mathbf{m}^*\biggr]
    \delta\bm{\rho} \, .
\end{align}

This is a reduced eigenvalue problem in the space of conserved densities, that is, of dimension $N_\mathrm{species} \times N_\mathrm{species}$.
It thus describes feedback in the mass redistribution of the different species induced by shifting local reactive equilibria, and expressed entirely in terms of nullcline-slope matrices.
A lateral instability (stationary or oscillatory) occurs when the matrix in square brackets has an eigenvalue with vanishing real part; the corresponding wavenumber $q_\mathrm{min}$ marks the lower edge of the band of unstable modes.
From the denominator of the second term, we read off $q_\mathrm{min}^2 \sim (\ell_c^\alpha)^{-2}$, consistent with the heuristic analysis of Sec.~\ref{sec:3c}.

\medskip
 
\textit{Regime of validity and mass-redistribution instabilities.}\;--- 
The requirement that mass redistribution be slower than local membrane reactions (assumption~3) restricts the approximation to wavenumbers satisfying ${q^2 D_{c,1}^{\alpha}/r\lesssim 1}$.
Since the onset wavenumber scales as $q_\mathrm{min}^2 \sim (\ell_c^\alpha)^{-2} = \lambda_\alpha^\mathrm{lin}/D_{c,2}^{\alpha}$, the instability falls within the regime of validity if the cytosolic conversion rates are comparable to or slower than the membrane reaction rates,
\begin{equation}\label{eq:cond-massRedInstab}
    {\lambda_\alpha^\mathrm{lin}\lesssim r}.
\end{equation}
When this condition holds, the reduced eigenvalue problem Eq.~\eqref{eq:sigma-approx} fully determines lateral stability: the instability is driven by self-amplifying mass redistribution through shifting local reactive equilibria, generalizing \emph{local equilibria theory}~\citep{Brauns.etal2020} to oscillatory onset.
For type-II (long-wavelength) instabilities, onset occurs at ${q = 0}$ and this condition is automatically satisfied.
If Eq.~\eqref{eq:cond-massRedInstab} is violated, the local reactive dynamics on the membrane, not only the reactive equilibria, become important for the instability.
 
The MinE-switch model for the Min system~\citep{Denk.etal2018,Ren.etal2025} illustrates the latter case: it exhibits a type-I instability in the experimentally relevant parameter regime, yet the weak recruitment rate of latent MinE violates $\lambda_\alpha^\mathrm{lin} \lesssim r$, and
 the reduced eigenvalue problem Eq.~\eqref{eq:sigma-approx} fails to predict any lateral instability in this regime.\footnote{To treat the switch model within our framework, we describe the reactive MinE state within a quasi-steady-state approximation, justified by its fast deactivation~\citep{Meindlhumer.etal2023}.}
The instability of the switch model therefore cannot be understood in terms of shifted local attachment-detachment equilibria alone.
The local reaction kinetics contribute significantly.
In contrast, the skeleton Min model~\citep{Huang.etal2003,Fange.Elf2006,Halatek.Frey2012} satisfies Eq.~\eqref{eq:cond-massRedInstab} and is well captured by the reduced eigenvalue problem, as we demonstrate in Sec.~\ref{sec:application} (Fig.~\ref{fig:numerics}).
 
\section{Application: Mechanisms leading to negative slope eigenvalues}
\label{sec:application}

In the preceding sections, we established that pattern-forming instabilities in multi-species McRD systems are governed by the eigenvalues of slope matrices---matrices of nullcline derivatives evaluated at the reactive equilibrium.
For type-II (long-wavelength) instabilities, the relevant criterion is an eigenvalue with a negative real part of the effective diffusion matrix $\mathbf{D}_c\partial_{\bm\rho}\bm\eta^*$ of the mass-redistribution potentials [Eq.~\eqref{eq:Deff-def}], while type-I (short-wavelength) instabilities are controlled by negative eigenvalues of the filtered slope matrices $\partial_{\bm\rho}\mathbf{c}^*_{1}$ of the cytosolic densities [Eq.~\eqref{eq:s-of-q}].
In this section, we analyze how negative eigenvalues of these slope matrices arise from the interplay of the underlying biochemical interactions.
We focus on the case of two conserved species, where the slope matrices are two-dimensional and their eigenvalues can be expressed in terms of the matrix trace and determinant.
This yields two distinct routes to instability: single-species feedback (negative trace) and inter-species coupling feedback (negative determinant).
We illustrate both mechanisms using two biological model systems---the Min system in \textit{E.~coli} (Sec.~\ref{sec:app-min}) and the PAR polarity system in \textit{C.~elegans} (Sec.~\ref{sec:app-par})---and show that they realize the two different instability mechanisms.

For a single conserved species, a negative slope implies that cytosolic densities decrease in regions of large total density, driving additional diffusive mass flux toward these regions and thus destabilizing the homogeneous steady state.
For multi-species systems, the analogous condition is a negative eigenvalue of the corresponding slope matrix, but the connection to specific biochemical interactions is less immediate.

To make this connection explicit, we consider two conserved species, labeled A and~B, with total densities $\rho_\mathrm{A}$ and $\rho_\mathrm{B}$.
The entry $\mathcal{S}_{\alpha\beta}$ of a slope matrix $\mathcal{S}$ is the derivative of a characteristic quantity of species~$\alpha$ (its mass-redistribution potential $\eta^*_\alpha$ or a cytosolic density $c^{\alpha*}_{1}$) with respect to the total density $\rho_\beta$ of species~$\beta$, evaluated at the reactive equilibrium:
\begin{equation}
    \mathcal{S}
    = \begin{pmatrix}
        \mathcal{S}_\mathrm{AA} & \mathcal{S}_\mathrm{AB} \\[4pt]
        \mathcal{S}_\mathrm{BA} & \mathcal{S}_\mathrm{BB}
    \end{pmatrix},
    \qquad
    \mathcal{S}^{\mathrm{I}}_{\alpha\beta}
    = \frac{\partial \eta^*_\alpha}{\partial \rho_\beta}
    \;\;\text{and}\;\;
    \mathcal{S}^{\mathrm{II}}_{\alpha\beta}
    = 
    \frac{\partial c^{\alpha*}_{1}}{\partial \rho_\beta} \,.\label{eq:slope_criterion_example}
\end{equation}
Here, $\mathcal{S^{\mathrm{II}}}$ governs type-II (long-wavelength) instabilities while $\mathcal{S^{\mathrm{I}}}$ governs type-I (short-wavelength) instabilities, assuming equal cytosolic diffusion coefficients and conversion rates for simplicity.
The diagonal entries $\mathcal{S}_\mathrm{AA}$ and $\mathcal{S}_\mathrm{BB}$ are the single-species nullcline slopes at fixed density of the other species---they reduce to the familiar single-species criterion $\partial_\rho \eta^*$ when the species are decoupled.
The off-diagonal entries $\mathcal{S}_\mathrm{AB}$ and $\mathcal{S}_\mathrm{BA}$ quantify the inter-species coupling: how the reactive equilibrium of one species shifts in response to a change in the total density of the other.

Under what conditions does $\mathcal{S}$ have a negative eigenvalue?
Given the trace ${\tau = \mathcal{S}_\mathrm{AA} + \mathcal{S}_\mathrm{BB}}$ and determinant ${\delta = \mathcal{S}_\mathrm{AA}\,\mathcal{S}_\mathrm{BB} - \mathcal{S}_\mathrm{AB}\,\mathcal{S}_\mathrm{BA}}$ of $\mathcal{S}$, its eigenvalues $\sigma_\pm$ are
\begin{equation}
    2\,\sigma_\pm
    = \tau \pm \sqrt{\tau^2 - 4\,\delta} \,.
\end{equation}
At least one eigenvalue has a negative real part if the trace and/or determinant are negative.
Since the trace is the sum of the single-species nullcline slopes, $\tau = \mathcal{S}_\mathrm{AA} + \mathcal{S}_\mathrm{BB}$, a negative trace indicates that both species are individually unstable, or that the pattern-forming tendency of one species (negative diagonal entry) outweighs the stabilizing tendency of the other (positive diagonal entry).
Even if the trace is non-negative, one species may still have a negative diagonal entry while the other's positive entry dominates the sum.
In this case, the diagonal entries have opposite signs, so their product $\mathcal{S}_\mathrm{AA}\,\mathcal{S}_\mathrm{BB}$ is negative, and the determinant $\delta = \mathcal{S}_\mathrm{AA}\,\mathcal{S}_\mathrm{BB} - \mathcal{S}_\mathrm{AB}\,\mathcal{S}_\mathrm{BA}$ is negative unless the inter-species coupling provides a sufficiently strong stabilizing contribution ($\mathcal{S}_\mathrm{AB}\,\mathcal{S}_\mathrm{BA} < 0$ with $|\mathcal{S}_\mathrm{AB}\,\mathcal{S}_\mathrm{BA}| > |\mathcal{S}_\mathrm{AA}\,\mathcal{S}_\mathrm{BB}|$).
These constitute instabilities driven by the feedback of a single species.

\begin{figure}
	\includegraphics{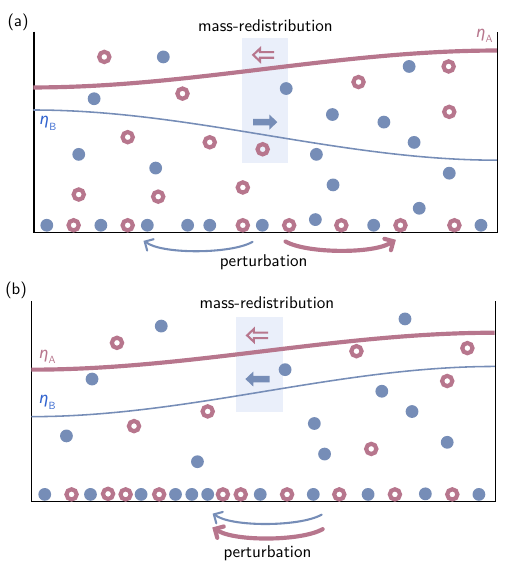}
	\caption{(a)~Coupling-driven instability mechanism for positive coupling characterized by $\mathcal{S}_\mathrm{AB},\, \mathcal{S}_\mathrm{BA} > 0$: A local increase in the total density of species~A (thick red circles) enhances the mass-redistribution potential $\eta_\mathrm{B}$ of species~B (blue dots), thereby increasing its cytosolic density $c_\mathrm{B}$.
    Because the local total amount of species~B remains conserved, this leads to its detachment and a higher $c_\mathrm{B}$ concentration on the perturbed side, driving a diffusive flux of~B toward the opposite side.
    As a consequence, species~B accumulates on that opposite side (blue curve line), increasing the mass-redistribution potential $\eta_\mathrm{A}$ there.
    In turn, the cytosolic~A density $c_\mathrm{A}$ increases on that side, and species~A diffuses back, closing a feedback loop that leads to spatial segregation: species~A accumulates on one side, while species~B accumulates on the other.
    (b) Same mechanism for negative coupling $\mathcal{S}_\mathrm{AB},\, \mathcal{S}_\mathrm{BA} < 0$: A local  increase in $\rho_\mathrm{A}$ decreases $\eta_\mathrm{B}$, leading to increased attachment of~B rather than detachment.
    Consequently, both A and~B accumulate on the same side.
    }
	\label{fig:coupling}
\end{figure}

Beyond such single-species-driven instabilities, that is, if none of the single species shows a negative slope and the diagonal entries are positive (implying a positive trace), an instability can still arise due to positive feedback in the coupling of the species.
Specifically, the determinant still becomes negative if
\begin{equation}\label{eq:positive-cycle-condition}
    \mathcal{S}_\mathrm{AB}\,\mathcal{S}_\mathrm{BA}
    \;>\;
    \mathcal{S}_\mathrm{AA}\,\mathcal{S}_\mathrm{BB}
    \;>\; 0 \,.
\end{equation}
Positivity of the product of the coupling terms $\mathcal{S}_\mathrm{AB}\,\mathcal{S}_\mathrm{BA}$ implies a positive feedback cycle between the two species:
If both are positive, i.e., $\mathcal{S}_\mathrm{AB}, \mathcal{S}_\mathrm{BA} > 0$, a density increase in species~A increases the mass-redistribution potential $\eta_\mathrm{B}$ (or cytosolic density, depending on the slope matrix analyzed), leading to diffusive depletion of the density of species~B.
Because species~B also increases $\eta_\mathrm{A}$, this depletion leads to a decrease of $\eta_\mathrm{A}$.
As a result, the mass-redistribution potentials $\eta_\mathrm{A,B}$ are decreased in regions of large densities $\rho_\mathrm{A,B}$ (see Fig.~\ref{fig:coupling}).
Similar feedback occurs if both coupling terms $\mathcal{S}_\mathrm{AB}, \mathcal{S}_\mathrm{BA}$ are negative.
The condition Eq.~\eqref{eq:positive-cycle-condition} ensures that this positive feedback is sufficiently strong compared to the stabilizing single-species processes mediated by the diagonal entries $\mathcal{S}_\mathrm{AA}, \mathcal{S}_\mathrm{BB}$, and induces a lateral instability.

Taken together, the coupled two-species system can become laterally unstable due to single-species feedback (negative diagonal entry in the slope matrix) or due to a destabilizing positive feedback loop between species with sufficiently large off-diagonal entries satisfying $\mathcal{S}_\mathrm{AB}\,\mathcal{S}_\mathrm{BA} > 0$).
In more-species systems, the negative eigenvalues can be related to destabilizing cycles using graph theory, refer to Ref.~\citep{Diego.etal2018}.
Importantly, the advantage of performing this analysis on the slope matrix instead of the full Jacobian of the linear stability analysis is that its dimension is given by the number of conservation laws, i.e., the number of species which, for proteins, is typically much smaller than the total number of components (different protein conformational states and complexes), which sets the dimension of the Jacobian.

We now apply this framework to two biological model systems.
For each, the figures show the instability regions in a parameter plane and compare the full numerical linear stability analysis (black dots and circles) with the slope-matrix criteria (shaded area: multiple-species criteria; contour lines: reduced single-species criteria).

\subsection{Min system: single-species-driven instability}
\label{sec:app-min}

\begin{figure*}
	\includegraphics[width=\linewidth]{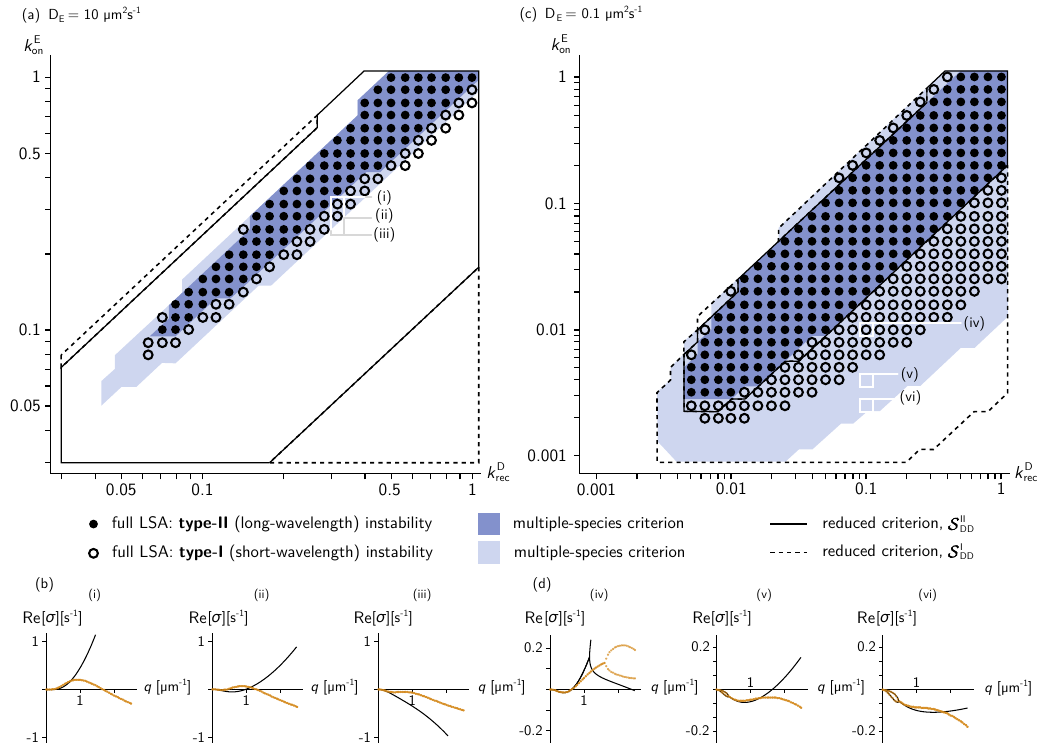}
	\caption{
    Linear stability analysis of the skeleton Min model, comparing the full numerical analysis (dots) with the reduced eigenvalue problem Eq.~\eqref{eq:sigma-approx} (shades and contour lines). The full parameter grid was scanned numerically by LSA. In the simplified representation, only unstable grid points are plotted; stable grid points, although analyzed, are left blank.
    (a)~Phase diagram in the plane of the two reaction rates $(k^{\mathrm{E}}_\mathrm{on},\, k^{\mathrm{D}}_\mathrm{rec})$ for physiological MinE diffusion $D_\mathrm{E} = \SI{10}{\micro m^2/s}$ (cf.\ Appendix~\ref{app:model}).
    Black dots: type-II (long-wavelength) instability from the full linear stability analysis; black circles: type-I (short-wavelength) instability.
    The instability regions predicted by the reduced eigenvalue problem Eq.~\eqref{eq:sigma-approx} are shown as shaded regions (dark blue: type-II, light blue: type-I instability).
    Black solid contour: region where the MinD nullcline slope of the mass-redistribution potential is negative (${\mathcal{S}_\mathrm{DD}^{\mathrm{II}}=\partial_{\rho_\mathrm{D}} \eta^*_\mathrm{D} < 0}$), corresponding to the single-species type-II instability criterion at fixed MinE density.
    Black dashed contour: region where the MinD nullcline slope of the binding-competent cytosolic density is negative (${\mathcal{S}_\mathrm{DD}^{\mathrm{I}}=\partial_{\rho_\mathrm{D}} c^*_{\mathrm{D,1}} < 0}$), corresponding to the single-species type-I instability criterion at fixed MinE density.
    The full two-species instability regions (dots and shades) are contained within these single-species regions, showing that MinE redistribution counteracts the instability induced by MinD.
    (c)~Real part of dispersion relations $\sigma(q)$ at the parameter points marked in panel~(a); the upper, middle, and lower markers correspond to the left, center, and right plots, respectively.
    A single branch denotes a complex-conjugate pair with nonzero imaginary part, omitted here.
    orange dots: real part from the full linear stability analysis.
    black solid lines: prediction from the reduced eigenvalue problem Eq.~\eqref{eq:sigma-approx}.
    Because the model has two conserved species, the reduced problem yields two branches that capture the leading eigenvalues well, particularly at small to intermediate wavenumbers.
    (b,d)~Same analysis as in (a,c) for reduced MinE diffusion $D_\mathrm{E} = \SI{0.1}{\micro m^2/s}$.
    The model equations and other parameters are given in Appendix~\ref{app:model}.
	}
	\label{fig:numerics}
\end{figure*}

We show as an example the skeleton model of the Min system in Fig.~\ref{fig:numerics}.
This model has two conservation laws, one for MinD and one for MinE [see Fig.~\ref{fig:model-systems}(b)].
MinD has two cytosolic components (MinD-ATP and MinD-ADP), which allows the model to exhibit both type-I (short-wavelength) and type-II (long-wavelength) instabilities.
The phase diagrams in Fig.~\ref{fig:numerics}(a,b) show the instability regions in the $(k^{\mathrm{E}}_\mathrm{on},\, k^{\mathrm{D}}_\mathrm{rec})$ plane for two values of the MinE cytosolic diffusion coefficient, comparing the full linear stability analysis with the predictions from the approximate dispersion relation Eq.~\eqref{eq:sigma-approx}.
The close agreement confirms that the instabilities of the skeleton model can be understood as mass-redistribution instabilities governed by the slope matrices.
The corresponding dispersion relations $\sigma(q)$ at selected parameter points are shown in Fig.~\ref{fig:numerics}(c,d), where the approximate dispersion relation well describes the two leading branches: Because MinD and MinE are the two conserved species, this system has two zero eigenvalues at ${q=0}$, and the approximate dispersion relation determined from the $2 \times 2$-dimensional slope matrices traces these two eigenvalues as $q$ increases away from zero.

Both the long- and short-wavelength instability regions lie within the contours marking negative diagonal entries ${\partial_{\rho_\mathrm{D}}\eta_\mathrm{D}^*<0}$ and ${\partial_{\rho_\mathrm{D}}c_{\mathrm{D}^{\mathrm{ATP}}}^*<0}$, respectively [see Fig.~\ref{fig:numerics}].
This suggests that MinE redistribution counteracts the instability induced by MinD, reducing the parameter regime that shows instability.
The counteracting effect of MinE redistribution becomes clear by comparing Fig.~\ref{fig:numerics}(a) and~(b): reducing the MinE diffusion coefficient from $D_\mathrm{E} = \SI{10}{\micro m^2/s}$ to $D_\mathrm{E} = \SI{0.1}{\micro m^2/s}$ slows MinE redistribution and expands the instability region.
This shows that the MinD dynamics under a fixed uniform MinE distribution drives the pattern-forming feedback.
The diagonal entry $\mathcal{S}_\mathrm{DD}$ (the MinD nullcline slope) is negative and drives the instability, while $\mathcal{S}_\mathrm{EE}$ (the MinE nullcline slope) is positive and stabilizing.

\begin{figure}
	\includegraphics{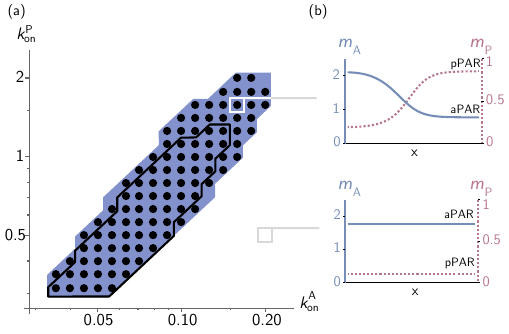}
	\caption{
    Linear stability analysis of the PAR model, comparing the full numerical analysis (dots) with the reduced instability criteria derived from the slope matrix. The full parameter grid was scanned numerically by LSA. In the simplified representation, only unstable grid points are plotted; stable grid points, although analyzed, are left blank.
    (a)~Phase diagram in the plane of the linear attachment rates $(k_{\mathrm{on}}^\mathrm{A},\, k_{\mathrm{on}}^\mathrm{P})$ (cf.\ Appendix~\ref{app:model}).
    Black dots: type-II (long-wavelength) instability from the full linear stability analysis;
    blue shading: instability regions predicted by the reduced slope-matrix analysis.
    Black solid contour: instability region obtained when the total pPAR density is kept fixed, corresponding to the effective single-species criterion for an aPAR-driven type-II instability.
    For the chosen parameter values, no pPAR-driven type-II instability is found.
    The area of blue shading outside the black solid contour indicates the region where the determinant becomes negative due to the inter-species coupling [Eq.~\eqref{eq:positive-cycle-condition}].
    (b)~Simulation of spatial profiles on a domain with no-flux boundary conditions and a length $L=50~\mu\mathrm{m}$ shows aPAR and pPAR accumulate on opposite sides.
    The model equations and other parameters are given in Appendix~\ref{app:model}.}
	\label{fig:par}
\end{figure}

\subsection{PAR system: coupling-driven instability}
\label{sec:app-par}
In contrast, the PAR polarity system realizes the coupling-driven instability mechanism, as we now discuss based on the results shown in Fig.~\ref{fig:par}.
The PAR system (``PAR'' for \emph{partitioning-defective}) is a conserved polarity module carefully studied in \textit{C.~elegans}.
The model introduced in Ref.~\citep{Goehring.etal2011} considers only a single cytosolic component for each aPAR (anterior PAR) and pPAR (posterior PAR) protein species [see Fig.~\ref{fig:model-systems}(a)].
Both species linearly attach onto and detach from the membrane, with their detachment enhanced in the presence of the other species (mutual detachment).
Pattern formation is driven by the mutual interaction of the proteins, and this system forms stationary patterns in which aPAR and pPAR proteins occupy distinct regions on the membrane [see Fig.~\ref{fig:par}(b)].

Because all species have only a single cytosolic component, the model only shows type-II (long-wavelength) stationary instabilities (numerically no oscillatory instabilities are found for the analyzed parameter values).
Figure~\ref{fig:par}(a) shows that only a part of the instability region is explained by a negative nullcline slope of aPAR keeping the total density of pPAR constant [negative diagonal entry ${\partial_{\rho_\mathrm{A}}\eta_\mathrm{A}^*<0}$ in slope matrix;~see Fig.~\ref{fig:par}(d), solid contour].
The negative diagonal term results from large aPAR densities inducing the detachment of pPAR, which thus induces less detachment of aPAR, resulting in an accumulation of aPAR on the membrane and reduced cytosolic densities (reduced mass-redistribution potential).
Due to asymmetric parameter values chosen (see Appendix~\ref{app:par}) \cite{Goehring.etal2011}, the pPAR nullcline slope at constant total aPAR density never becomes negative.
Outside this region of negative aPAR nullcline slope, the positive feedback cycle between the two species induced by $\mathcal{S}_\mathrm{AB}\,\mathcal{S}_\mathrm{BA} > 0$ destabilizes the system [see Fig.~\ref{fig:par}(d), dashed contour].
This criterion for the off-diagonal coupling terms well captures the additional region of instability.
Nonetheless, the same molecular interactions drive pattern formation in both regions.
Thus, this model highlights an important point for interpretation: the two instability mechanisms identified in the slope matrix need not stem from different molecular interactions.
However, together, they provide a framework for interpreting the molecular interactions.

\subsection{Data-driven reconstruction of instability regions}\label{sec:app-data}

A key practical advantage of the slope-matrix criterion is that it can be evaluated without knowledge of the underlying reaction network or rate constants.
The entries of $\mathcal{S}$ are derivatives of reactive-equilibrium quantities with respect to total species densities, and these can be estimated directly from measurements of the well-mixed (spatially homogeneous) steady state at different total protein concentrations.

To demonstrate this, we performed a numerical concentration scan for the PAR model: for each point $(\rho_\mathrm{A},\, \rho_\mathrm{P})$ on a grid in the total-density plane, we computed the spatially homogeneous reactive equilibrium and recorded the concentrations of all four components (membrane-bound and cytosolic aPAR and pPAR).

\begin{figure}[t]
\includegraphics{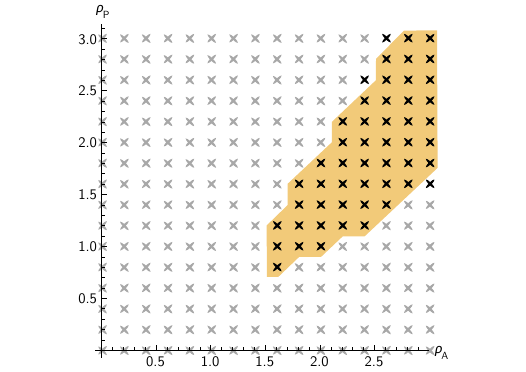}
	\caption{
    Comparison of theoretical and data-driven instability predictions for the PAR model (with $k^{\mathrm{P}}_{on}=1/\mathrm{s}$ and $k^{\mathrm{A}}_{on} = 0.1/\mathrm{s}$).
    The orange-shaded area indicates the unstable region obtained from the full linear stability analysis (the reference method).
    Each point represents an equilibrium concentration pair $(\rho_\mathrm{A},\, \rho_\mathrm{P})$.
    The crosses (gray vs.\ black) shows whether the slope-based estimate (calculated with 4 nearest neighbors) classifies the system as stable or unstable. The model equations and other parameters are given in Appendix~\ref{app:model}.
    }
	\label{fig:concentration-prediction}
\end{figure}

We then estimated the entries of $\mathcal{S}$ by numerical finite differences: for each grid point, we identified nearest neighbors along each coordinate direction and computed the partial derivatives $\partial \eta^*_\alpha / \partial \rho_\beta$ by central differences.
Evaluating the trace and determinant of the resulting $2 \times 2$ slope matrix at each point classifies the homogeneous steady state as laterally stable or unstable.
The result is shown in Fig.~\ref{fig:concentration-prediction}, where the data-driven classification (crosses) is compared with the unstable region obtained from the full linear stability analysis (orange-shaded area) in the $(\rho_\mathrm{A},\, \rho_\mathrm{P})$ plane.
The agreement demonstrates that the slope criterion can be applied in a model-agnostic manner, requiring only the dependence of reactive-equilibrium concentrations on total protein amounts.
Experimentally, this dependence is accessible by titrating total protein concentrations in a well-mixed bulk assay and measuring the resulting steady-state partitioning between membrane-bound and cytosolic states---without the need to resolve spatial structure or to know the reaction rates.

\section{Discussion and Outlook}
\label{sec:conclusion}

We have developed a classification of pattern-forming instabilities in mass-conserving reaction-diffusion systems based on the local geometry of the reactive equilibrium manifold, reducing the linear stability problem from the full component space to the space of conserved species.
The geometry is encoded in slope matrices whose entries are derivatives of the mass-redistribution potentials and of the binding-competent cytosolic densities with respect to the conserved total species densities, evaluated at the homogeneous steady state.
Whether a homogeneous state is stable or unstable against lateral perturbations can be read off from the eigenvalue structure of these matrices, with no knowledge of individual reaction rates required.
This parallels the curvature criterion for spinodal decomposition in multicomponent equilibrium mixtures, where stability is set by the local curvature of the free-energy landscape encoded in its Hessian, that is, the slopes of the local chemical potentials.
Here the role of the local chemical potentials is played by the reactive equilibrium manifold, and the role of the Hessian by the slope matrices.
The criterion thus extends a familiar equilibrium concept to nonequilibrium reaction--diffusion networks.
Other than in phase separation (Model B dynamics) where the instability always arises at long wavelengths, we here show that steady-state criteria exist for both stationary long- and short-wavelength dynamic instabilities.
A practical consequence is that the sufficient conditions for instability depend only on the eigenvalues of the slope matrices, accessible from steady-state titration without any kinetic fitting and knowledge of the complete reaction network.

Beyond the stability criterion, the analysis identifies a unifying physical mechanism: pattern formation in a broad class of intracellular systems proceeds via a \textit{mass-redistribution instability}, a self-amplifying redistribution of protein mass driven by shifts in the local reactive equilibria, independently of the detailed local reaction dynamics.
This view  extends the local equilibria theory~\citep{Brauns.etal2020} from single-species to multi-species systems, and from long-wavelength to finite-wavelength instabilities.
At long wavelengths, instability is governed by the slope matrix of the mass-redistribution potentials: a negative eigenvalue of this matrix signals a type-II (long-wavelength, Cahn-Hilliard-like) instability \cite{Brauns.etal2021}. When a species carries multiple cytosolic components, a stationary finite-wavelength type-I (short-wavelength, conserved Turing) onset becomes accessible, controlled instead by the slope matrix of the binding-competent cytosolic densities. The exact criteria and the corresponding rate-independent sufficient conditions are collected in the classification table of Fig.~\ref{fig:classification}.

The mechanism for type-I instability is diffusive averaging: at scales below the cytosolic diffusion length $\ell_c = \sqrt{D_c/\lambda}$, gradients in the non-binding cytosolic components are smoothed out by diffusion before conversion can deplete them, removing their stabilising gradients and letting the destabilising autocatalytic feedback in the binding-competent component dominate.
The selected length scale of the resulting pattern is set by $\ell_c$ itself: it shrinks as the conversion rate increases or the cytosolic diffusivity decreases.

Application of the framework to two biological systems exposes two distinct biochemical routes to a negative eigenvalue of the slope matrix.
In the Min system [Fig.~\ref{fig:numerics}], instability is single-species driven: MinD self-recruitment is sufficient on its own to make the diagonal entry of the slope matrix negative, with MinE redistribution playing a stabilizing role.
The PAR system [Fig.~\ref{fig:par}] supports two regimes.
In the intermediate regime, the aPAR diagonal entry is negative, though caused by inter-species feedback between aPAR and pPAR.
Outside this regime, the  diagonal entries are positive and instability is induced by inter-species positive feedback, with the product of cross entries exceeding the product of diagonals.
For larger networks, the same eigenvalue analysis connects to the recent classification of destabilising feedback cycles in the reaction graph~\citep{Diego.etal2018}.
The advantage of the criteria based on slope-matrices is that only the interactions between the different species, but not all protein states separately, has to be analyzed, reducing the dimensionality of the eigenvalue problem significantly. 
Moreover, the required steady-state titration data are obtainable from cell-free \emph{in vitro} reconstitutions of purified components or from \emph{in vivo} perturbations of protein levels, making the slope-matrix criterion a model-agnostic diagnostic accessible from steady-state measurements alone.

The exact slope-matrix criteria require negligible membrane diffusion and a stationary instability.
Beyond this limit, we derived an approximate reduced eigenvalue problem [Eq.~\eqref{eq:sigma-approx}] that captures the full complex dispersion relation up to low wavenumbers of the band of unstable modes in the reduced space of conserved species, valid when cytosolic density variations are small compared to membrane ones and when cytosolic conversion is not faster than the membrane reaction rate.
When cytosolic conversion becomes faster than the membrane reaction rate, the local reactive dynamics on the membrane become essential and the slope-matrix classification fails.
The MinE-switch model~\citep{Denk.etal2018,Ren.etal2025} of the Min system is a concrete example of this kinetics-dependent regime, and identifying what replaces the geometric criterion in such systems remains an open problem.
Two further restrictions of the present framework should be noted.
First, our cytosolic topology assumes that only the binding-competent cytosolic component couples to the membrane reaction term; systems where multiple cytosolic states directly participate in membrane attachment, or where cytosolic reactions provide their own positive feedback, lie outside this scope.
Second, the criteria characterize the linear lateral instability, not the subsequent nonlinear dynamics; the final pattern amplitude and selected wavelength require a separate nonlinear analysis, see, e.g., Refs.~\citep{Brauns.etal2021,Brauns.etal2021b}.

An exciting future direction is an  extension of this approach to systems without strict mass conservation: in cells responding to externally imposed morphogen gradients, effective slope matrices describing the response of non-diffusing, cell-autonomous components to diffusing signals could yield analogous pattern-forming criteria without modelling the full intracellular network for developmental patterning systems \cite{Smith.Dalchau2018}.
On the experimental side, the data-driven reconstruction demonstrated for the PAR model is directly applicable to titration scans of other reconstituted systems, mapping the instability boundary in a model-agnostic manner and identifying which regions of concentration space support spatial self-organization.
We expect that this approach can also be used to design novel biochemical pattern-forming systems.
More broadly, structural and dynamical connections between non-equilibrium reaction-diffusion dynamics and phase-separating systems have also been found for their fully nonlinear dynamics \cite{Schlogl1972,Tateno.Ishihara2021,Weyer.etal2023,Miller.etal2023,Weyer.etal2025a,Zhou.Frey2026}, together deepening the link between nonequilibrium patterning and equilibrium phase separation. 
Taken together, the slope-matrix framework reduces the question of whether a given protein network can pattern to a property of its reactive equilibria, a quantity that is measurable, system-specific, and independent of kinetic detail.
It makes accessible not just whether patterning occurs in the Min and PAR systems and in future reconstituted networks, but which molecular interactions are responsible and how robustly they support protein pattern formation.

\begin{acknowledgments}
    
This work was funded by the Deutsche Forschungsgemeinschaft (DFG, German Research Foundation) through the Excellence Cluster ORIGINS under Germany's Excellence Strategy (EXC-2094 -- 390783311), the Excellence Cluster BioSystem under Germany's Excellence Strategy (EXC3092/1-533751719), the European Union (ERC, CellGeom, project number 101097810), and the Chan-Zuckerberg Initiative (CZI). C.Y.L. is a member of the Graduate School of Quantitative and Molecular Biosciences (QMB).
 
\end{acknowledgments}

\section*{Data Availability Statement}
The data and code that support the findings of this article are openly available at Ref.~\cite{weyer2026github}.

\appendix

\cleardoublepage

\section{Biological Model Systems}
\label{app:model}
Here we specify the two reaction-diffusion models used in the main text and list the parameter values employed in the numerical results.

\subsection{The MinDE system}

We consider a minimal model of the \textit{E.~coli} MinDE system \citep{Huang.etal2003,Fange.Elf2006,Halatek.Frey2012} that captures the core ATPase cycle of the MinD/MinE pair.
The system consists of two mass--conserving species, MinD and MinE, represented by five components (see Tab.~\ref{tab:skeleton_components}).

\begin{table}[h!]
\centering
\begin{tabular}{lcl}
\multicolumn{3}{c}{\textbf{Components in the MinDE model}} \\
\toprule
\textbf{Component} & \textbf{concentration} & \textbf{Localization} \\
\midrule
MinD--ADP              & $c_{\mathrm{D}^{\mathrm{ADP}}}$   & cytosolic \\
MinD--ATP              & $c_{\mathrm{D}^{\mathrm{ATP}}}$   & cytosolic \\
MinE                  & $c_{\mathrm{E}}$    & cytosolic \\
membrane-bound MinD   & $m_{\mathrm{D}}$    & membrane \\
MinDE complex         & $m_{\mathrm{DE}}$   & membrane \\
\bottomrule
\end{tabular}
\caption{Components of the MinDE model.}
\label{tab:skeleton_components}
\end{table}

Defining the state vector of cytosolic and membrane components as
\begin{align}
\mathbf u
:=
\big(c_{\mathrm{D}^{\mathrm{ADP}}},\,c_{\mathrm{D}^{\mathrm{ATP}}},\,c_{\mathrm{E}},\,m_{\mathrm{D}},\,m_{\mathrm{DE}}\big)^{\! \top},
\end{align}
the reaction--diffusion equations read
\begin{align}
\partial_t \mathbf u
=
\mathbf{D}\,\boldsymbol{\nabla}^2 \mathbf u
+\mathbf R(\mathbf u),
\end{align}
where the diffusion matrix is
\begin{align}
\mathbf{D}
:=
\mathrm{diag} \big(D_c^{\mathrm{D}},\,D_c^{\mathrm{D}},\,D_c^{\mathrm{E}},\,D_m^{\mathrm{D}},\,D_m^{\mathrm{DE}}\big),
\end{align}
and the reaction fluxes read
\begin{align}
\mathbf R
=
\begin{pmatrix}
-\lambda\, c_{\mathrm{D}^{\mathrm{ADP}}} + d_{\mathrm{de}}(m_{\mathrm{DE}})\\[0.15em]
\phantom{-}\lambda\, c_{\mathrm{D}^{\mathrm{ADP}}} - a_{\mathrm{D}}(m_{\mathrm{D}},m_{\mathrm{DE}})\,c_{\mathrm{D}^{\mathrm{ATP}}}\\[0.15em]
-\,a_{\mathrm E}(m_{\mathrm{D}},m_{\mathrm{DE}})\,c_{\mathrm{E}} + d_{\mathrm{de}}(m_{\mathrm{DE}})\\[0.15em]
\phantom{-}\,a_{\mathrm{D}}(m_{\mathrm{D}},m_{\mathrm{DE}})\,c_{\mathrm{D}^{\mathrm{ATP}}} - a_{\mathrm E}(m_{\mathrm{D}},m_{\mathrm{DE}})\,c_{\mathrm{E}}\\[0.15em]
\phantom{-}\,a_{\mathrm E}(m_{\mathrm{D}},m_{\mathrm{DE}})\,c_{\mathrm{E}} - d_{\mathrm{de}}(m_{\mathrm{DE}})
\end{pmatrix},\label{eq:skeleton_reaction_vector}
\end{align}
with
\begin{subequations}
\begin{align}
a_{\mathrm{D}}(m_{\mathrm{D}},m_{\mathrm{DE}}) &=  k_{\mathrm{on}}^{\mathrm{D}} + k_{\mathrm{rec}}^{\mathrm{D}}\,m_{\mathrm{D}},\\
a_{\mathrm E}(m_{\mathrm{D}},m_{\mathrm{DE}}) &= k_{\mathrm{on}}^{\mathrm{E}}\,m_{\mathrm{D}},\\
d_{\mathrm{de}}(m_{\mathrm{DE}}) &= k_{\mathrm{hyd}}\,m_{\mathrm{DE}}.
\end{align}
\end{subequations}

Cytosolic MinD--ADP ($c_{\mathrm{D}^{\mathrm{ADP}}}$) undergoes nucleotide exchange at rate $\lambda$ to form MinD--ATP ($c_{\mathrm{D}^{\mathrm{ATP}}}$), which attaches to the membrane with a basal rate $k_{\mathrm{on}}^{\mathrm{D}}$ and a self-recruitment rate $k_{\mathrm{rec}}^{\mathrm{D}}\,m_{\mathrm{D}}$ proportional to the membrane-bound MinD density.
Membrane-bound MinD recruits cytosolic MinE ($c_{\mathrm E}$) at rate $k_{\mathrm{on}}^{\mathrm{E}}\,m_{\mathrm{D}}$ to form MinDE complexes ($m_{\mathrm{DE}}$), which detach upon ATP hydrolysis at rate $k_{\mathrm{hyd}}$, releasing MinD--ADP and MinE back into the cytosol \citep{Raskin.DeBoer1999,Hu.Lutkenhaus1999,Hu.Lutkenhaus2001,Huang.etal2003,Halatek.Frey2012,Lutkenhaus2007,Heermann.etal2021}.

The parameter values (Tab.~\ref{tab:min_parameters}) are based on Ref.~\citep{Halatek.Frey2018}, with selected parameters varied to explore the instability phase diagram.

\begin{table}[h!]
\centering
\renewcommand{\arraystretch}{1.2}
\begin{tabular}{>{\centering\arraybackslash}p{2.5cm} 
                >{\centering\arraybackslash}p{2.5cm} 
                >{\centering\arraybackslash}p{2.5cm}}
\multicolumn{3}{c}{\textbf{Parameter values used in the MinDE model}} \\
\toprule
\textbf{Parameter} & \textbf{Value} & \textbf{Unit} \\
\midrule
$\rho_{\mathrm{D}}$              & 625     & $\mu\mathrm{m}^{-3}$ \\
$\rho_{\mathrm{E}}$              & 500     & $\mu\mathrm{m}^{-3}$ \\
$D_c^{\mathrm{D}}$                 & 16      & $\mu\mathrm{m}^2\!/\mathrm{s}$ \\
$D_c^{\mathrm{E}}$                 & 0.1 - 10      & $\mu\mathrm{m}^2\!/\mathrm{s}$ \\
$D_m^{\mathrm{D}}$                 & 0.013   & $\mu\mathrm{m}^2\!/\mathrm{s}$ \\
$D_m^{\mathrm{DE}}$                & 0.013   & $\mu\mathrm{m}^2\!/\mathrm{s}$ \\
$k_{\mathrm{on}}^{\mathrm{D}}$    & 0.4     & $1/\mathrm{s}$ \\
$k_{\mathrm{rec}}^{\mathrm{D}}$   & 0.001 - 1  & $\mu\mathrm{m}^2\!/\mathrm{s}$ \\
$k_{\mathrm{on}}^{\mathrm{E}}$        & 0.001 - 1  & $\mu\mathrm{m}^2\!/\mathrm{s}$ \\
$k_{\mathrm{hyd}}$                 & 1       & $1/\mathrm{s}$ \\
$\lambda$                        & 6       & $1/\mathrm{s}$ \\
\bottomrule
\end{tabular}
\caption{Parameter values for the MinDE model.}
\label{tab:min_parameters}
\end{table}

\subsection{PAR system}
\label{app:par}

As a second example, we consider a model of the PAR polarity system in C.~elegans zygotes, based on the formulation of Goehring \textit{et al.}~\citep{Goehring.etal2011} with the addition of explicit cytosolic diffusion \citep{Tostevin.Howard2008}.
The system consists of two mass--conserving species, anterior PAR (aPAR) and posterior PAR (pPAR), each comprising a cytosolic and a membrane--bound component (see Tab.~\ref{tab:Par_components}).

\begin{table}[t!]
\centering
\begin{tabular}{lcl}
\multicolumn{3}{c}{\textbf{Components in the PAR model}} \\
\toprule
\textbf{Component} & \textbf{concentration} & \textbf{Localization} \\
\midrule
cytosolic aPAR        & $c_\mathrm{A}$   & cytosolic \\
cytosolic pPAR        & $c_\mathrm{P}$   & cytosolic \\
membrane-bound aPAR   & $m_\mathrm{A}$   & membrane \\
membrane-bound pPAR   & $m_\mathrm{P}$   & membrane \\
\bottomrule
\end{tabular}
\caption{Components of the PAR model.}
\label{tab:Par_components}
\end{table}

Defining the state vector as
\begin{align}
\mathbf u
:=
\big(c_{\mathrm{A}},\,c_{\mathrm{P}},\,m_{\mathrm{A}},\,m_{\mathrm{P}}\big)^{\!\top},
\end{align}
the reaction--diffusion equations take the same form as above, with diffusion matrix
\begin{align}
\mathbf{D}
:=
\mathrm{diag}\big(D_c^{\mathrm{A}},\,D_c^{\mathrm{P}},\,D_m^{\mathrm{A}},\,D_m^{\mathrm{P}}\big)
\end{align}
and reaction fluxes
\begin{align}
\mathbf R
=
\begin{pmatrix}
-\,k_{\mathrm{on}}^{\mathrm{A}}\,c_{\mathrm{A}} + d_{\mathrm{A}}(m_{\mathrm{A}},m_{\mathrm{P}})\\[0.15em]
-\,k_{\mathrm{on}}^{\mathrm{P}}\,c_{\mathrm{P}} + d_{\mathrm{P}}(m_{\mathrm{A}},m_{\mathrm{P}})\\[0.15em]
\phantom{-}\,k_{\mathrm{on}}^{\mathrm{A}}\,c_{\mathrm{A}} - d_{\mathrm{A}}(m_{\mathrm{A}},m_{\mathrm{P}})\\[0.15em]
\phantom{-}\,k_{\mathrm{on}}^{\mathrm{P}}\,c_{\mathrm{P}} - d_{\mathrm{P}}(m_{\mathrm{A}},m_{\mathrm{P}})
\end{pmatrix}.
\label{eq:par_reaction_vector}
\end{align}

Cytosolic aPAR and pPAR proteins bind to the membrane with rates $k_{\mathrm{on}}^{\mathrm{A}}$ and $k_{\mathrm{on}}^{\mathrm{P}}$ and detach spontaneously with rates $k_{\mathrm{off}}^{\mathrm{A}}$ and $k_{\mathrm{off}}^{\mathrm{P}}$.
Mutual antagonism is implemented through cross-detachment:~membrane-bound aPAR is removed at an additional rate $k_{\mathrm{AP}}\,(m_{\mathrm{P}})^{\alpha}$ that grows with the local pPAR density, and vice versa for pPAR at rate $k_{\mathrm{PA}}\,(m_{\mathrm{A}})^{\beta}$.
The full detachment fluxes are
\begin{subequations}
\begin{align}
d_{\mathrm{A}}(m_{\mathrm{A}},m_{\mathrm{P}}) &= k_{\mathrm{off}}^{\mathrm{A}}\,m_{\mathrm{A}} + k_{\mathrm{AP}}\,m_{\mathrm{A}}\,(m_{\mathrm{P}})^{\alpha},\\
d_{\mathrm{P}}(m_{\mathrm{A}},m_{\mathrm{P}}) &= k_{\mathrm{off}}^{\mathrm{P}}\,m_{\mathrm{P}} + k_{\mathrm{PA}}\,m_{\mathrm{P}}\,(m_{\mathrm{A}})^{\beta},
\end{align}
\end{subequations}
where the exponents $\alpha$ and $\beta$ set the cooperativity of the antagonistic interaction \citep{Goehring.etal2011,Tostevin.Howard2008}.

The parameter values (Tab.~\ref{tab:par_parameters}) follow Ref.~\citep{Goehring.etal2011}.
Unless specified otherwise, these values are used as baseline parameters for the numerical phase diagrams.
\begin{table}[h!]
\centering
\renewcommand{\arraystretch}{1.2}
\begin{tabular}{>{\centering\arraybackslash}p{2.7cm} 
                >{\centering\arraybackslash}p{2.7cm} 
                >{\centering\arraybackslash}p{2.7cm}}
\multicolumn{3}{c}{\textbf{Parameter values used in the PAR model}} \\
\toprule
\textbf{Parameter} & \textbf{Value} & \textbf{Unit} \\
\midrule
$\rho_{\mathrm{A}}$        & 2.0      & $\mu\mathrm{m}^{-3}$ \\
$\rho_{\mathrm{P}}$        & 1.4      & $\mu\mathrm{m}^{-3}$ \\
system size                  & 50        & $\mu\mathrm{m}$ \\
$D_c^{\mathrm{A}},\,D_c^{\mathrm{P}}$  & 1.5  & $\mu\mathrm{m}^2/\mathrm{s}$ \\
$D_m^{\mathrm{A}}$                    & 0.28 & $\mu\mathrm{m}^2/\mathrm{s}$ \\
$D_m^{\mathrm{P}}$                    & 1.0  & $\mu\mathrm{m}^2/\mathrm{s}$ \\
$k_{\mathrm{on}}^{\mathrm{A}}$           & 0.03 - 0.35 (0.1)  & $1/\mathrm{s}$ \\
$k_{\mathrm{on}}^{\mathrm{P}}$           & 0.3 - 3.5 (1) & $1/\mathrm{s}$ \\
$k_{\mathrm{off}}^{\mathrm{A}}$          & $5.4 \times 10^{-3}$ & $1/\mathrm{s}$ \\
$k_{\mathrm{off}}^{\mathrm{P}}$          & $7.3 \times 10^{-3}$ & $1/\mathrm{s}$ \\
$k_{\mathrm{AP}}$          & 0.19     & $\mu\mathrm{m}^2/\mathrm{s}$ \\
$k_{\mathrm{PA}}$          & 2.0      & $\mu\mathrm{m}^4/\mathrm{s}$ \\
$\alpha$                   & 1        & -- \\
$\beta$                    & 2        & -- \\
\bottomrule
\end{tabular}
\caption{Parameter values for the PAR model. Number in bracket is used only in Sec.~\ref{sec:app-data}.}
\label{tab:par_parameters}
\end{table}

\section{Type-I instability in McRD systems with multiple cytosolic components per species}
\label{app:multi-cyt-components}

Here we analyze stationary short-wavelength (conserved Turing, type-I) instabilities in multi-species McRD systems where each species may carry a cytosolic cascade that converts non-binding intermediates stepwise into the binding-competent protein conformation.

The analysis proceeds in the same way as for the single-species case treated in Sec.~\ref{sec:3c}: because membrane components do not diffuse, marginal modes (${\sigma=0}$) at any wavenumber $q$ must satisfy local reactive balance on the membrane, which constrains the perturbation to lie in the null space of the membrane-reaction Jacobian. Using the nullcline parametrization by the conserved total densities~$\bm\rho$, the marginal eigenvector can be expressed entirely in terms of density shifts $\widetilde{\delta\bm\rho}$ weighted by the slopes of the reactive-equilibrium manifold, with non-binding cytosolic components subject to a $q$-dependent low-pass filter set by their respective screening lengths.

Beginning with the two-cytosolic-component case (Sec.~\ref{app:2c}) and then generalizing to arbitrary cytosolic series (Sec.~\ref{app:cytSeries}), this yields a compact slope-based marginality condition at a finite wavenumber $q_\mathrm{min}$. 
In the special case where detachment feeds only the last cytosolic level, the condition reduces to
\begin{align}
\label{eq:app-typeI-overview}
\det\!\big(\mathbf Q(q_\mathrm{min})\,\partial_{\bm\rho}\mathbf c_1^{*}+\partial_{\bm\rho}\bm\eta^{*}\big)=0\,.
\end{align}
Physically, the matrix $\mathbf{Q}(q)$ captures the low-pass filtering of non-binding cytosolic components by the conversion cascade: it vanishes as ${q\to 0}$ (recovering the LQSS limit) and grows monotonically with $q$, so that short-wavelength onset is controlled by the binding-competent slope matrix $\partial_{\bm\rho}\mathbf c_1^{*}$.

\subsection{Two cytosolic components per species}
\label{app:2c}

We consider a multi-species McRD system with a membrane-bound sector and a cytosolic sector. 
For the \emph{cytosolic} states of each species $\alpha$, we distinguish two diffusing states: a \emph{binding-competent} state $c_1^\alpha$ and a \emph{non-binding} (purely cytosolic) state $c_2^\alpha$. 
Detachment of proteins from the membrane feeds the non-binding pool~\footnote{
A detachment term into the binding-competent species could be added and would not change the following derivation because such a term would only change the mathematical form of ${\mathbf R}_m(\mathbf{m},\mathbf{c}_1)$ but the membrane-reaction term would still only depend on the binding-competent cytosolic components.} via a reactive flux $d_\alpha(\mathbf{m})$, and attachment draws from the binding-competent pool at rate $a_\alpha(\mathbf{m})\,c_1^\alpha$.
Reactivation in the cytosol converts the non-binding states $c_2^\alpha$ back into the binding-competent state $c_1^\alpha$ through a potentially nonlinear reaction~\footnote{For example, such a density dependence may emerge for cytosolic dimerization of MinD upon nucleotide exchange (cf.\ discussion in Appendix~\ref{app:cytSeries}).}
\begin{align}\label{eq:conversion-reaction}
    c_2^\alpha 
    \xrightarrow{\ \lambda_\alpha (c_2^\alpha)\ } c_1^\alpha 
    \, ,
    \qquad
    \partial_{c_2^\alpha} \big[\lambda_\alpha (c_2^\alpha)\,c_2^\alpha\big] > 0 
    \, .
\end{align}
The resulting diffusion-reaction equations for the cytosolic densities read
\begin{subequations}\label{eq:RDS-2cyt}
\begin{align}
    \partial_t c^\alpha_2  
    &= D_{c,2}^\alpha \boldsymbol{\nabla}^2 c^\alpha_2 - \lambda_\alpha (c^\alpha_2) \, c^\alpha_2 + d_\alpha(\mathbf{m})
    \, ,
    \label{eq:inactive-species}
    \\
    \partial_t c^\alpha_1 
    &= D_{c,1}^\alpha \boldsymbol{\nabla}^2 c^\alpha_1 + \lambda_\alpha(c^\alpha_2) \, s_{c,2}^\alpha \, c^\alpha_2 - a_\alpha(\mathbf{m}) \, c^\alpha_1
    \, ,
\label{eq:active-species}
\end{align}
\end{subequations}
where $s_{c,2}^\alpha$ is a stoichiometric factor that counts how many molecules of species $\alpha$ are contained in one unit of the non-binding state $c_2^\alpha$. The corresponding factor for the binding-competent state $c_1^\alpha$ has been set to unity by our choice of normalization [cf.\ Eq.~\eqref{eq:def-tot-dens}]; without this convention, only the ratio $s_{c,2}^\alpha/s_{c,1}^\alpha$ would appear.
The membrane dynamics are specified by the reaction term ${\mathbf R}_m(\mathbf{m},\mathbf{c}_1)$.

The two-cytosolic-state systems considered here are special cases of the general McRD dynamics, Eq.~\eqref{eq:RDS}. Consequently, their long-wavelength (type-II) instability is governed by the slope criterion given in Sec.~\ref{sec:long-wavelength-limit}. Here we address the complementary question: the onset of a \emph{stationary} finite-wavelength (type~I, conserved Turing) instability. The derivation parallels Sec.~\ref{sec:3c} and extends it to multiple species with two cytosolic states per species.

\subsubsection*{Linear stability analysis via marginal mode perturbations}

We now linearize the dynamics at the homogeneous steady state to characterize marginal (${\sigma=0}$) modes at finite $q$. 
As a first step we derive the linear response of the non-binding cytosolic pool $c_2^\alpha$ and identify the species-dependent screening length $\ell_c^\alpha$. 
Next, using the absence of membrane diffusion, we impose local reactive balance on the membrane and exploit the nullcline parametrization by total densities $\rho^\alpha$ to express all perturbations via density shifts $\widetilde{\delta \rho}^\alpha$. 
This yields the compact reparameterization of marginal modes in terms of nullcline slopes and a $q$-dependent low-pass factor on the non-binding pool.

\smallskip

\emph{Linear response of the non-binding cytosolic pool.---}
Linearizing the equation for $c_2^\alpha$ [Eq.~\eqref{eq:inactive-species}] at the homogeneous steady state (hss), inserting an eigenmode $e^{\sigma t + i\mathbf q \cdot \mathbf x}$, and evaluating at a zero of the dispersion relation (${\sigma = 0}$) gives (cf.\ Sec.~\ref{sec:lsa})
\begin{align}
    \left(
    \lambda_\alpha^{\mathrm{lin}}
    + D_{c,2}^\alpha \,q^2
    \right)
    \delta c_{2}^\alpha
    =
    \partial_{\mathbf m} d_\alpha |_{\mathrm{hss}} \cdot
    \delta\mathbf m
    \, ,
\label{eq:suppression-inactive-cytosolic-1}
\end{align}
with the linearized conversion rate
\begin{align}\label{eq:lin-conversion-rate}
    \lambda_\alpha^{\mathrm{lin}}
    \equiv
    \partial_{c_2^\alpha}
    \big( \lambda_\alpha (c_2^\alpha) \, c_2^\alpha 
    \big) \big|_{\mathrm{hss}}>0 
    \, .
\end{align}
Defining the \emph{screening lengths} 
\begin{align}\label{eq:screening-length}
    \ell_c^\alpha
    \equiv
    \sqrt{D^\alpha_{c,2}/\lambda_\alpha^{\mathrm{lin}}}\,,
\end{align}
we rewrite Eq.~\eqref{eq:suppression-inactive-cytosolic-1} as
\begin{align}\label{eq:c2-lowpass}
    \delta c_{2}^\alpha
    =
    \frac{1}{\lambda_\alpha^{\mathrm{lin}}} \;   \,
    \frac{1}{1+(q\ell_c^\alpha)^2} \;
    \partial_{\mathbf m} d_\alpha
    |_{\mathrm{hss}} \cdot \delta\mathbf m \,.
\end{align}
This shows that the non-binding cytosolic protein pool responds to membrane perturbations with a screened (low-pass filtered) amplitude.
While long-wavelength modes (${q \ell_c^\alpha \ll 1}$) pass essentially unattenuated, short-wavelength modes (${q \ell_c^\alpha \gg 1}$) are suppressed by a factor  ${\sim 1/(q\ell_c^\alpha)^2}$.
This screening originates from diffusion of the non-binding cytosolic pool: diffusion smooths short-wavelength cytosolic fluctuations on the timescale ${(D_{c,2}^\alpha \, q^2)^{-1}}$, faster than the conversion reaction can read them out when ${q\ell_c^\alpha \gg 1}$.

\smallskip

\emph{Reactive equilibrium on the membrane.---}
Because membrane diffusion is neglected, at a zero of the dispersion relation
(${\sigma=0}$), the membrane block of the eigenvalue problem reduces to
\begin{align}
    0
    = 
    \big[\partial_{\mathbf m}{\mathbf R}_m\big]_{\mathrm{hss}} \; \delta\mathbf m
    +  
    \big[\partial_{\mathbf c_1}{\mathbf R}_m\big]_{\mathrm{hss}} \; \delta\mathbf c_{1}
    \, .
\label{eq:multiSpecies-membrane-eq}
\end{align}
where only the binding-competent cytosolic components $\delta\mathbf{c}_1$ appear because, by assumption, ${\mathbf R}_m$ does not depend on any non-binding component $c_{i\geq 2}^\alpha$.
Geometrically, this condition requires the marginal perturbation $(\delta\mathbf m, \delta\mathbf c_1)^\top$ to be tangent to the manifold of membrane-reaction equilibria ${\mathbf R}_m = 0$ at the homogeneous steady state, or equivalently, to lie in the null space of the membrane-reaction Jacobian $[\partial_{(\mathbf m, \mathbf c_1)} {\mathbf R}_m]_{\mathrm{hss}}$.

\emph{Parametrization by nullcline slopes.---}
A natural candidate basis for this null space is
provided by the nullcline tangent vectors
$(\partial_{\rho_\alpha}\mathbf m^*,
\partial_{\rho_\alpha}\mathbf c_1^*)^\top$.
These lie in the null space because
$\mathbf R_m(\mathbf m^*(\bm\rho),
\mathbf c_1^*(\bm\rho)) = 0$ holds for all
$\bm\rho$; differentiating with respect to
$\rho_\alpha$ gives
$[\partial_{\mathbf m}\mathbf R_m]_\mathrm{hss}\,
\partial_{\rho_\alpha}\mathbf m^*
+ [\partial_{\mathbf c_1}\mathbf R_m]_\mathrm{hss}\,
\partial_{\rho_\alpha}\mathbf c_1^* = 0$,
which is exactly the null-space condition
Eq.~\eqref{eq:multiSpecies-membrane-eq}.

However, since the nullcline tangent space satisfies the stronger condition $\mathbf R = 0$ (not just $\mathbf R_m = 0$), its projection onto the $(\mathbf m, \mathbf c_1)$-subspace could in principle be smaller than the full null space.
Whether the nullcline tangent vectors are sufficient depends on a comparison of dimensions.
The membrane-reaction Jacobian $[\partial_{(\mathbf m,\mathbf c_1)} {\mathbf R}_m]_{\mathrm{hss}}$ has $N_m$ rows and $N_m + N_{m,c}$ columns, where $N_{m,c}$ is the number of cytosolic components on which ${\mathbf R}_m$ depends.
Generically, this matrix has full row rank $N_m$, so by the rank--nullity theorem\footnote{For any matrix with full row rank, the dimension of its null space equals the number of columns minus the number of rows.} its null space has dimension $N_{m,c}$.
The nullcline tangent space, on the other hand, has dimension $N_\mathrm{species}$ (one parameter $\rho_\alpha$ per conservation law).
Our assumption that exactly one cytosolic component per species is binding-competent ensures ${N_{m,c} = N_\mathrm{species}}$, so the $N_\mathrm{species}$ projected nullcline tangent vectors generically span the full null space (see Sec.~\ref{sec:RDS}); they fail to do so only if two or more projections are exactly parallel.
Consequently, every admissible marginal perturbation $(\delta\mathbf m, \delta\mathbf c_1)^\top$ can be written as a linear combination of nullcline tangent vectors.

Consequently, we write
\begin{align}\label{eq:marginal-m-c1}
    \begin{pmatrix} 
    \delta\mathbf{m} \\[6pt] 
    \delta\mathbf{c}_{1} 
    \end{pmatrix} 
    &= \sum_\alpha 
    \begin{pmatrix} 
    \partial_{\rho_\alpha}\mathbf{m}^* 
    \big|_{\mathrm{hss}} \\[6pt] 
    \partial_{\rho_\alpha}\mathbf{c}_1^* 
    \big|_{\mathrm{hss}} 
    \end{pmatrix}
    \delta\widetilde{\rho}_\alpha 
    \nonumber \\
    &= 
    \begin{pmatrix} 
    \big[\partial_{\bm\rho}\mathbf{m}^*
    \big]_{\mathrm{hss}} \\[6pt] 
    \big[\partial_{\bm\rho}\mathbf{c}_1^*
    \big]_{\mathrm{hss}} 
    \end{pmatrix} 
    \widetilde{\delta\bm\rho} 
    \, .
\end{align}
The coefficient vector $\widetilde{\delta\bm\rho}$---the \emph{nullcline displacement}---is uniquely determined by the perturbations in the binding-competent components $\delta\mathbf{c}_1$ via invertibility of  $[\partial_{\bm\rho}\mathbf{c}_1^*]_{\mathrm{hss}}$:\footnote{Due to the assumption of a single binding-competent cytosolic component per species, the Jacobian $[\partial_{\bm\rho}\mathbf{c}_1^*]_{\mathrm{hss}}$ is a square matrix and generically invertible; it becomes singular only on a codimension-one subset where ${\det\left[\partial_{\bm{\rho}} \mathbf{c}_1^{*}\right]_{\mathrm{hss}}=0}$.}
\begin{equation}\label{eq:def-total-density-variations}
    \widetilde{\delta\bm{\rho}} 
    = 
    \big[\partial_{\bm{\rho}}\mathbf{c}_1^{*}
    \big]_{\mathrm{hss}}^{-1} 
    \; \delta \mathbf{c}_{1} 
    \, .
\end{equation}
It measures how far one must move along the nullcline manifold $\mathbf{u}^*(\bm\rho)$ to reproduce the binding-competent perturbation $\delta\mathbf{c}_{1}$.
We emphasize that $\widetilde{\delta\bm\rho}$ is a coordinate on the nullcline manifold, not the physical total-density perturbation $\delta\bm\rho = \mathbf{S}\,\delta\mathbf{u}$.
The two coincide in the long-wavelength limit $q \to 0$, where the full eigenmode is tangent to the nullcline, but differ at finite $q$ because the non-binding components $\delta\mathbf{c}_{2}$ are filtered (see Sec.~\ref{app:perturbation_rho_detail}).
For simplicity of notation, we omit the subscript ``hss'' and the bracket notation $[\cdot]_{\mathrm{hss}}$ in the following, as all nullcline slopes are evaluated at the homogeneous steady state throughout.

\smallskip

\emph{Remark (algebraic verification).---}
The relation $\delta\mathbf{m} = \partial_{\bm\rho}\mathbf{m}^*\, \widetilde{\delta\bm\rho}$ can also be verified by direct calculation without referring to nullcline tangency of the perturbation vector.
Differentiating the equilibrium identity $\mathbf{R}_m(\mathbf{m}^*(\bm\rho), \mathbf{c}_1^*(\bm\rho)) = 0$ with respect to $\bm\rho$ gives
\begin{align}\label{eq:nullcline-deriv}
    \partial_{\mathbf m}\mathbf{R}_m \; 
    \partial_{\bm\rho}\mathbf{m}^* 
    = -\,\partial_{\mathbf c_1}\mathbf{R}_m \; 
    \partial_{\bm\rho}\mathbf{c}_1^* 
    \, .
\end{align}
Multiplying both sides by $\widetilde{\delta\bm\rho}$ and using $\partial_{\bm\rho}\mathbf{c}_1^*\, \widetilde{\delta\bm\rho} = \delta\mathbf{c}_{1}$ [Eq.~\eqref{eq:def-total-density-variations}] yields
\begin{align}\label{eq:slaved-comp}
    \partial_{\mathbf m}\mathbf{R}_m \; 
    \partial_{\bm\rho}\mathbf{m}^* \; 
    \widetilde{\delta\bm\rho} 
    = -\,\partial_{\mathbf c_1}\mathbf{R}_m \; 
    \delta\mathbf{c}_{1} 
    \, .
\end{align}
The right-hand side is identical to that of the marginal membrane balance Eq.~\eqref{eq:multiSpecies-membrane-eq}, so subtracting the two gives
\begin{align}\label{eq:tangency-difference}
    \partial_{\mathbf m}\mathbf{R}_m \; 
    \big(\delta\mathbf{m} 
    - \partial_{\bm\rho}\mathbf{m}^* \; 
    \widetilde{\delta\bm\rho}\big) 
    = \mathbf{0} 
    \, .
\end{align}
Since $\partial_{\mathbf m}\mathbf{R}_m$ is an $N_m \times N_m$ matrix and generically invertible, the desired relation follows.

\smallskip

\emph{Non-binding cytosolic components.---}
It remains to determine $\delta c_{2}^\alpha$ in terms of $\widetilde{\delta\bm\rho}$.
In Eq.~\eqref{eq:suppression-inactive-cytosolic-1} the right-hand side involves the membrane perturbation $\delta\mathbf{m}$.
Replacing $\delta\mathbf{m}$ using Eq.~\eqref{eq:marginal-m-c1} gives
\begin{align}\label{eq:c2-step1}
    \left(\lambda_\alpha^{\mathrm{lin}} 
    + D_{c,2}^\alpha \, q^2 \right) 
    \delta c_{2}^\alpha
    = \partial_{\mathbf m} d_\alpha 
    \cdot \partial_{\bm\rho}\mathbf m^* \;
    \widetilde{\delta\bm\rho} 
    \, .
\end{align}
To express the right-hand side in terms of nullcline slopes, we differentiate the reactive-equilibrium condition for the non-binding pool, $d_\alpha(\mathbf m^*) = \lambda_\alpha(c_2^{\alpha*})\,c_2^{\alpha*}$, with respect to the total densities $\bm\rho$:
\begin{align}\label{eq:reactive_equilibrium_non-binding}
    \partial_{\mathbf{m}} d_\alpha \;
    \partial_{\bm\rho}\mathbf{m}^*
    = \lambda_\alpha^\mathrm{lin} \,
    \partial_{\bm\rho} c^{\alpha*}_2 
    \, .
\end{align}
Inserting this identity into Eq.~\eqref{eq:c2-step1}
and dividing by $\lambda_\alpha^{\mathrm{lin}} + D_{c,2}^\alpha q^2 = \lambda_\alpha^{\mathrm{lin}}(1 + (q\ell_c^\alpha)^2)$ gives the compact result
\begin{equation}\label{eq:suppression-inactive-cytosolic}
    \delta c_{2}^\alpha 
    = \frac{1}{1+(q\ell_c^\alpha)^2}\;
    \partial_{\bm\rho} c^{\alpha*}_2 
    \cdot \widetilde{\delta\bm\rho}
    \, .
\end{equation}
This shows that the non-binding cytosolic component responds to nullcline displacements with a low-pass filtered version of the nullcline slope: long-wavelength perturbations ($q\ell_c^\alpha \ll 1$) follow the reactive equilibrium, while short-wavelength perturbations ($q\ell_c^\alpha \gg 1$) are suppressed.
The LQSS limit (cf.\ Sec.~\ref{sec:qss}) is recovered for $q \to 0$.

\emph{Full marginal perturbation vector.---}
Taken together, the full marginal perturbation vector can be written as the product of a nullcline-slope matrix and the nullcline displacement $\widetilde{\delta\bm{\rho}}$:
\begin{equation}\label{eq:2c-reparameterization}
    \delta \mathbf{u} 
    = 
    \begin{pmatrix} 
    \partial_{\bm{\rho}}\mathbf{m}^{*} \\[6pt] 
    \partial_{\bm{\rho}}\mathbf{c}_1^{*} \\[6pt] 
    \mathsf H(q)\,
    \partial_{\bm{\rho}}\mathbf{c}^{*}_2 
    \end{pmatrix} 
    \widetilde{\delta\bm{\rho}} 
    \, ,
\end{equation}
with the diagonal low-pass filter matrix 
\begin{equation}\label{eq:low-pass-filter}
    \mathsf H(q) 
    \equiv \mathrm{diag}_\alpha\!\left(
    \frac{1}{1+(q\ell_c^\alpha)^2}\right).
\end{equation}
This is the central result of the marginal-mode construction: the entire eigenvector is parametrized by the single $N_\mathrm{species}$-dimensional nullcline displacement $\widetilde{\delta\bm\rho}$, weighted by the slopes of the reactive-equilibrium manifold.
The $(\mathbf{m},\mathbf{c}_1)$-components are given directly by these slopes, while the non-binding $\mathbf{c}_2$-components are attenuated by $\mathsf{H}(q)$.
Defining the nullcline tangent vector
\begin{align}
    \mathbf t 
    = \partial_{\bm\rho}\mathbf u^* \;
    \widetilde{\delta\bm\rho},
\end{align}
the marginal model is an anisotropically filtered version of this tangent:
\begin{align}
\label{eq:eigenmode-tangentvector-relation}
    \delta\mathbf u 
    &= \mathsf{F}(q)\,\mathbf t \, ,
\\
\label{eq:filtering-matrix-F}
    \mathsf{F}(q) 
    &\equiv \mathrm{block\text{-}diag}\,
    \big(\mathsf I_m,\
    \mathsf I_{c_1},\
    \mathsf H(q)\big) \, .
\end{align}
For $q\ell_c^\alpha \ll 1$, the eigenvector is tangent to the nullcline manifold; for $q\ell_c^\alpha \gg 1$, the $\mathbf{c}_2$-components are suppressed and the eigenvector is effectively projected onto the $(\mathbf{m},\mathbf{c}_1)$-subspace (cf.\ Fig.~\ref{fig:marginal-stability-geometry}).

\subsubsection*{Total-density variations}
\label{app:perturbation_rho_detail}

The nullcline displacement
$\widetilde{\delta\bm\rho}$ defined in
Eq.~\eqref{eq:def-total-density-variations} and
the physical total-density perturbations
$\delta\bm\rho = \mathbf{S}
\delta\mathbf{u}$ are conceptually distinct
quantities that coincide only in the
long-wavelength limit.
We now make this precise.
If the eigenmode were exactly a nullcline
tangent vector, $\delta\mathbf{u} = \mathbf{t}
= \partial_{\bm\rho}\mathbf{u}^*\,
\widetilde{\delta\bm\rho}$, the two would agree.
This follows from the definition of the total
densities Eq.~\eqref{eq:def-tot-dens}:
\begin{equation}\label{eq:rho-tilde-equals-rho}
    \delta\rho_\alpha 
    = \mathbf{s}_\alpha \cdot \mathbf{t} 
    = \partial_{\bm\rho}
    (\mathbf{s}_\alpha \cdot \mathbf{u}^*) \;
    \widetilde{\delta\bm\rho} 
    = \partial_{\bm\rho}\rho_\alpha \;
    \widetilde{\delta\bm\rho} 
    = \widetilde{\delta\rho}_\alpha 
    \, .
\end{equation}
In the long-wavelength limit ${q\to 0}$, the marginal mode becomes tangent to the nullcline and $\widetilde{\delta\rho}_\alpha=\delta\rho_\alpha$ two coincide.
However, at finite wavenumber $q>0$ the eigenmode is not a nullcline tangent vector but a filtered version of it, ${\delta\mathbf{u} = \mathsf{F}(q)\,\mathbf{t}}$ [Eq.~\eqref{eq:eigenmode-tangentvector-relation}],
where the filter $\mathsf{F}(q)$ suppresses the
non-binding $c_2^\alpha$-components, and one finds
\begin{equation}
\label{eq:rho-tilde-neq-rho}
    \delta\rho_\alpha 
    = \mathbf{s}_\alpha \cdot
    \mathsf{F}(q)\,\mathbf{t} 
    = \partial_{\bm\rho} \big(
    \mathbf{s}_\alpha \cdot
    \mathsf{F}(q)\,\mathbf{u}^*\big) \;
    \widetilde{\delta\bm\rho} 
    \neq \widetilde{\delta\rho}_\alpha 
    \, .
\end{equation}
In summary, $\widetilde{\delta\bm\rho}$
measures how far one has moved along the nullcline, whereas $\delta\bm\rho$ quantifies the actual displaced total mass.
Accordingly, in the treatment of long-wavelength
instabilities (cf.\ Sec.~\ref{sec:long-wavelength-limit}) one may
freely identify
${\widetilde{\delta\bm\rho}
\approx\delta\bm\rho}$, whereas in the treatment
of short-wavelength instabilities
(cf.\ Sec.~\ref{sec:3c}) one must track the
filter factors explicitly and one must distinguish $\widetilde{\delta\bm\rho}$ from
$\delta\bm\rho$.
Nonetheless, under certain assumptions, the
difference is negligible, which we critically use
in Sec.~\ref{sec:approx-dispRel-oscillatory}.

\subsubsection*{Continuity equation and marginality condition}
\label{sec:continuity-marginality}

For each species $\alpha$, conservation of the total density $\rho_\alpha$ gives the linearized continuity equation in Fourier space
\begin{align}\label{eq:lsa-eta-scalar}
    \sigma\,\delta\rho_\alpha 
    = -\,D_{c,1}^\alpha\,q^2\,\delta\eta_\alpha \,,
\end{align}
with the mass-redistribution potentials
\begin{align}
   \eta_\alpha 
    \equiv c_1^\alpha 
    + \frac{D_{c,2}^\alpha}{D_{c,1}^\alpha}\, 
    s_{c,2}^\alpha\, c_2^\alpha 
    \, . 
\end{align}
Using the marginal mode vector, Eq.~\eqref{eq:2c-reparameterization}, the perturbation of $\eta_\alpha$ for the marginal mode becomes
\begin{align}\label{eq:delta-eta-1}
    \delta\eta_\alpha 
    = \Bigg( 
    \partial_{\bm\rho} c_1^{\alpha *}
    + \frac{D_{c,2}^\alpha}{D_{c,1}^\alpha}\,
    \frac{s_{c,2}^\alpha}{1+(q\ell_c^\alpha)^2}\,
    \partial_{\bm\rho} c_2^{\alpha *} 
    \Bigg) 
    \cdot \widetilde{\delta\bm{\rho}} 
    \, ,
\end{align}
where we recall that the nullcline slopes are evaluated at the homogeneous steady state.
This relation can equivalently be written as
\begin{align}\label{eq:delta-eta-2}
    \delta\eta_\alpha 
    = \frac{1}{1+(q\ell_c^\alpha)^2}\,
    \Big( 
    (q \ell_c^\alpha)^2\, 
    \partial_{\bm\rho} c_1^{\alpha *} 
    + \partial_{\bm\rho} \eta_\alpha^{*} 
    \Big) 
    \cdot \widetilde{\delta\bm{\rho}} 
    \, .
\end{align}
Inserting Eq.~\eqref{eq:delta-eta-2} into the continuity equation~\eqref{eq:lsa-eta-scalar} gives
\begin{align}\label{eq:self-consistency}
    \sigma \, \delta \rho_\alpha 
    = \frac{-D_{c,1}^\alpha \, q^2}
    {1+(q\ell_c^\alpha)^2} \, 
    \Big( 
    (q \ell_c^\alpha)^2\, 
    \partial_{\bm\rho} c_1^{\alpha *} 
    + \partial_{\bm\rho} \eta_\alpha^{*} 
    \Big) 
    \cdot \widetilde{\delta\bm{\rho}} 
    \, .
\end{align}
Self-consistency with the assumption that the analyzed eigenmode is marginal, that is, it fulfills ${\sigma=0}$, thus requires
\begin{align}\label{eq:self-consistency_0}
    0 
    = \frac{-D_{c,1}^\alpha \, q^2}
    {1+(q\ell_c^\alpha)^2} \, 
    \Big( 
    (q \ell_c^\alpha)^2\, 
    \partial_{\bm\rho} c_1^{\alpha *} 
    + \partial_{\bm\rho} \eta_\alpha^{*} 
    \Big) 
    \cdot \widetilde{\delta\bm{\rho}} 
    \, .
\end{align}
In other words, marginality demands that all gradients in the mass-redistribution potentials vanish (${\delta\bm\eta=0}$) because any gradients in $\delta\bm\eta$ would induce mass-redistribution and thus dynamics (${\sigma\neq 0}$).

The self-consistency equations Eq.~\eqref{eq:self-consistency} can be written in vector form by defining
the diagonal matrix
\begin{align}\label{eq:matrix-defs}
    \mathsf L 
    &\equiv \mathrm{diag}_\alpha 
    \big( \ell_c^\alpha \big) 
    \, ,
\end{align}
and
\begin{align}\label{eq:eta-slope-def}
    \partial_{\bm\rho}\boldsymbol\eta^{*} 
    \equiv \partial_{\bm\rho}\mathbf c_1^{*} 
    + \mathrm{diag}_\alpha 
    \Big(
    \frac{D_{c,2}^\alpha}{D_{c,1}^\alpha}\,
    s_{c,2}^\alpha
    \Big)\,
    \partial_{\bm\rho}\mathbf c_2^{*} 
    \, .
\end{align}
Then Eq.~\eqref{eq:self-consistency} takes the form, assuming ${q>0}$,
\begin{align}\label{eq:typeI-onset-vector-prefactor}
    \Big(\, 
    q^2 \, \mathsf L^2\, 
    \partial_{\bm\rho}\mathbf c_1^{*} 
    + \partial_{\bm\rho}\boldsymbol\eta^{*} 
    \Big)\, 
    \widetilde{\delta\bm\rho} 
    = \mathbf 0 
    \, .
\end{align}
This self-consistency equation determines the wavenumber(s) $q$ of the marginal mode, i.e., the wavenumber at which the dispersion relation crosses zero.

\subsubsection*{Finite--wavelength instability}
\label{sec:finite-wavelength-instability}

The analysis in the previous section shows that besides the trivial root at ${q=0}$ [generically following from mass conservation, cf.\ Eq.~\eqref{eq:self-consistency}], a \emph{finite} zero of the dispersion relation arises at ${q_{\min}>0}$ precisely when the marginality condition~\eqref{eq:typeI-onset-vector-prefactor} admits a nontrivial solution ${q>0}$.
Equivalently, the matrix multiplying $\widetilde{\delta\bm\rho}$ must be singular, i.e.,
\begin{align}\label{eq:type-I-instab-condition}
    0 
    = \det 
    \big(\, 
    q_\mathrm{min}^2 \, \mathsf L^2 \, 
    \partial_{\bm{\rho}} \mathbf{c}_1^{*} 
    + \partial_{\bm{\rho}} \bm{\eta}^{*} 
    \big) 
    \, .
\end{align}
If some species $\alpha$ has only a single cytosolic component, we set ${\ell_c^\alpha=0}$ since the limit ${\lambda_\alpha^{\mathrm{lin}} \to \infty}$ corresponds to an instantaneous conversion of non-binding into binding-competent cytosolic states.
Such species then enter Eq.~\eqref{eq:type-I-instab-condition} only through the slope matrix $\partial_{\bm{\rho}}\bm{\eta}^{*}$, while contributing no $q$-dependent correction term.

Assuming long-wavelength stability, a positive solution ${q^2_{\min}>0}$ of Eq.~\eqref{eq:type-I-instab-condition} signals the existence of a stationary type-I (conserved Turing) instability.
This condition is the generalization of the scalar solvability condition Eq.~\eqref{eq:solvability} for the single-species case derived in the main text.
Specifically, for a single species, ${N_{\mathrm{species}}=1}$, Eq.~\eqref{eq:typeI-onset-vector-prefactor} reduces to
\begin{align}\label{eq:single-species-typeI}
    (q\ell_c)^2\, \partial_\rho c_1^* + \partial_\rho \eta^* = 0 \, .
\end{align}
Also in the multi-species generalization, the instability is determined by nullcline slopes.
A short-wavelength instability arises because,
at large $q$, the relative contribution of the
binding-competent component to mass redistribution
grows compared to that of the non-binding
components, whose gradients are suppressed by
the low-pass filter.
In multi-species systems, instabilities can
additionally arise from cross-coupling between
the species; this is captured by extending the
scalar solvability condition of the single-species
case to the determinant condition for the full
species-slope matrix.

Paralleling the solution of the scalar single-species condition, we note that $\partial_{\bm{\eta}} \mathbf{c}^*_1 = \partial_{\bm{\rho}} \mathbf{c}^*_1 (\partial_{\bm{\rho}}\bm{\eta}^*)^{-1}$ describes the change of the reactive-equilibrium densities of the binding-competent components when changing the mass-redistribution potentials (assuming the ${N_\mathrm{species}\times N_\mathrm{species}}$-dimensional matrix $\partial_{\bm{\rho}}\bm{\eta}^*$ is invertible).
Then, factoring $\det[\partial_{\bm\rho}\bm\eta^*]$ out of Eq.~\eqref{eq:type-I-instab-condition} and using the determinant product rule gives
\begin{equation}
\label{eq:type-I-instab-condition-cetaslope}
    0 = \det\!\big(q^2 \,\mathsf L^2\, \partial_{\bm{\eta}} \mathbf{c}^*_1 + \mathsf I\big) \, .
\end{equation}
This condition requires $-1$ to be an eigenvalue of $q^2\mathsf L^2\,\partial_{\bm\eta}\mathbf{c}_1^*$.
Consequently, $q_\mathrm{min}^2$ is determined by the most negative eigenvalue $\mu_\mathrm{min}$ of $\mathsf L^2\,\partial_{\bm\eta}\mathbf{c}_1^*$ via $q_\mathrm{min}^2 = -1/\mu_\mathrm{min}$.
In other words, an exact criterion for type-I instability is that the matrix $\mathsf L^2\, \partial_{\bm{\eta}} \mathbf{c}^*_1$ has at least one negative eigenvalue, while $\mathbf{D}_1\partial_{\bm{\rho}}\bm{\eta}$ has only eigenvalues with positive real parts (ensuring long-wavelength stability).

Taken together, as in the single-species case, the instability is determined entirely by nullcline slopes.
A short-wavelength instability arises because, at large $q$, the low-pass filter $\mathsf{H}(q)$ suppresses gradients in the non-binding components, so that the relative contribution of the binding-competent component to mass redistribution grows.
In multi-species systems, instabilities can additionally arise from cross-coupling between species; this is captured by extending the scalar solvability condition of the single-species case to the determinant condition~\eqref{eq:type-I-instab-condition} for the full species-slope matrix.

\subsubsection*{Special case: single species with two cytosolic states}
\label{sec:special-case-single-species}

In the special case that only a single species $\alpha$ has two cytosolic states and all other species ($\beta \neq \alpha$) have only a single (binding-competent) cytosolic state, the short-wavelength instability criterion Eq.~\eqref{eq:type-I-instab-condition} can be solved more explicitly.
The skeleton model of the Min protein system (cf.\ Sec.~\ref{app:model}) is an example.

For notational convenience, we abbreviate
\begin{align}\label{eq:AB-def}
    \mathsf A 
    \equiv \partial_{\bm\rho}\boldsymbol\eta^*\,,
    \qquad
    \mathsf B 
    \equiv \mathsf L^2\,
    \partial_{\bm\rho}\mathbf c_1^*\,,
\end{align}
so that the onset condition Eq.~\eqref{eq:type-I-instab-condition} becomes $\det(\mathsf A+q^2\mathsf B)=0$.
Since ${\ell_c^{\beta}=0}$ for $\beta\neq\alpha$, $\mathsf B$ is a \emph{rank-one} matrix with only the $\alpha$-row nonzero:
\begin{align}\label{eq:B-rank-one}
    \mathsf L^2\,\partial_{\bm\rho}\mathbf c_1^*
    = \ell^2 \,
    \boldsymbol e_\alpha\,
    \big(
    \boldsymbol e_\alpha^{\! \top}\, 
    \partial_{\bm\rho}\mathbf c_1^*
    \big)
    \equiv \mathbf v\,\mathbf w^{\! \top} \,,
\end{align}
where $\boldsymbol e_\alpha$ is the unit vector in direction $\alpha$, $\ell \equiv \ell_c^\alpha$ denotes the screening length of the species with two cytosolic states, and
\begin{align}\label{eq:vw-def}
    \mathbf v
    \equiv \ell^2\,\boldsymbol e_\alpha \, ,
    \quad
    \mathbf w^{\! \top}
    \equiv \boldsymbol e_\alpha^{\! \top}\, 
    \partial_{\bm\rho}\mathbf c_1^* \, .
\end{align}
Assuming $\mathsf A$ to be invertible, the matrix--determinant lemma yields
\begin{align}\label{eq:det-lemma-result}
    \det(\mathsf A+q^2\mathsf B)
    &= \det(\mathsf A)\,
    \big(
    1+q^{2}\,\mathbf w^{\! \top}
    \mathsf A^{-1}\mathbf v
    \big) \, .
\end{align}
Using ${\mathsf A^{-1} = (\partial_{\bm\rho}\bm\eta^*)^{-1}}$ and ${\partial_{\bm\eta}\mathbf c_1^* = \partial_{\bm\rho}\mathbf c_1^*\,(\partial_{\bm\rho}\bm\eta^*)^{-1}}$, one gets ${\mathbf w^{\! \top}\mathsf A^{-1}\mathbf v = \ell^2\,\big(\partial_{\bm\eta}\mathbf c_1^{*}\big)_{\alpha\alpha}}$.
Inserting this expression back, one arrives at the short-wavelength instability condition
\begin{align}
\label{eq:det-condition-alpha}
    0 
    &= \det(\mathsf A + q^{2}\mathsf B)
    \nonumber \\
    &= \det(\mathsf A)\,
    \big(
    1 + q^{2}\,\ell^2 \,
    \big(\partial_{\bm\eta}\mathbf c_1^{*}
    \big)_{\alpha\alpha}
    \big) \, .
\end{align}
Since $\det(\mathsf A) \neq 0$ by assumption, the condition reduces to the vanishing of the second factor.
Thus, the homogeneous steady state is unstable due to a short-wavelength instability if a positive root $q_\mathrm{min}>0$ fulfilling
\begin{align}\label{eq:qmin-alpha}
    q_{\min}^{2}
    = -\,\frac{1}{\ell^2\,
    \big(\partial_{\bm\eta}\mathbf c_1^{*}
    \big)_{\alpha\alpha}}
\end{align}
exists.
Then, a band of unstable modes ${q>q_\mathrm{min}}$ exists.
Since $\ell^2>0$, this solution exists if and only if
\begin{align}\label{eq:alpha-criterion}
    \big(\partial_{\bm\eta}\mathbf c_1^{*}
    \big)_{\alpha\alpha} < 0 
    \, .
\end{align}
Moreover, at ${q=q_{\min}}$, the right nullvector is proportional to $\mathsf A^{-1}\mathbf v$:
\begin{align}\label{eq:nullvector-alpha}
    \widetilde{\delta\bm\rho} 
    \ \propto\ \mathsf A^{-1} \, \mathbf v
    \propto 
    \big(\partial_{\bm\rho}\boldsymbol\eta^{*}
    \big)^{-1}\,\boldsymbol e_\alpha \,,
\end{align}
i.e., the $\alpha$-th column of $(\partial_{\bm\rho}\boldsymbol\eta^{*})^{-1}$, as can be shown by insertion into Eq.~\eqref{eq:typeI-onset-vector-prefactor} and using the result for $q_\mathrm{min}^2$.

\subsubsection*{Sufficient instability criterion}

The full type-I onset condition involves the cytosolic reactivation rates through the screening lengths in the matrix $\mathsf L = \mathrm{diag}_\alpha(\ell_c^\alpha)$; see Eq.~\eqref{eq:type-I-instab-condition}.
However, one can obtain a \emph{sufficient}, conversion--independent criterion that depends only on the geometry (slopes) of the reactive--equilibrium (nullcline) surface.
The key is to examine the large--$q$ behavior of the real function
\begin{align}\label{eq:sq-def}
    s(q) 
    &= \det \Big(
    \partial_{\bm\rho}\boldsymbol\eta^* 
    + q^2\,\mathsf L^2\,
    \partial_{\bm\rho}\mathbf c_1^* 
    \Big) 
    \nonumber \\
    &= \det \big(\mathsf A + q^2\mathsf B\big) \,,
\end{align}
where we again use the short-hand notation ${\mathsf A\equiv \partial_{\bm\rho}\boldsymbol\eta^*}$ and ${\mathsf B\equiv \mathsf L^2\,\partial_{\bm\rho}\mathbf c_1^*}$.
Because long-wavelength stability implies that $\mathsf A$ has only eigenvalues with positive real part and $s(0) = \det(\mathsf A) > 0$, a negative sign of $s(q)$ as ${q\to\infty}$ guarantees a finite-$q$ root of $s(q)$ and thus a type-I instability.
In the following, we show that the sign sign of $s(q)$ as ${q\to\infty}$ is controlled by $\det(\partial_{\bm\rho}\mathbf c_1^*)$ alone.

To show this, we reorder the species so that the $N_2$ species with two cytosolic states come first, then the $N_1$ species with only a single cytosolic state.
The matrices then take the $2\times2$ block form:
\begin{subequations}
\label{eq:block-form}
\begin{align}
    \mathsf A 
    &= 
    \begin{pmatrix}
    \mathsf A_{22} & \mathsf A_{21}\\
    \mathsf A_{12} & \mathsf A_{11}
    \end{pmatrix} 
    \, , \\
    \mathsf B 
    &= \mathsf L^2\,\partial_{\bm\rho}\mathbf c_1^* 
    =
    \begin{pmatrix}
    \mathsf B_{22} & \mathsf B_{21}\\
    \mathbf 0 & \mathbf 0
    \end{pmatrix} 
    \, ,
\end{align}
\end{subequations}
where we used that the entries of $\mathsf L$ vanish for the single-state species.
Introducing the rescaling matrix
\begin{align}\label{eq:rescaling-matrix}
    \mathsf R(q) 
    \equiv \mathrm{diag} \big(
    q^{-2}\mathsf I_{N_2},\,
    \mathsf I_{N_1}\big) 
\end{align}
and using $\det(\mathsf R) = q^{-2N_2}$, one obtains
\begin{align}\label{eq:rescaled-det}
    &\det(\mathsf A + q^2\mathsf B) 
    = \det\!\big(\mathsf R(q)\big)^{-1}\,
    \det\!\Big(
    \mathsf R(q)\mathsf A 
    + \mathsf R(q)\,q^2\mathsf B
    \Big)
    \nonumber \\
    &= q^{2N_2}\,
    \det\!
    \begin{pmatrix}
    q^{-2}\mathsf A_{22} + \mathsf B_{22} 
    & \;\;q^{-2}\mathsf A_{21} + \mathsf B_{21}\\
    \mathsf A_{12} 
    & \;\;\mathsf A_{11}
    \end{pmatrix} .
\end{align}
Taking the limit $q\to\infty$ gives
\begin{align}\label{eq:large-q-limit}
    q^{-2N_2}\,\det(\mathsf A + q^2\mathsf B)
    \;\xrightarrow[q\to\infty]{}\;
    \det\!
    \begin{pmatrix}
    \mathsf B_{22} & \mathsf B_{21}\\
    \mathsf A_{12} & \mathsf A_{11}
    \end{pmatrix} .
\end{align}
For single-state species one has ${\bm\eta\equiv \mathbf c_1}$, implying $\mathsf A_{11}=(\partial_{\bm\rho}\mathbf c_1^*)_{I_1,I_1}$ and $\mathsf A_{12}=(\partial_{\bm\rho}\mathbf c_1^*)_{I_1,I_2}$, where $I_1$ and $I_2$ denote the index sets of single-state and two-state species, respectively.
Consequently, the limiting block matrix factors as
\begin{align}\label{eq:block-factorization}
    \begin{pmatrix}
    \mathsf B_{22} & \mathsf B_{21}\\
    \mathsf A_{12} & \mathsf A_{11}
    \end{pmatrix}
    = \underbrace{\mathrm{diag}\!\big(
    \mathsf L^2_{I_2,I_2},\,
    \mathsf I_{N_1}\big)}_{\equiv\mathsf G_0}\;
    \big(\partial_{\bm\rho}\mathbf c_1^*\big) \, .
\end{align}
Hence
\begin{align}\label{eq:det-large-q}
    \det(\mathsf A + q^2\mathsf B) 
    &\xrightarrow[q\to\infty]{} 
    q^{2N_2}\,\det(\mathsf G_0)\,
    \det\!\big(\partial_{\bm\rho}\mathbf c_1^*\big) 
    \nonumber \\ 
    &= \Big(\prod_{\alpha\in I_2} 
    (q\ell_c^{\alpha})^2 \Big)\,
    \det\!\big(\partial_{\bm\rho}\mathbf c_1^*\big) \, .
\end{align}
As a result, under long-wavelength stability ($s(0) = \det\mathsf A > 0$), the sign of $s(q)$ for large $q$ is determined by the sign of ${\det(\partial_{\bm\rho}\mathbf c_1^*)}$.
Consequently, if 
\begin{align}\label{eq:sufficient-criterion}
    \det\!\big(\partial_{\bm\rho}\mathbf c_1^*\big) 
    < 0 \,,
\end{align}
then $s(q)$ must change sign between ${q = 0}$ and ${q \to \infty}$, and a ${q_\mathrm{min}>0}$ with ${s(q_\mathrm{min})=0}$ exists, i.e., a finite-$q$ root and thus a type-I instability.
Equivalently, an odd number of negative real eigenvalues of ${\partial_{\bm\rho}\mathbf c_1^*}$ guarantees this sign change.
This provides a \emph{sufficient} (not necessary) criterion for the short-wavelength (type-I) instability, provided that the long-wavelength instability is absent.
This sufficient criterion generalizes the exact (necessary and sufficient) result that a stationary type-I instability in a single-species system occurs under long-wavelength stability ${\partial_\rho \eta^*>0}$ and ${\partial_\rho c_1^*<0}$ [cf.\ Eq.~\eqref{eq:3c-qmin}].
Importantly, this criterion is independent of the conversion rates $\lambda_\alpha^{\mathrm{lin}}$ and depends only on the reactive equilibria.

\subsubsection*{Summary}
\label{sec:2c-summary}

This section derived the conditions for stationary short-wavelength (type-I) instabilities in multi-species McRD systems with two cytosolic components per species.
The central idea is that the marginal eigenmode at $\sigma = 0$ can be constructed entirely from nullcline slopes.

The construction rests on two observations.
First, because membrane components do not diffuse, the marginal perturbation $(\delta\mathbf m, \delta\mathbf c_1)^\top$ must lie in the null space of the membrane-reaction Jacobian, which---under the assumption of one binding-competent cytosolic component per species---is spanned by the nullcline tangent vectors and can therefore be parametrized by a nullcline displacement $\widetilde{\delta\bm\rho}$ [Eq.~\eqref{eq:marginal-m-c1}].
Second, the non-binding components $\delta c_2^\alpha$ follow the reactive equilibrium only at long wavelengths and are progressively suppressed at short wavelengths by a low-pass filter with characteristic scale $\ell_c^\alpha$ [Eq.~\eqref{eq:suppression-inactive-cytosolic}].

Inserting the resulting eigenvector [Eq.~\eqref{eq:2c-reparameterization}] into the continuity equation yields the self-consistency condition [Eq.~\eqref{eq:typeI-onset-vector-prefactor}], whose solvability requires
\begin{align*}
    \det\!\Big(
    q^2\,\mathsf L^2\,
    \partial_{\bm\rho}\mathbf c_1^*
    + \partial_{\bm\rho}\bm\eta^*
    \Big) = 0 \, .
\end{align*}
A positive root $q_\mathrm{min}^2 > 0$ signals a type-I instability.
Equivalently, $q_\mathrm{min}^2 = -1/\mu_\mathrm{min}$, where $\mu_\mathrm{min}$ is the most negative eigenvalue of $\mathsf L^2\,\partial_{\bm\eta}\mathbf c_1^*$ [Eq.~\eqref{eq:type-I-instab-condition-cetaslope}].
A sufficient, rate-independent criterion is [Eq.~\eqref{eq:sufficient-criterion}]
\begin{align*}
    \det(\partial_{\bm\rho}\mathbf c_1^*) < 0 \, .
\end{align*}
It guarantees a sign change of the determinant between $q = 0$ and $q \to \infty$, independent of the linearized conversion rates $\lambda_\alpha^\mathrm{lin}$.
For the special case that only a single species $\alpha$ has a conversion reaction, the criterion reduces to ${(\partial_{\bm\eta}\mathbf c_1^*)_{\alpha\alpha} < 0}$ [Eq.~\eqref{eq:alpha-criterion}].

\subsection{Series of cytosolic components}
\label{app:cytSeries}

\begin{figure}[t]
\centering
\includegraphics[width=\columnwidth]{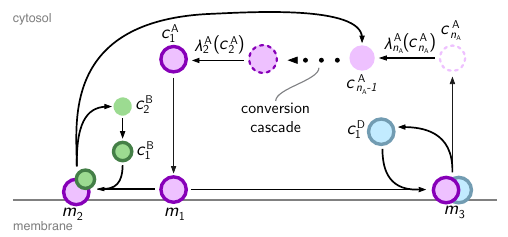}
\caption{
Exemplary membrane-cytosol reaction network with cytosolic conversion cascades for three species with different cascade depths.
Species~A (long cascade, magenta) possesses $n_\mathrm{A}$ cytosolic components converting stepwise $c_{n_\mathrm{A}}^\mathrm{A} \to \cdots \to c_1^\mathrm{A}$ at rate functions $\lambda_i^\mathrm{A}(c_i^\mathrm{A})$;
species~B (two-component cascade, green) has one non-binding and one binding-competent cytosolic state;
species~D (no cascade, cyan) has a single cytosolic component.
Arrows between cytosolic and membrane levels denote attachment (from $c_1^\alpha$ at rate $a_\alpha(\mathbf{m})\,c_1^\alpha$) and detachment [with fluxes $d_i^\alpha(\mathbf{m})$].
Detachment can feed into any cascade level: here, the membrane complex $m_2$ dissociates into $c_2^\mathrm{B}$ and $c_{n_\mathrm{A}-1}^\mathrm{A}$, after which the remaining conversion steps regenerate the respective binding-competent states $c_1^\mathrm{B}$ and $c_1^\mathrm{A}$.}
\label{fig:cartoonNetwork_cascades}
\end{figure}

We consider multi-species McRD systems in which each species $\alpha$ possesses a \emph{series} of cytosolic components $c_1^\alpha,\dots,c_{n_\alpha}^\alpha$ (see Fig.~\ref{fig:cartoonNetwork_cascades} for an illustration).
The binding-competent state is $c_1^\alpha$; the most upstream state $c_{n_\alpha}^\alpha$ is called the \emph{entry level} of the cascade.
Non-binding components convert stepwise toward the binding-competent state:
\begin{align}\label{eq:cytosolic-cascade}
    c_{n_\alpha}^\alpha
    \to
    c_{n_\alpha-1}^\alpha
    \to
    \;\cdots\;
    \to
    c_2^\alpha
    \to
    c_1^\alpha \,,
\end{align}
where each conversion step $c_{i+1}^\alpha \to c_i^\alpha$ is described by a rate function $\lambda_{i+1}^\alpha(c_{i+1}^\alpha)$ that may depend on the local concentration of the converting component and increases monotonously with the concentration [$\partial_{c_i^\alpha}[\lambda_i^\alpha(c_i^\alpha)\,c_i^\alpha] > 0$; cf.\ Eq.~\eqref{eq:conversion-reaction}].
Proteins detaching from the membrane can enter the cascade at any level; the chain then carries them through the remaining conversion steps to regenerate the membrane-binding-competent state.

Such multi-step cytosolic processing arises naturally when proteins must undergo nucleotide exchange, conformational changes, or oligomerization in the cytosol before they can rebind to the membrane.
A well-studied example is the Min protein system in \textit{E.\ coli}.
Upon MinE-stimulated ATP hydrolysis on the membrane, MinD detaches as an ADP-bound monomer~\citep{Hu.etal2002,Lackner.etal2003,Lutkenhaus2007}.
In the cytosol, nucleotide exchange converts MinD--ADP into MinD--ATP, which is a prerequisite for dimerization~\citep{Hu.etal2003,Lackner.etal2003}.
The MinD--ATP dimer stably binds the membrane through its C-terminal amphipathic helices while a single one is not sufficient for stable binding~\citep{Hu.Lutkenhaus2003,Szeto.etal2002,Szeto.etal2003}.
The number of cytosolic components in the conversion cascade depends on whether dimerization occurs in the cytosol or on the membrane after monomer attachment:
if MinD--ATP dimerizes in the cytosol before attaching, the cascade comprises three components ($c_3 = [\text{MinD--ADP}]$, $c_2 = [\text{MinD--ATP monomer}]$, $c_1 = [(\text{MinD--ATP})_2]$) [Fig.~\ref{fig:model-systems}c];
if instead MinD--ATP monomers bind the membrane directly and dimerize there, the cascade reduces to two components (${c_2 = [\text{MinD--ADP}]}$, ${c_1 = [\text{MinD--ATP monomer}]}$).

Combining the conversion cascade Eq.~\eqref{eq:cytosolic-cascade} with membrane attachment $a_\alpha(\mathbf{m})\,c_1^\alpha$ of the binding-competent state and detachment fluxes ${d_i^\alpha(\mathbf{m})}$ into the cytosolic states ${c_i^\alpha}$, the cytosolic reaction fluxes ${{\mathbf{R}}_c^\alpha=\big(R_{c,1}^\alpha,\dots,R_{c,n_\alpha}^\alpha\big)^{\!\top}}$ for each species $\alpha$ take the following form.
The binding-competent component is populated by conversion from level~2 and loses them by membrane attachment:
\begin{subequations}\label{eq:cytSeries-Rc}
\begin{equation}\label{eq:cytSeries-bindingCompetent}
R_{c,1}^\alpha
= -a_\alpha(\mathbf{m})\,c_1^\alpha
+ s_{c,2}^\alpha\,\lambda_2^\alpha(c_2^\alpha)\,c_2^\alpha .
\end{equation}
Each intermediate non-binding component (${2\le i\le n_\alpha{-}1}$) receives detachment from the membrane, converts toward the next higher level, and is fed by the next upstream level:
\begin{equation}\label{eq:cytSeries-nonBinding-1}
R_{c,i}^\alpha
= d_i^\alpha(\mathbf{m})
-\lambda_i^\alpha(c_i^\alpha)\,c_i^\alpha
+\frac{s_{c,i+1}^\alpha}{s_{c,i}^\alpha}\,
\lambda_{i+1}^\alpha(c_{i+1}^\alpha)\,c_{i+1}^\alpha .
\end{equation}
The entry level $i=n_\alpha$ has no upstream component:
\begin{equation}\label{eq:cytSeries-nonBinding-2}
R_{c,n_\alpha}^\alpha
= d_{n_\alpha}^\alpha(\mathbf{m})
-\lambda_{n_\alpha}^\alpha(c_{n_\alpha}^\alpha)\,c_{n_\alpha}^\alpha ,
\end{equation}
\end{subequations}
which can be included in the relation Eq.~\eqref{eq:cytSeries-nonBinding-1} by defining ${\lambda_{n_\alpha+1}^\alpha,c_{n_\alpha}^\alpha=0}$. 
The membrane reaction term ${\mathbf{R}}_m(\mathbf{m},\mathbf{c}_1)$ is defined as before.
The stoichiometric factors $s_{c,i}^\alpha$ account for multi-merization processes (e.g., cytosolic dimerization of MinD); we normalize these factors such that $s_{c,1}^\alpha = 1$.
Together with the diffusion matrix $\mathbf{D}_c^\alpha \equiv \mathrm{diag}(D_{c,1}^\alpha,\dots,D_{c,n_\alpha}^\alpha)$, the reaction-diffusion equations for the cytosolic components of species $\alpha$ read
\begin{equation}\label{eq:cytSeries-RD}
    \partial_t \mathbf{c}^\alpha
    = 
    \mathbf{D}_c^\alpha \,
    \bm{\nabla}^2 
    \mathbf{c}^\alpha
    + 
    {\mathbf{R}}_c^\alpha ( \mathbf{m}, \mathbf{c}^\alpha ) \,.
\end{equation}

\subsubsection*{Linearized non-binding chain}

As in the two-component case (Sec.~\ref{app:2c}), we linearize the reaction-diffusion equations~\eqref{eq:cytSeries-RD} about the homogeneous steady state and insert a Fourier eigenmode $\propto e^{\sigma t + i\mathbf{q}\cdot\mathbf{x}}$ (cf.\ Sec.~\ref{sec:lsa}).
The Laplacian $\bm\nabla^2$ is thereby replaced by $-q^2$, and we evaluate at a marginal mode ($\sigma = 0$), so the time derivative drops out.
Moreover, we write ${\delta d_i^\alpha \equiv \partial_{\mathbf{m}} d_i^\alpha|_\mathrm{hss}\cdot\delta\mathbf{m}}$ for the linearized detachment flux and define the linearized conversion rates
\begin{equation}\label{eq:cytSeries-lin-conversion}
    \lambda_{i,\mathrm{lin}}^\alpha 
    \equiv 
    \left.\frac{\mathrm d}{\mathrm d c}\big[\lambda_i^\alpha(c)\,c\big]\right|_{c=c_{i,\mathrm{hss}}^{\alpha}} > 0\,,
\end{equation}
simplifying to $\lambda_i^\alpha \equiv \lambda_{i,\mathrm{lin}}^\alpha$ in the following.
The non-binding components of each species $\alpha$ then satisfy, for $2\le i\le n_\alpha{-}1$,
\begin{equation}\label{eq:cytSeries-lin-nonBinding-general}
    \big(\lambda_i^\alpha + D_{c,i}^\alpha q^2\big) \delta c^\alpha_i 
    = \delta d_i^\alpha + \frac{s_{c,i+1}^\alpha}{s_{c,i}^\alpha} \lambda_{i+1}^\alpha \delta c^\alpha_{i+1}\,,
\end{equation}
and for the entry level $i = n_\alpha$,
\begin{equation}\label{eq:cytSeries-lin-nonBinding-terminal}
    \big(\lambda_{n_\alpha}^\alpha + D_{c,n_\alpha}^\alpha q^2\big) \delta c^\alpha_{n_\alpha} = \delta d_{n_\alpha}^\alpha\,.
\end{equation}
Dividing by ${\lambda_i^\alpha}$ and introducing the cytosolic diffusion lengths
\begin{equation}\label{eq:cytSeries-diff-length}
    \ell_{c,i}^{\alpha}
    \equiv
    \sqrt{D_{c,i}^\alpha\big/\lambda_i^\alpha}
\end{equation}
recasts the chain, for $2\le i\le n_\alpha{-}1$, as
\begin{equation}\label{eq:cytSeries-normalized-chain}
    \big[1 + (\ell_{c,i}^{\alpha} q)^2\big]\, 
    \delta c_i^\alpha
    = \frac{\delta d_i^\alpha}{\lambda_i^\alpha}
    + \frac{\lambda_{i+1}^\alpha s_{c,i+1}^\alpha}{\lambda_i^\alpha s_{c,i}^\alpha}\,\delta c_{i+1}^\alpha\,,
\end{equation}
and for the entry level,
\begin{equation}\label{eq:cytSeries-normalized-terminal}
    \big[1 + (\ell_{c,n_\alpha}^\alpha q)^2\big]\, 
    \delta c_{n_\alpha}^\alpha
    = \frac{\delta d_{n_\alpha}^\alpha}{\lambda_{n_\alpha}^\alpha}\,.
\end{equation}
The entry-level relation can again be included in Eq.~\eqref{eq:cytSeries-lin-nonBinding-general} under our definition ${\lambda_{n_\alpha+1}^\alpha,c_{n_\alpha}^\alpha=0}$.
These equations form a recursion relation for the cytosolic perturbations $\delta c_i^\alpha$ in terms of the detachment perturbations $\delta d_i^\alpha$.
Defining the \emph{cumulative filtering factor} for ${k\ge i\ge 2}$,\footnote{Note that the filtering factor also depends on the species index $\alpha$, $P_i^{k,\alpha}(q)$, but we suppress this index in order to simplify the notation.}
\begin{equation}\label{eq:cumulative-filter-def}
P_i^k(q)
\equiv \prod_{l=i}^{k}\Big[1+(\ell_{c,l}^{\alpha} q)^2\Big]^{-1} ,
\end{equation}
the explicit solution of the recursion is
\begin{equation}\label{eq:cytSeries-detachment-recursionSol}
\delta c_i^\alpha
= \frac{1}{\lambda_i^\alpha\,s_{c,i}^\alpha}
\sum_{k=i}^{n_\alpha} s_{c,k}^\alpha\,\delta d_k^\alpha\,P_{i}^k(q)\,,
\qquad i\ge 2.
\end{equation}

Thus, each conversion step acts as a diffusive low-pass filter with cutoff set by ${\ell_{c,l}^\alpha}$.
The product ${P_{i}^k(q)}$ collects ${k-i+1}$ such filters (from level $l=i$ to $l=k$), describing how detachment into upstream state ${k}$ is attenuated before reaching ${c_i^\alpha}$.
The limiting behaviors are
\begin{subequations}\label{eq:P-asymptotics}
\begin{align}
&P_{i}^k(q)\xrightarrow[q\to 0]{} 1\,,
\\
&P_{i}^k(q)\sim \Bigl[\prod_{l=i}^{k}(\ell_{c,l}^{\alpha})^{-2}\Bigr]\,
q^{-2(k-i+1)}
\;\; (q\to\infty)\,.
\end{align}   
\end{subequations}
At large ${q}$, the chain response is thus dominated by the nearest non-binding levels; far-upstream states contribute only algebraically small corrections.
Increasing any ${\ell_{c,l}^\alpha}$ (stronger diffusion or slower reactivation at level ${l}$) strengthens filtering and shifts the onset of type-I instability to longer wavelengths (smaller ${q_\mathrm{min}}$), as derived below.
In the long-wavelength limit ${q\to 0}$, ${P_{i}^k\to 1}$ and the recursion solution Eq.~\eqref{eq:cytSeries-detachment-recursionSol} becomes $q$-independent: $\delta c_i^\alpha = (\lambda_i^\alpha s_{c,i}^\alpha)^{-1}\sum_k s_{c,k}^\alpha\,\delta d_k^\alpha$.
In this limit, all non-binding levels respond as if conversion were instantaneous relative to diffusion, and the chain is in LQSS which will become clear in the following.

Inserting the recursion solution Eq.~\eqref{eq:cytSeries-detachment-recursionSol} into the definition of the mass-redistribution potential $\eta_\alpha$ (cf.\ Sec.~\ref{sec:3c}), ${D_{c,1}^\alpha\,\delta\eta_\alpha=\sum_{j=1}^{n_\alpha}D_{c,j}^\alpha s_{c,j}^\alpha\,\delta c_j^\alpha}$, gives
\begin{align}
    D_{c,1}^\alpha\,\delta \eta_\alpha
    &= 
    D_{c,1}^\alpha 
    \delta c_1^\alpha
\nonumber \\
&\quad+ \sum_{i=2}^{n_\alpha}\frac{D_{c,i}^\alpha}{\lambda_i^\alpha}
  \sum_{k=i}^{n_\alpha} s_{c,k}^\alpha\,\delta d_k^\alpha\,P_i^k(q)\,.
\label{eq:general-deltaEta-cytosolicSeries-ellP}
\end{align}
Here, the binding-competent component (${i=1}$) has been separated out. 

\subsubsection*{Link to nullcline slopes}

As in the two-component case [Eq.~\eqref{eq:marginal-m-c1}], we parametrize the binding-competent components by a nullcline displacement $\widetilde{\delta\bm\rho}$:
\begin{equation}
    \delta c_1^\alpha = (\partial_{\bm\rho} c_1^{\alpha*})\,\cdot\widetilde{\delta\bm\rho}\, .
\end{equation}
Because the structure of the reaction terms for the membrane-bound components remains the same, we then have by the same arguments as in the two-component case
\begin{equation}\label{eq:cytSeries-delta-m-param}
    \delta m_i^\alpha =
    \partial_{\bm \rho} \mathbf{m}^*\,\widetilde{\delta \bm\rho}\,.
\end{equation}
With these parametrizations, we now construct the marginal-mode components of the non-binding states.

\smallskip
\paragraph*{Eliminating detachment fluxes in favor of nullcline slopes.---}
Our goal is to rewrite the inner sum in Eq.~\eqref{eq:general-deltaEta-cytosolicSeries-ellP} so that all detachment variations $\delta d_k^\alpha$ are expressed through the nullcline shifts ${\delta c_k^{\alpha*}}$.
To this end, we evaluate the homogeneous steady state of the non-binding components given by ${R_{c,i}^\alpha = 0}$.
We differentiate this steady-state balance with respect to the conserved total densities $\bm\rho$ [cf.\ Eqs.~\eqref{eq:cytSeries-nonBinding-1},~\eqref{eq:cytSeries-nonBinding-2}].
Evaluating the derivative along the nullcline displacement $\widetilde{\delta\bm\rho}$ gives, for $i\geq 2$, the linearized conversion--detachment flux balance
\begin{equation}\label{eq:cytSeries-chemEquilibria}
\lambda_i^\alpha\,s_{c,i}^\alpha\,\delta c_i^{\alpha*}
-
\lambda_{i+1}^\alpha\,s_{c,i+1}^\alpha\,\delta c_{i+1}^{\alpha*}
=
s_{c,i}^\alpha\,\delta d_i^\alpha\,,
\end{equation}
where
\begin{equation}\label{eq:def-delta-ci-star}
\delta c_i^{\alpha *} \equiv \partial_{\bm\rho} c_i^{\alpha*}\big|_\text{hss} \cdot \widetilde{\delta\bm\rho}
\end{equation}
denotes the nullcline shift of component $c_i^\alpha$ and we used Eq.~\eqref{eq:cytSeries-delta-m-param} in
\begin{equation}
    \delta d_i^\alpha = \partial_{\mathbf{m}} d_i^\alpha\big|_\mathrm{hss}\,
    \partial_{\bm \rho} \mathbf{m}^*\,\widetilde{\delta \bm\rho}\,.
\end{equation}

Substituting the flux balance Eq.~\eqref{eq:cytSeries-chemEquilibria} (with the entry-level convention $\lambda_{n_\alpha+1}^\alpha=\delta c_{n_\alpha+1}^{\alpha*}=0$) into the filtered inner sum
\begin{equation}\label{eq:Fi-def}
F_i
\equiv \sum_{k=i}^{n_\alpha} s_{c,k}^\alpha\,\delta d_k^\alpha\,P_{i}^{\,k}(q),
\end{equation}
and shifting the summation index in the second term, gives
\begin{align}\label{eq:Fi-shifted}
F_i
&= \sum_{k=i}^{n_\alpha}\lambda_k^\alpha\,s_{c,k}^\alpha\,\delta c_k^{\alpha*}\,P_{i}^{\,k}(q)
\nonumber \\
&\quad-\sum_{k=i+1}^{n_\alpha}\lambda_k^\alpha\,s_{c,k}^\alpha\,\delta c_k^{\alpha*}\,P_{i}^{\,k-1}(q).
\end{align}
Using the product identity $P_{i}^{\,k-1}=P_{i}^{\,k}\,(1+ (q\ell_{c,k}^\alpha)^2)$ [cf.\ Eq.~\eqref{eq:cumulative-filter-def}] and $\lambda_k^\alpha(\ell_{c,k}^\alpha q)^2 = D_{c,k}^\alpha q^2$, the $k \ge i+1$ terms partially cancel, leaving
\begin{align}\label{eq:Fi-simplified-step1}
F_i
&= \lambda_i^\alpha\,s_{c,i}^\alpha\,\delta c_i^{\alpha*}\,P_i^i(q)
\nonumber \\
&\quad - q^2 \sum_{k=i+1}^{n_\alpha} D_{c,k}^\alpha\,s_{c,k}^\alpha\,\delta c_k^{\alpha*}\,P_{i}^{\,k}(q)\,.
\end{align}

Extending the sum to include $k=i$ and using $(\lambda_i^\alpha+D_{c,i}^\alpha q^2)\,P_i^i = \lambda_i^\alpha$ simplifies this to
\begin{equation}\label{eq:Fi-final}
F_i
= \lambda_i^\alpha\,s_{c,i}^\alpha\,\delta c_i^{\alpha*}
- q^2 \sum_{k=i}^{n_\alpha} D_{c,k}^\alpha\,s_{c,k}^\alpha\,\delta c_k^{\alpha*}\,P_{i}^{\,k}(q).
\end{equation}
Inserting $F_i$ into the outer sum of Eq.~\eqref{eq:general-deltaEta-cytosolicSeries-ellP} and swapping the summation order $\sum_{i=2}^{n_\alpha}\sum_{k=i}^{n_\alpha}=\sum_{k=2}^{n_\alpha}\sum_{i=2}^{k}$ yields
\begin{align}\label{eq:bracket}
&\sum_{i=2}^{n_\alpha}
\frac{D_{c,i}^\alpha}{\lambda_i^\alpha} \, F_i
= \sum_{k=2}^{n_\alpha} D_{c,k}^\alpha\,s_{c,k}^\alpha\,\delta c_k^{\alpha*}
\nonumber \\
&\qquad\times
\Bigg[1-q^2\sum_{i=2}^{k}\ell_{c,i}^\alpha\,P_{i}^{\,k}(q)\Bigg].
\end{align}
The bracket can be collapsed using the algebraic identity (verified by induction)
\begin{equation}\label{eq:prod-identity}
\prod_{m=2}^{k}\frac{1}{1+a_m}
= 1-\sum_{i=2}^{k}a_i\prod_{m=i}^{k}\frac{1}{1+a_m}\,,
\end{equation}
with ${a_m\equiv(\ell_{c,m}^\alpha q)^2}$.
So, the bracket equals $P_2^k(q)$.
Adding the binding-competent contribution from Eq.~\eqref{eq:general-deltaEta-cytosolicSeries-ellP} then gives the compact result
\begin{align}
\label{eq:deta-cytSeries-final}
D_{c,1}^\alpha\,\delta \eta_\alpha
&= \bigg[ D_{c,1}^\alpha \partial_{\bm\rho}c_1^{\alpha*}
\nonumber \\
&\quad + \sum_{i=2}^{n_\alpha} D_{c,i}^\alpha\,s_{c,i}^\alpha\,
\partial_{\bm{\rho}} c_i^{\alpha*}\, P_2^i (q)\bigg]
\widetilde{\delta\bm\rho}\,.
\end{align}
The binding-competent component contributes its full nullcline slope $\partial_{\bm\rho}c_1^{\alpha*}$ as previously in the two-component case because it couples directly to the membrane without intervening conversion steps.
Each non-binding level ${i \ge 2}$, by contrast, is attenuated by the cumulative filter
\[
P_2^i(q)=\prod_{k=2}^{i}\bigl[1+(q\ell_{c,k}^\alpha)^2\bigr]^{-1},
\]
which collects one low-pass factor per conversion step from level~2 to level~$i$.
At finite $q$, this cumulative filter decreases with cascade depth,
$P_2^{i+1}(q)<P_2^i(q)$; hence, the deeper a component sits in the cascade, the more filtering steps suppress its contribution.
In the LQSS limit $q\to 0$, $P_2^i \to 1$ and all levels contribute with their full nullcline slopes, so Eq.~\eqref{eq:deta-cytSeries-final} reduces to the definition of $\partial_{\bm\rho}\eta_\alpha^*$.
For $q\to\infty$, the non-binding contributions are progressively suppressed with $P_2^i \sim q^{-2(i-1)}$, so that $\delta\eta_\alpha$ is increasingly dominated by the binding-competent slope alone.

\smallskip
\paragraph*{Self-consistency (marginality) condition.---}
At marginality ($\sigma = 0$), the mass-redistribution potential must be spatially uniform, $\delta\eta_\alpha = 0$, because any spatial variations would induce mass transport and, thus, dynamics [cf.\ Eq.~\eqref{eq:self-consistency}].
Using $\partial_{\bm\rho}\eta_\alpha^*
= \partial_{\bm\rho}c_1^{\alpha*}
+ \sum_{i=2}^{n_\alpha}(D_{c,i}^\alpha/D_{c,1}^\alpha)\,s_{c,i}^\alpha\,\partial_{\bm\rho}c_i^{\alpha*}$ and the expression for the mass-redistribution potential of the marginal mode Eq.~\eqref{eq:deta-cytSeries-final}, ${\delta\eta_\alpha = 0}$ results in the self-consistency equation
\begin{align}
    0
    = \Bigg[
    \partial_{\bm\rho}\eta_\alpha^*
    &- \sum_{i=2}^{n_\alpha}\frac{D_{c,i}^\alpha}{D_{c,1}^\alpha}\,s_{c,i}^\alpha\,
    \big[1-P_2^{\,i}(q)\big]
    \nonumber \\
    &\quad\times\partial_{\bm\rho}c_i^{\alpha*}
    \Bigg]\widetilde{\delta\bm\rho}\,.
\label{eq:neutrality-eta-singled}
\end{align}
To write this in vectorial form across all species, we define the diagonal matrices $\mathbf{D}_i \equiv \mathrm{diag}(D_{c,i}^{\mathrm A}, D_{c,i}^{\mathrm B}, \dots)$ collecting the diffusion constants of cytosolic level $i$ across species, the filtering matrix ${\mathbf{P}_{2}^{\,i}(q)=\mathrm{diag}(P_{2}^{\,i,\mathrm A},P_{2}^{\,i,\mathrm B},\dots)}$, and the stoichiometric matrix
$\mathbf{S}_{c,i}=\mathrm{diag}(s_{c,i}^{\mathrm A},s_{c,i}^{\mathrm B},\dots)$
(with $s_{c,i}^\alpha = 0$ whenever species $\alpha$ does not possess level $i$).
Stacking species, Eq.~\eqref{eq:neutrality-eta-singled} becomes
\begin{align}\label{eq:cytSeries-marginalityCond-vectorial}
0=\Bigg[\,
\partial_{\bm\rho}\bm\eta^{*}
 -\sum_{i=2}^{N_c}\mathbf{D}_i\mathbf{D}_1^{-1}\,\mathbf{S}_{c,i}\,
 &\big[\mathbf{I}-\mathbf{P}_{2}^{\,i}(q)\big]
 \nonumber \\
 &\times\partial_{\bm\rho}\mathbf{c}_i^{*}
 \Bigg]\widetilde{\delta\bm\rho}\,.
 \end{align}
A nontrivial solution $\widetilde{\delta\bm\rho} \neq 0$ exists if and only if the matrix in brackets is singular.
The lower boundary $q_\mathrm{min}$ of the band of unstable modes is therefore the smallest positive wavenumber solving
\begin{align}\label{eq:cytSeries-typeI-detCond}
0=\det\Bigg(\,
\partial_{\bm\rho}\bm\eta^{*}
 -\sum_{i=2}^{N_c}\mathbf{D}_i\mathbf{D}_1^{-1}\,&\mathbf{S}_{c,i}\,
 \big[\mathbf{I}-\mathbf{P}_{2}^{\,i}(q_\mathrm{min})\big]
 \nonumber \\
 &\times\partial_{\bm\rho}\mathbf{c}_i^{*}
 \Bigg).
 \end{align}

Equation~\eqref{eq:cytSeries-typeI-detCond} is the exact type-I onset condition for arbitrary detachment into non-binding levels.
Its structure is worth examining.
The leading term $\partial_{\bm\rho}\bm\eta^*$ is the slope of the mass-redistribution potential, which governs the LQSS stability at $q=0$.
The sum corrects this slope at finite $q$ by subtracting a $q$-dependent contribution for each non-binding level $i \ge 2$.
The correction for level $i$ involves three factors: the diffusion-ratio matrix $\mathbf{D}_i\mathbf{D}_1^{-1}$ comparing level-$i$ to binding-competent diffusion, the stoichiometric matrix $\mathbf{S}_{c,i}$, and the \emph{filter defect} $\mathbf{I} - \mathbf{P}_2^{\,i}(q)$.
At $q=0$, every filter defect vanishes ($\mathbf{P}_2^{\,i} \to \mathbf{I}$), so the sum drops out and the condition reduces to $\det(\partial_{\bm\rho}\bm\eta^*) = 0$, the LQSS condition.
At finite $q$, the filter defect grows: each non-binding level's contribution to $\partial_{\bm\rho}\bm\eta^*$ is partially subtracted again because diffusion suppresses gradients in the non-binding components.
The deeper a cytosolic components sits in the cascade (larger $i$), the more conversion steps separate it from the binding-competent state, the larger its filter defect, and the more strongly its contribution is filtered out as $q$ increases.

Instead of subtracting contributions from the non-binding components, the instability criterion for the two-cytosolic-component system Eq.~\eqref{eq:type-I-instab-condition} was formulated by adding an additional contribution of the binding-competent state, which increases with $q$.
We arrive at the analogous expression here by multiplying the component-wise relation for $q_\mathrm{min}$, Eq.~\eqref{eq:neutrality-eta-singled}, by ${\prod_{k=2}^{n_\alpha}[1+(q \ell_{c,k}^\alpha)^2]}$.
Stacking species, this yields
\begin{equation}\label{eq:cytSeries-typeI-detCond-Qi}
0
=
\det\Biggl(
\partial_{\bm\rho}\bm\eta^*
+
\sum_{i=1}^{N_c}
\mathbf{Q}_i(q_\mathrm{min})\,\partial_{\bm\rho}\mathbf{c}_i^*
\Biggr) ,
\end{equation}
with the diagonal level-resolved gain matrices
\begin{equation}\label{eq:cytSeries-Qi-def}
\mathbf{Q}_i(q)
=
\mathbf{D}_i\mathbf{D}_1^{-1}
\Biggl[
\prod_{k=i+1}^{N_c}
\bigl(\mathbf{I}+\mathbf{D}_k\bm\lambda_k^{-1}q^2\bigr)
-
\mathbf{I}
\Biggr]
\mathbf{S}_{c,i}\,,
\end{equation}
where $\bm\lambda_i \equiv \mathrm{diag}(\lambda_i^{\mathrm A},\lambda_i^{\mathrm B},\dots)$.
For species lacking cytosolic level $i$, we set $D_{c,i}^\alpha = 0$, $s_{c,i}^\alpha = 1$, and $\lambda_i^\alpha = \infty$, so that the corresponding diagonal entry of $\mathbf{Q}_i(q)$ vanishes.

The two forms Eqs.~\eqref{eq:cytSeries-typeI-detCond} and \eqref{eq:cytSeries-typeI-detCond-Qi} are exactly equivalent. While the contributions from the non-binding components are suppressed at finite $q$ in Eq.~\eqref{eq:cytSeries-typeI-detCond}, the contribution of the binding-competent component (and components closer to it in the conversion chain) is increased at finite $q$ in Eq.~\eqref{eq:cytSeries-typeI-detCond-Qi}.
Including only a single non-binding level $c_2^\alpha$, the second expression Eq.~\eqref{eq:cytSeries-typeI-detCond-Qi} recovers the result for the two-component case, Eq.~\eqref{eq:type-I-instab-condition}.
By the same arguments as for the two-component case, a slope matrix $\partial_{\bm\rho}\mathbf{c}_1^*$ with an odd number of eigenvalues with a negative real part is a sufficient condition for short-wavelength instability (given long-wavelength stability) also in the general case of a cytosolic conversion cascade.

\subsubsection*{Detachment only into the entry level of the cytosolic cascade}

Our starting point is the general relation Eq.~\eqref{eq:cytSeries-marginalityCond-vectorial}, written with the shorthand $\mathbf{R}_i\equiv\mathbf{D}_1^{-1}\,\mathbf{D}_i\,\mathbf{S}_{c,i}$:
\begin{align}
0
= \Bigg[
\partial_{\bm\rho}\bm\eta^{*}
- \sum_{i=2}^{N_c} \mathbf{R}_i\,\big(\mathbf{I}-\mathbf{P}_{2}^{\,i}(q)\big)\,\partial_{\bm\rho}\mathbf{c}_i^{*}
\Bigg]\widetilde{\delta\bm\rho}\,.
\label{eq:neutral-compact-start}
\end{align}
If detachment only occurs into the entry level of the cascade, the flux balance Eq.~\eqref{eq:cytSeries-chemEquilibria} of the homogeneous steady state has a vanishing right-hand side for all but the entry level.
All non-binding nullcline slopes are then proportional for $i,j\ge2$:
\begin{equation}\label{eq:inactive-prop}
\partial_{\bm\rho}\mathbf{c}_i^{*}
= \bm\Gamma_{i2}\,\partial_{\bm\rho}\mathbf{c}_2^{*},
\qquad
\bm\Gamma_{i2} \equiv \bm\lambda_2\,\bm\lambda_i^{-1}\,\mathbf{S}_{c,2}\,\mathbf{S}_{c,i}^{-1},
\end{equation}
where $\bm\lambda_i\equiv\mathrm{diag}(\lambda_i^{\mathrm A},\lambda_i^{\mathrm B},\dots)$.
Substituting into the expression for $\partial_{\bm\rho}\bm\eta^*$ yields
\begin{equation}\label{eq:eta-slope-A0}
\partial_{\bm\rho}\bm\eta^{*}
= \partial_{\bm\rho}\mathbf{c}_1^{*}
+ \mathbf{A}(0)\,\partial_{\bm\rho}\mathbf{c}_2^{*},
\end{equation}
where $\mathbf{A}(0) \equiv \sum_{i=2}^{N_c}\mathbf{R}_i\,\bm\Gamma_{i2}$.
Defining the $q$-dependent diagonal matrix
\begin{equation}\label{eq:A-q-def}
\mathbf{A}(q) \equiv \sum_{i=2}^{N_c} \mathbf{R}_i\,\bm\Gamma_{i2}\,\mathbf{P}_{2}^{\,i}(q)\,,
\end{equation}
and using Eqs.~\eqref{eq:inactive-prop},~\eqref{eq:eta-slope-A0} to eliminate $\partial_{\bm\rho}\mathbf{c}_i^*$ and $\partial_{\bm\rho}\mathbf{c}_2^*$ from Eq.~\eqref{eq:neutral-compact-start}, one arrives at the simplified condition

\begin{equation}\label{eq:final-Q}
0
= \Big[\partial_{\bm\rho}\bm\eta^{*}
+ \mathbf{Q}(q)\,\partial_{\bm\rho}\mathbf{c}_1^{*}\Big]\widetilde{\delta\bm\rho}\,,
\end{equation}
where $\mathbf{Q}(q) \equiv \mathbf{A}(0)\,\mathbf{A}(q)^{-1}-\mathbf{I}$.
Since all matrices involved are diagonal, $\mathbf{Q}(q)$ is diagonal with per-species entries $Q_{\alpha\alpha}(q)$.
Resolving the intermediate definitions ($\mathbf{R}_i = \mathbf{D}_1^{-1}\mathbf{D}_i\mathbf{S}_{c,i}$, $\bm\Gamma_{i2} = \bm\lambda_2\bm\lambda_i^{-1}\mathbf{S}_{c,2}\mathbf{S}_{c,i}^{-1}$) and noting that the common prefactor $\lambda_2^\alpha s_{c,2}^\alpha/D_{c,1}^\alpha$ cancels in the ratio $A_{\alpha\alpha}(0)/A_{\alpha\alpha}(q)$, one obtains the explicit per-species formula
\begin{equation}\label{eq:Q-explicit}
Q_{\alpha\alpha}(q)
= \frac{\displaystyle\sum_{i=2}^{n_\alpha}(\ell_{c,i}^\alpha)^2}
{\displaystyle\sum_{i=2}^{n_\alpha}(\ell_{c,i}^\alpha)^2\,P_2^i(q)}
- 1\,,
\end{equation}
where $\ell_{c,i}^\alpha = \sqrt{D_{c,i}^\alpha/\lambda_i^\alpha}$ is the diffusion length of level $i$ [Eq.~\eqref{eq:cytSeries-diff-length}] and $P_2^i(q) = \prod_{k=2}^{i}[1+(\ell_{c,k}^\alpha q)^2]^{-1}$ is the cumulative filter [Eq.~\eqref{eq:cumulative-filter-def}].
At $q=0$, all filters equal unity and the two sums coincide, giving $Q_{\alpha\alpha}=0$, recovering the LQSS condition for an onset of instability.
As $q$ increases, the filters suppress the deeper levels ($P_2^i < 1$), the denominator shrinks, and $Q_{\alpha\alpha}(q)$ grows.
This leads, as in the cases analyzed before, to increased contributions of the binding-competent and close-by components.
If no direct detachment occurs into the intermediate components of the cascade, their collective effect can be described by increasing the contribution of the binding-competent component alone (because the component variations are proprotional to each other).
For species with $n_\alpha = 1$, the sums are empty and we set $Q_{\alpha\alpha}(q) \equiv 0$.

In the case of no intermediate detachment, the determinant condition for a type-I instability thus reduces to the existence of a solution $q_\mathrm{min}>0$ of
\begin{equation}\label{eq:Q-neutrality}
0
= \det\!\Big( \mathbf{Q}(q_\mathrm{min})\,\partial_{\bm\rho}\mathbf{c}_1^{*}
+ \partial_{\bm\rho}\bm\eta^{*} \Big).
\end{equation}
Again, as $\mathbf{Q}(q)$ monotonously increases with $q$, a sufficient criterion for short-wavelength instability (assuming long-wavelength stability) is the existence of an odd number of eigenvalues with a negative real part of the slope matrix ${\partial_{\bm\rho}\mathbf{c}_1^{*}}$ of the binding-competent state.

\subsection{Self-recruitment and counteracting cytosolic gradients}
\label{app:self-recruitment}
To make the role of self-recruitment for short-wavelength instabilities explicit, we work out, for a minimal reaction kinetics of a three-component model, how self-recruitment produces cytosolic profiles in $c_1$ and $c_2$ with opposite signs in the LQSS limit.
In the main text (see Sec.~\ref{sec:3c}), we argued that this is the condition under which the three-component model can show a short-wavelength instability.

Consider the well-mixed kinetics
\begin{subequations}
\label{eq:wellmixed-app}
\begin{align}
    \partial_t m   &= a(m)\,c_1 - d(m)\,m \, ,\\
    \partial_t c_1 &= \lambda\,c_2 - a(m)\,c_1 \, ,\\
    \partial_t c_2 &= d(m)\,m - \lambda\,c_2 \, ,
\end{align}
\end{subequations}
with attachment rate $a(m)$, detachment rate $d(m)$, and conversion rate $\lambda$ for $c_2 \to c_1$.
At a reactive equilibrium $(m^*,c_1^*,c_2^*)$,
\begin{equation}
\label{eq:reactive-equilibrium-conditions-app}
\lambda\,c_2^* = a(m^*)\,c_1^* \, ,
\qquad
d(m^*)\,m^* = \lambda\,c_2^* \, ,
\end{equation}
which fix the cytosolic densities as
\begin{align}
c_1^*=\frac{d(m^*)}{a(m^*)}\,m^* \, ,
\qquad
c_2^*=\frac{d(m^*)}{\lambda}\,m^* \, .
\end{align}
We assume that, on the branch of reactive equilibria considered (Sec.~\ref{sec:heuristics_2cmcrd}), one has $\partial_\rho m^*>0$ because we assume changes in the membrane density to be large compared to changes in the cytosolic densities.
Then, the signs of $\partial_\rho c_1^*$ and $\partial_\rho c_2^*$ are determined by the dependence of $c_1^*/m^*$ and $c_2^*/m^*$ on $m^*$.

Specializing to a constant detachment rate $d(m)\equiv d$,
\begin{subequations}
\begin{align}
    \partial_\rho c_2^* &= \frac{d}{\lambda}\,\partial_\rho m^* > 0 \, ,\\ 
    \partial_\rho c_1^* &= \frac{d}{a(m^*)^2}\,\big[a(m^*)-m^* a'(m^*)\big]\,\partial_\rho m^* \, .
\end{align}
\end{subequations}
The sign of $\partial_\rho c_1^*$ is therefore set by the bracket: $\partial_\rho c_1^*<0$ when $m^* a'(m^*) > a(m^*)$, i.e., when attachment is sufficiently self-recruiting (the attachment rate $a(m)$ grows superlinearly with $m$ at the equilibrium). Under this condition, the cytosolic profiles in $c_1$ and $c_2$ have opposite signs in the LQSS limit, as sketched in Fig.~\ref{fig:3c-mass-redistribution}(a).

Whether the homogeneous state is stable at long wavelength is set by the relative magnitudes of these two contributions in 
$D_\mathrm{eff} = D_{c_1}\partial_\rho c_1^* + D_{c_2}\partial_\rho c_2^*$. 
The stabilizing $c_2$ term grows with $d/\lambda$, so for sufficiently large $d/\lambda$ the long-wavelength instability is suppressed even when $\partial_\rho c_1^*<0$.
This is the regime, long-wavelength stable despite $\partial_\rho c_1^*<0$, in which short-wavelength instabilities arise (Sec.~\ref{sec:3c}).

\section{Bulk--boundary coupled systems}
\label{app:bbc}

Membrane attachment and detachment couple the reaction-diffusion dynamics in the three-dimensional cytosolic bulk to those on the two-dimensional membrane.
The sections above have neglected the perpendicular extent of the cytosol and treated it as a compartment of the same geometry as the membrane.
Protein gradients perpendicular to the membrane can critically affect the dynamics and couple the patterns to the system geometry~\citep{Brauns.etal2021b,Halatek.etal2018,Thalmeier.etal2016,Wurthner.etal2022,Burkart.etal2022a,Burkart.Mueller.Frey2024}.

\begin{figure}
\centering
\includegraphics[width=\columnwidth]{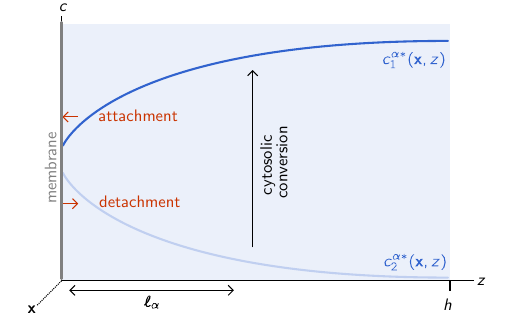}
\caption{
    The cytosolic density profiles of the laterally homogenenous steady state.
    The steady-state density profiles of the active $c_1^{\alpha*}$ (thick dark blue curve) and inactive $c_2^{\alpha*}$ (thin light blue curve) cytosolic components of a species $\alpha$ are shown in a system with bulk-boundary coupling.
	The membrane (gray) is the boundary of the cytosolic bulk volume (light blue shade) that extend to a height $z=h$ perpendicular to the membrane.
    We assume a no-flux boundary condition at $z=h$.
    Proteins detach from the membrane into the inactive component (horizontal right arrow), are converted into the active state inside the cytosol (black upward arrow) and reattach onto the membrane in the active state (horizontal left arrow).
    As a result, gradients form in the cytosolic species on the diffusive length scale $\ell_c=\sqrt{D_{c_2^\alpha}/\lambda_\alpha}$ determined by the diffusion coefficient of the inactive cytosolic component $D_{c_2^\alpha}$ and the linear conversion rate $\lambda_\alpha$.
	}
	\label{fig:2cyt-bbc}
\end{figure}

To study this bulk-boundary coupling, we consider a flat membrane bounding a cytosolic bulk of finite height $h$, with the perpendicular coordinate $z$ running from $z=0$ at the membrane to a closed boundary at $z=h$ (cf.\ Fig.~\ref{fig:2cyt-bbc}).
As throughout the paper, we set $D_m=0$, so the membrane evolves under reactions alone, with cytosolic densities entering the reaction kinetics through their values at $z=0$.
The membrane dynamics depend only on the cytosolic densities at the membrane [cf.\ Eq.~\eqref{eq:RDS-mem}],
\begin{equation}
    \partial_t \mathbf{m} = \mathbf{f}_m(\mathbf{m},\mathbf{c}|_{z=0}),
\end{equation}
while the cytosolic components fulfil
\begin{equation}
    \partial_t \mathbf{c} = \mathbf{D}_c\boldsymbol{\nabla}^2\mathbf{c} + \mathbf{f}_c(\mathbf{c}),
\end{equation}
where $\boldsymbol{\nabla}^2$ is the three-dimensional Laplacian and $\mathbf{f}_c$ contains only bulk reactions.
Attachment and detachment occur only at the boundary $z=0$ and enter as a reactive boundary condition~\citep{Levine.Rappel2005,Halatek.Frey2012,
Halatek.Frey2018,Frey.Brauns2022},
\begin{equation}\label{eq:reac-boundary-condition}
    -\mathbf{D}_c\partial_z\mathbf{c}|_{z=0} = \mathbf{r}(\mathbf{m},\mathbf{c}|_{z=0}).
\end{equation}
The boundary term $\mathbf{r}$ contains the attachment and detachment fluxes between cytosol and membrane.
Redefining $\mathbf{f} = (\mathbf{r},\mathbf{f}_m)$, mass conservation of species $\alpha$ is ensured by $\mathbf{s}_\alpha\cdot \mathbf{f}=0$ (across membrane and boundary) and $\mathbf{s}^\alpha_c\cdot \mathbf{f}_c=0$ (within the bulk).
 
To account for the different dimensionalities of membrane and bulk, we redefine the total density of species $\alpha$ as its membrane contribution plus the column-integrated cytosolic contribution,
\begin{equation}\label{eq:tot-density-bbc}
    \rho_\alpha \equiv \mathbf{s}_m^\alpha\cdot\mathbf{m}+\int_0^h\mathrm{d}z\, \mathbf{s}_c^\alpha\cdot\mathbf{c}.
\end{equation}
This density satisfies a lateral continuity equation $\partial_t \rho_\alpha = \boldsymbol{\nabla}_{2\mathrm{d}}^2 \int_0^h\mathrm{d}z\, \mathbf{s}_c^\alpha\cdot\mathbf{D}_c\mathbf{c}$, obtained by combining $\mathbf{s}_\alpha\cdot \mathbf{f}=0$, the no-flux condition at $z=h$, and the reactive boundary condition at $z=0$.

\subsection{One cytosolic component per species}
If each species has only a single cytosolic component, the bulk reaction term $\mathbf{f}_c$ is absent and the bulk dynamics are purely diffusive.
The laterally uniform steady state then has constant cytosolic densities.

To analyse the linear stability of this homogeneous steady state, we look for zeros of the dispersion relation. At such a zero, $\partial_t\delta\mathbf{c}=0$ and $\delta\mathbf{c}$ satisfies the Laplace equation
\begin{equation}
    0 = \boldsymbol{\nabla}^2\delta\mathbf{c}.
\end{equation}
The membrane dynamics also vanish, $\mathbf{f}_m = 0$, so the mass-conservation constraint $\mathbf{s}_\alpha\cdot \mathbf{f}=0$ forces $\mathbf{r} = 0$. The reactive boundary condition at $z=0$ then becomes a no-flux condition.
 
For a lateral mode with wavenumber $q$, we apply the separation ansatz $\delta c^\alpha(x,y,z) = X(x,y)Z(z)$ to each scalar cytosolic component independently, The perpendicular component $Z(z)$ satisfies $\partial_z^2 Z = q^2 Z$ with general solution $Z(z) = A\cosh(qz) + B\sinh(qz)$. No-flux at $z=0$ gives $B=0$, and no-flux at $z=h$ then requires $Aq\sinh(qh)=0$, which for $q>0$ forces $A=0$ and hence $\delta\mathbf{c}\equiv 0$. The dispersion relation thus crosses zero only at $q=0$, and systems with one cytosolic component per species exhibit only stationary type-II instabilities, as without bulk-boundary coupling (cf.\ Sec.~\ref{sec:single-cyt-comp}).

Instability is determined in the long-wavelength limit $q\to 0$. There, the LQSS approximation applies: local cytosolic densities are fixed by the reactive equilibria $\mathbf{c}^*(\bm{\rho})$. Because bulk reactions are absent, these densities are constant perpendicular to the membrane at the homogeneous steady state, and long-wavelength perturbations ($2\pi/q\gg h$) remain nearly flat in $z$.
A mass-redistribution instability then destabilises the homogeneous state if the slope matrix $\partial_{\bm{\rho}}\mathbf{c}^*(\bm{\rho})$ has a negative eigenvalue, with $\bm{\rho}$ now the column-integrated total density of Eq.~\eqref{eq:tot-density-bbc}.
This criterion recovers the result derived for systems without bulk-boundary coupling.

\subsection{Two cytosolic components for some or all species}

To extend the results to a series of cytosolic components in the presence of bulk-boundary coupling, we consider a species $\alpha$ with two cytosolic components and a linear conversion reaction. Its cytosolic dynamics read
\begin{subequations}\label{eq:2c-bbc}
\begin{align}
    \partial_t c_1^\alpha &= D_{c_1^\alpha}\boldsymbol{\nabla}^2 c_1^\alpha + \lambda_\alpha s_{c,2}^\alpha c_2^\alpha,\\
    \partial_t c_2^\alpha &= D_{c_2^\alpha}\boldsymbol{\nabla}^2 c_2^\alpha - \lambda_\alpha c_2^\alpha.\label{eq:2c-bbc-inactive}
\end{align}
\end{subequations}
We again normalize the stoichiometric coefficients by setting ${s_{c,1}^\alpha=1}$ 
The membrane components $\mathbf{m}$ follow the dynamics
\begin{equation}\label{eq:2c-bbc-membrane-dyn}
    \partial_t \mathbf{m} = \mathbf{f}_m(\mathbf{m},\mathbf{c}_1|_{z=0}),
\end{equation}
where we defined the vector of active cytosolic components $\mathbf{c}_1=(c_1^\mathrm{A},c_1^\mathrm{B},\dots)$.
 
Proteins detach from the membrane into the inactive cytosolic state, diffuse into the bulk, convert into the active state, and reattach. The inactive density therefore decreases with distance $z$ from the membrane and the active density increases, even in the laterally homogeneous steady state (see Fig.~\ref{fig:2cyt-bbc})~\citep{Halatek.Frey2012,Halatek.etal2018,Denk.etal2018,Frey.Brauns2022}.

For a linear conversion reaction, these steady-state profiles can be calculated explicitly.
The equation for the stationary profile of the inactive component $c_2^{\alpha*}$ reads [see Eq.~\eqref{eq:2c-bbc-inactive}]
\begin{equation}
    0 = (D_{c_2^\alpha} \partial_z^2 - \lambda_\alpha) c_2^{\alpha*}(z),
\end{equation}
which is solved, considering the boundary condition at $z=h$, by
\begin{equation}\label{eq:cosh-profile}
    c_2^{\alpha*}(z) = c_2^{\alpha*}(0) \frac{\cosh\left((h-z)/\ell_\alpha\right)}{\cosh\left(h/\ell_\alpha\right)},
\end{equation}
with the diffusive length scale $(\ell_\alpha)^2=D_{c_2^\alpha}/\lambda_\alpha$ of the bulk gradient.
The profile of the active component can be obtained from the mass-redistribution potential $\eta_\alpha = c_1^\alpha+(D_{c_2^\alpha}/D_{c_1^\alpha}) s_{c,2}^\alpha c_2^\alpha$. Summing the two equations \eqref{eq:2c-bbc} gives the closed diffusion equation $\partial_t \eta_\alpha = D_{c_1^\alpha}\boldsymbol{\nabla}^2\eta_\alpha$, so at the laterally homogeneous steady state $\partial_z^2 \eta_\alpha = 0$. With the no-flux condition at $z=h$, $\eta_\alpha$ is then independent of $z$.

Stationary lateral instability is now determined by the zeros of the dispersion relation in the presence of bulk gradients in both the homogeneous state and the perturbation modes. We use the separation ansatz $\delta c_i^\alpha(x,y,z,t) = e^{\sigma t} X(x,y) Z_i(z)$, with $X(x,y)$ a Fourier mode of wavenumber $q$. At a zero of the dispersion relation, $\sigma=0$, and the perpendicular profile $Z_2(z)$ satisfies [see Eq.~\eqref{eq:2c-bbc-inactive}]
\begin{equation}
    0 = (D_{c_2^\alpha} \partial_z^2 - \lambda_\alpha - D_{c_2^\alpha} q^2) Z_2(z),
\end{equation}
which is again a $\cosh$-profile of the form Eq.~\eqref{eq:cosh-profile} but with $\ell_\alpha$ replaced by $\ell_\alpha(q)^2 \equiv D_{c_2^\alpha}/(\lambda_\alpha+D_{c_2^\alpha} q^2)$.

The amplitude of $Z_2$ is fixed by the reactive boundary condition Eq.~\eqref{eq:reac-boundary-condition}, which for a species with two cytosolic components reads
\begin{subequations}\label{eq:boundary-flux-bbc}
\begin{align}
    - D_{c^\alpha_2} \partial_z c^\alpha_2|_{z=0} &= d_\alpha(\mathbf{m}),\label{eq:inactive-species-boundary-flux-bbc}\\
    - D_{c,1}^\alpha \partial_z c^\alpha_1|_{z=0} &=
    - a_\alpha(\mathbf{m}) c^\alpha_1|_{z=0}.\label{eq:active-species-boundary-flux-bbc}
\end{align}
\end{subequations}
The first equation describes detachment into the inactive cytosolic state, the second the attachment of the active component onto the membrane.
Inserting the separation ansatz with the $\cosh$-profile into the first equation yields
\begin{equation}
    \frac{D_{c_2^\alpha}}{\ell_\alpha(q)}X(x,y)Z_2(0)\tanh(h/\ell_\alpha(q))=\delta d_\alpha,
\end{equation}
where $\delta d_\alpha = (\partial_{\mathbf{m}}d_\alpha)\cdot\delta\mathbf{m}$ is the linearised detachment term.
The perturbation of the inactive cytosolic density is therefore
\begin{equation}\label{eq:deltaC-bbc-profile}
    \delta c_2^\alpha =  \frac{\delta d_\alpha}{D_{c_2^\alpha}Z(\lambda_\alpha/D_{c_2^\alpha},q)} \frac{\cosh\left((h-z)/\ell_\alpha(q)\right)}{\cosh\left(h/\ell_\alpha(q)\right)}.
\end{equation}
with the bulk--boundary geometric factor
\begin{equation}
    Z\left(\frac{\lambda}{D},q\right) = \sqrt{\frac{\lambda}{D}+q^2}\tanh\left(h\sqrt{\frac{\lambda}{D}+q^2}\right),
\end{equation}

As without bulk-boundary coupling, the perturbation $\delta d_\alpha$ can be expressed in terms of variations along the reactive equilibrium manifold.
At the zero of the dispersion relation, the membrane dynamics also vanish.
Since the membrane equation \eqref{eq:2c-bbc-membrane-dyn} has the same structure as without bulk--boundary coupling, the variations of $\mathbf{m}$ and $\mathbf{c}_1|_{z=0}$ can again be parameterised by variations $\widetilde{\delta\bm\rho}$ along the nullcline surfaces [cf.\ Eq.~\eqref{eq:2c-reparameterization}]:
\begin{subequations}
\begin{align}
    \delta\mathbf{m} &= ( \partial_{\bm{\rho}}\mathbf{m}^*)\widetilde{\delta\bm\rho},\\
    \delta\mathbf{c}_1|_{z=0} &= ( \partial_{\bm{\rho}}\mathbf{c}_1^*(0))\widetilde{\delta\bm\rho}.\label{eq:2c-bbc-active-species}
\end{align}
\end{subequations}
The stationary equation $0=\mathbf{f}_m$ parameterises the membrane densities in terms of $\mathbf{c}_1|_{z=0}$, and we reparameterize these solutions in terms of $\widetilde{\delta\bm\rho}$. This is well defined for small variations around the homogeneous steady state whenever the square matrix $\partial_{\bm{\rho}}\mathbf{c}_1^*|_{z=0}$ is invertible, which holds up to codimension-one subsets of the nullcline surfaces.

Because $d_\alpha$ depends only on the membrane components, the parameterisation $\delta\mathbf{m}(\widetilde{\delta\bm\rho})$ gives $\delta d_\alpha = (\partial_{\bm{\rho}}d^*_\alpha)\cdot\widetilde{\delta\bm\rho}$.

As without bulk-boundary coupling, the variations $\delta d_\alpha$ can be translated into variations of the stationary cytosolic concentrations [cf.\ Eq.~\eqref{eq:suppression-inactive-cytosolic}].
In the spatially homogeneous steady state, the inactive cytosolic species satisfies [see Eq.~\eqref{eq:2c-bbc-inactive}]
\begin{equation}
     d_\alpha^* = c_2^{\alpha*}(0) D_{c_2^\alpha}Z(\lambda_\alpha/D_{c_2^\alpha},0).
\end{equation}
It follows that
\begin{equation}
    \delta d_\alpha = D_{c_2^\alpha}Z(\lambda_\alpha/D_{c_2^\alpha},0) \, \partial_{\bm{\rho}}c_2^{\alpha*}(0) \cdot \widetilde{\delta\bm\rho}.
\end{equation}

Inserting this into Eq.~\eqref{eq:deltaC-bbc-profile} gives 
\begin{equation}
\label{eq:2c-bbc-inactive-suppression}
    \delta c_2^\alpha|_{z=0} = \frac{Z(\lambda_\alpha/D_{c_2^\alpha},0)}{Z(\lambda_\alpha/D_{c_2^\alpha},q)} \, \partial_{\bm{\rho}}c_2^{\alpha*}(0) \cdot \widetilde{\delta\bm\rho}.
\end{equation}
The prefactor ${Z(\lambda_\alpha/D_{c_2^\alpha},0)/Z(\lambda_\alpha/D_{c_2^\alpha},q)<1}$ suppresses variations in the inactive species at large wavenumbers $q$, and Eq.~\eqref{eq:2c-bbc-inactive-suppression} is the bulk--boundary coupling analogue of Eq.~\eqref{eq:suppression-inactive-cytosolic}.
 
To obtain the condition for a zero of the dispersion relation from mass redistribution, the perpendicular profile of the mass-redistribution potential ${\delta\eta_\alpha(x,y,z) = X(x,y) Z_\eta(z)}$ is needed.
Summing Eqs.~\eqref{eq:2c-bbc} at the zero of the dispersion relation yields $\boldsymbol{\nabla}^2 \delta \eta_\alpha = 0$, so the perpendicular profile satisfies $(\partial_z^2 - q^2)Z_\eta = 0$.
With the no-flux condition at $z = h$, this gives
\begin{equation}\label{eq:2c-bcc-eta-profile}
    \delta\eta = \delta\eta|_{z=0} \frac{\cosh((h-z) q)}{\cosh(h q)}.
\end{equation}
 
The lateral continuity equation for $\rho_\alpha$ derived in the setup [following Eq.~\eqref{eq:tot-density-bbc}] reads, in the present case,
\begin{equation}
    \partial_t\rho_\alpha =D_{c_1^\alpha} \boldsymbol{\nabla}_{2\mathrm{d}}^2\int_0^h\mathrm{d}z\, \eta_\alpha.
\end{equation}
At linear order with lateral wavenumber $q$,
\begin{equation}
    \partial_t\delta\rho_\alpha = - D_{c_1^\alpha} q^2 \int_0^h\mathrm{d}z\, \delta\eta_\alpha.
\end{equation}
Inserting the profile Eq.~\eqref{eq:2c-bcc-eta-profile} gives $\int_0^h\mathrm{d}z\, \delta\eta_\alpha = \delta\eta|_{z=0}\tanh(hq)/q$, so at a zero of the dispersion relation
\begin{equation}
    0 = - D_{c_1^\alpha} Z(0,q_\mathrm{min}) \delta\eta|_{z=0},
\end{equation}
using $Z(0,q) = q\tanh(hq)$.

Inserting the expressions for the density variations of the active and inactive components, Eqs.~\eqref{eq:2c-bbc-active-species} and \eqref{eq:2c-bbc-inactive-suppression}, into $\delta\eta|_{z=0}$ then yields
\begin{align}
    0 &= - D_{c_1^\alpha} Z(0,q_\mathrm{min}) \\
    &\quad \left[\partial_{\bm{\rho}}c_1^{\alpha*}(0) + \frac{Z(\lambda_\alpha/D_{c_2^\alpha},0)}{Z(\lambda_\alpha/D_{c_2^\alpha},q)}\frac{D_{c_2^\alpha}}{D_{c_1^\alpha}} s_{c,2}^\alpha \partial_{\bm{\rho}}c_2^{\alpha*}(0)\right]\cdot\delta\bm{\rho}. \nonumber
\end{align}
For a nontrivial solution at ${q_\mathrm{min}>0}$, the bracket must vanish, which after using ${\partial_{\bm{\rho}}\eta^*_\alpha = \partial_{\bm{\rho}}c_1^{\alpha*} + (D_{c_2^\alpha}/D_{c_1^\alpha})s_{c,2}^\alpha\,\partial_{\bm{\rho}}c_2^{\alpha*}}$ rearranges to
\begin{equation}
    0 =\left[\left(\frac{Z(\lambda_\alpha/D_{c_2^\alpha},q)}{Z(\lambda_\alpha/D_{c_2^\alpha},0)}-1\right)\partial_{\bm{\rho}}c_1^{\alpha*}(0) + \partial_{\bm{\rho}}\eta^*_\alpha(0)\right]\cdot\delta\bm{\rho}.
\end{equation}
Collecting these conditions across all species, a stationary type-I instability occurs whenever
\begin{equation}\label{eq:2c-bbc-det-condition}
    0 = \det\left(\mathbf{Q}_\mathrm{bbc}(q_\mathrm{min})\partial_{\bm{\rho}}\mathbf{c}_1^*(0)+ \partial_{\bm{\rho}}\bm{\eta}^*(0)\right)
\end{equation}
admits a solution ${q_\mathrm{min}>0}$, provided $\partial_{\bm{\rho}}\bm{\eta}^*(0)$ has only eigenvalues with a positive real part. The diagonal matrix $\mathbf{Q}_\mathrm{bbc}$ is defined by
\begin{align}
    \mathbf{Q}_\mathrm{bbc}(q) = \mathrm{diag}\Bigg(&\frac{Z(\lambda_\mathrm{A}/D_{c_2^\mathrm{A}},q)}{Z(\lambda_\mathrm{A}/D_{c_2^\mathrm{A}},0)}-1,\nonumber\\
    &\frac{Z(\lambda_\mathrm{B}/D_{c_2^\mathrm{B}},q)}{Z(\lambda_\mathrm{B}/D_{c_2^\mathrm{B}},0)}-1,\dots\Bigg).
\end{align}
For species with only one cytosolic component the corresponding entry is defined to vanish, consistent with taking the limit $\lambda_\alpha/D_{c_2^\alpha}\to\infty$. For species with two cytosolic components the diagonal entries are positive and vanish as ${q\to 0}$. Equation~\eqref{eq:2c-bbc-det-condition} generalises Eq.~\eqref{eq:type-I-instab-condition} to the case with bulk--boundary coupling. In the limit $h\to 0$ where the perpendicular extent of the bulk vanishes,
\begin{equation}
    \mathbf{Q}_\mathrm{bbc}(q)\to \mathbf{D}_2\bm{\lambda}^{-1} q^2,
\end{equation}
and Eq.~\eqref{eq:2c-bbc-det-condition} reduces to Eq.~\eqref{eq:type-I-instab-condition}.


\begin{thebibliography}{95}%
\makeatletter
\providecommand \@ifxundefined [1]{%
 \@ifx{#1\undefined}
}%
\providecommand \@ifnum [1]{%
 \ifnum #1\expandafter \@firstoftwo
 \else \expandafter \@secondoftwo
 \fi
}%
\providecommand \@ifx [1]{%
 \ifx #1\expandafter \@firstoftwo
 \else \expandafter \@secondoftwo
 \fi
}%
\providecommand \natexlab [1]{#1}%
\providecommand \enquote  [1]{``#1''}%
\providecommand \bibnamefont  [1]{#1}%
\providecommand \bibfnamefont [1]{#1}%
\providecommand \citenamefont [1]{#1}%
\providecommand \href@noop [0]{\@secondoftwo}%
\providecommand \href [0]{\begingroup \@sanitize@url \@href}%
\providecommand \@href[1]{\@@startlink{#1}\@@href}%
\providecommand \@@href[1]{\endgroup#1\@@endlink}%
\providecommand \@sanitize@url [0]{\catcode `\\12\catcode `\$12\catcode `\&12\catcode `\#12\catcode `\^12\catcode `\_12\catcode `\%12\relax}%
\providecommand \@@startlink[1]{}%
\providecommand \@@endlink[0]{}%
\providecommand \url  [0]{\begingroup\@sanitize@url \@url }%
\providecommand \@url [1]{\endgroup\@href {#1}{\urlprefix }}%
\providecommand \urlprefix  [0]{URL }%
\providecommand \Eprint [0]{\href }%
\providecommand \doibase [0]{https://doi.org/}%
\providecommand \selectlanguage [0]{\@gobble}%
\providecommand \bibinfo  [0]{\@secondoftwo}%
\providecommand \bibfield  [0]{\@secondoftwo}%
\providecommand \translation [1]{[#1]}%
\providecommand \BibitemOpen [0]{}%
\providecommand \bibitemStop [0]{}%
\providecommand \bibitemNoStop [0]{.\EOS\space}%
\providecommand \EOS [0]{\spacefactor3000\relax}%
\providecommand \BibitemShut  [1]{\csname bibitem#1\endcsname}%
\let\auto@bib@innerbib\@empty
\bibitem [{\citenamefont {Litschel}\ \emph {et~al.}(2018)\citenamefont {Litschel}, \citenamefont {Ramm}, \citenamefont {Maas}, \citenamefont {Heymann},\ and\ \citenamefont {Schwille}}]{Litschel.etal2018}%
  \BibitemOpen
  \bibfield  {author} {\bibinfo {author} {\bibfnamefont {T.}~\bibnamefont {Litschel}}, \bibinfo {author} {\bibfnamefont {B.}~\bibnamefont {Ramm}}, \bibinfo {author} {\bibfnamefont {R.}~\bibnamefont {Maas}}, \bibinfo {author} {\bibfnamefont {M.}~\bibnamefont {Heymann}},\ and\ \bibinfo {author} {\bibfnamefont {P.}~\bibnamefont {Schwille}},\ }\bibfield  {title} {\bibinfo {title} {Beating {{Vesicles}}: {{Encapsulated Protein Oscillations Cause Dynamic Membrane Deformations}}},\ }\href {https://doi.org/10.1002/anie.201808750} {\bibfield  {journal} {\bibinfo  {journal} {Angewandte Chemie International Edition}\ }\textbf {\bibinfo {volume} {57}},\ \bibinfo {pages} {16286} (\bibinfo {year} {2018})}\BibitemShut {NoStop}%
\bibitem [{\citenamefont {{Reverte-L{\'o}pez}}\ \emph {et~al.}(2024)\citenamefont {{Reverte-L{\'o}pez}}, \citenamefont {Kanwa}, \citenamefont {Qutbuddin}, \citenamefont {Belousova}, \citenamefont {Jasnin},\ and\ \citenamefont {Schwille}}]{Reverte-Lopez.etal2024}%
  \BibitemOpen
  \bibfield  {author} {\bibinfo {author} {\bibfnamefont {M.}~\bibnamefont {{Reverte-L{\'o}pez}}}, \bibinfo {author} {\bibfnamefont {N.}~\bibnamefont {Kanwa}}, \bibinfo {author} {\bibfnamefont {Y.}~\bibnamefont {Qutbuddin}}, \bibinfo {author} {\bibfnamefont {V.}~\bibnamefont {Belousova}}, \bibinfo {author} {\bibfnamefont {M.}~\bibnamefont {Jasnin}},\ and\ \bibinfo {author} {\bibfnamefont {P.}~\bibnamefont {Schwille}},\ }\bibfield  {title} {\bibinfo {title} {Self-organized spatial targeting of contractile actomyosin rings for synthetic cell division},\ }\href {https://doi.org/10.1038/s41467-024-54807-9} {\bibfield  {journal} {\bibinfo  {journal} {Nature Communications}\ }\textbf {\bibinfo {volume} {15}},\ \bibinfo {pages} {10415} (\bibinfo {year} {2024})}\BibitemShut {NoStop}%
\bibitem [{\citenamefont {Fu}\ \emph {et~al.}(2023)\citenamefont {Fu}, \citenamefont {Burkart}, \citenamefont {Maryshev}, \citenamefont {Franquelim}, \citenamefont {{Merino-Salom{\'o}n}}, \citenamefont {{Reverte-L{\'o}pez}}, \citenamefont {Frey},\ and\ \citenamefont {Schwille}}]{Fu.etal2023}%
  \BibitemOpen
  \bibfield  {author} {\bibinfo {author} {\bibfnamefont {M.}~\bibnamefont {Fu}}, \bibinfo {author} {\bibfnamefont {T.}~\bibnamefont {Burkart}}, \bibinfo {author} {\bibfnamefont {I.}~\bibnamefont {Maryshev}}, \bibinfo {author} {\bibfnamefont {H.~G.}\ \bibnamefont {Franquelim}}, \bibinfo {author} {\bibfnamefont {A.}~\bibnamefont {{Merino-Salom{\'o}n}}}, \bibinfo {author} {\bibfnamefont {M.}~\bibnamefont {{Reverte-L{\'o}pez}}}, \bibinfo {author} {\bibfnamefont {E.}~\bibnamefont {Frey}},\ and\ \bibinfo {author} {\bibfnamefont {P.}~\bibnamefont {Schwille}},\ }\bibfield  {title} {\bibinfo {title} {Mechanochemical feedback loop drives persistent motion of liposomes},\ }\bibfield  {journal} {\bibinfo  {journal} {Nature Physics}\ }\href {https://doi.org/10.1038/s41567-023-02058-8} {10.1038/s41567-023-02058-8} (\bibinfo {year} {2023})\BibitemShut {NoStop}%
\bibitem [{\citenamefont {Chaikin}\ and\ \citenamefont {Lubensky}(1995)}]{Chaikin.Lubensky1995}%
  \BibitemOpen
  \bibfield  {author} {\bibinfo {author} {\bibfnamefont {P.~M.}\ \bibnamefont {Chaikin}}\ and\ \bibinfo {author} {\bibfnamefont {T.~C.}\ \bibnamefont {Lubensky}},\ }\href {https://doi.org/10.1017/CBO9780511813467} {\emph {\bibinfo {title} {Principles of {{Condensed Matter Physics}}}}},\ \bibinfo {edition} {1st}\ ed.\ (\bibinfo  {publisher} {Cambridge University Press},\ \bibinfo {address} {Cambridge, UK},\ \bibinfo {year} {1995})\BibitemShut {NoStop}%
\bibitem [{\citenamefont {Mao}\ \emph {et~al.}(2019)\citenamefont {Mao}, \citenamefont {Kuldinow}, \citenamefont {Haataja},\ and\ \citenamefont {Ko{\v s}mrlj}}]{Mao.etal2019}%
  \BibitemOpen
  \bibfield  {author} {\bibinfo {author} {\bibfnamefont {S.}~\bibnamefont {Mao}}, \bibinfo {author} {\bibfnamefont {D.}~\bibnamefont {Kuldinow}}, \bibinfo {author} {\bibfnamefont {M.~P.}\ \bibnamefont {Haataja}},\ and\ \bibinfo {author} {\bibfnamefont {A.}~\bibnamefont {Ko{\v s}mrlj}},\ }\bibfield  {title} {\bibinfo {title} {Phase behavior and morphology of multicomponent liquid mixtures},\ }\href {https://doi.org/10.1039/C8SM02045K} {\bibfield  {journal} {\bibinfo  {journal} {Soft Matter}\ }\textbf {\bibinfo {volume} {15}},\ \bibinfo {pages} {1297} (\bibinfo {year} {2019})}\BibitemShut {NoStop}%
\bibitem [{\citenamefont {Satnoianu}\ \emph {et~al.}(2000)\citenamefont {Satnoianu}, \citenamefont {Menzinger},\ and\ \citenamefont {Maini}}]{Satnoianu.etal2000}%
  \BibitemOpen
  \bibfield  {author} {\bibinfo {author} {\bibfnamefont {R.~A.}\ \bibnamefont {Satnoianu}}, \bibinfo {author} {\bibfnamefont {M.}~\bibnamefont {Menzinger}},\ and\ \bibinfo {author} {\bibfnamefont {P.~K.}\ \bibnamefont {Maini}},\ }\bibfield  {title} {\bibinfo {title} {Turing instabilities in general systems},\ }\href {https://doi.org/10.1007/s002850000056} {\bibfield  {journal} {\bibinfo  {journal} {Journal of Mathematical Biology}\ }\textbf {\bibinfo {volume} {41}},\ \bibinfo {pages} {493} (\bibinfo {year} {2000})}\BibitemShut {NoStop}%
\bibitem [{\citenamefont {{Villar-Sep{\'u}lveda}}\ and\ \citenamefont {Champneys}(2023)}]{Villar-Sepulveda.Champneys2023}%
  \BibitemOpen
  \bibfield  {author} {\bibinfo {author} {\bibfnamefont {E.}~\bibnamefont {{Villar-Sep{\'u}lveda}}}\ and\ \bibinfo {author} {\bibfnamefont {A.~R.}\ \bibnamefont {Champneys}},\ }\bibfield  {title} {\bibinfo {title} {General conditions for {{Turing}} and wave instabilities in reaction -diffusion systems},\ }\href {https://doi.org/10.1007/s00285-023-01870-3} {\bibfield  {journal} {\bibinfo  {journal} {Journal of Mathematical Biology}\ }\textbf {\bibinfo {volume} {86}},\ \bibinfo {pages} {39} (\bibinfo {year} {2023})}\BibitemShut {NoStop}%
\bibitem [{\citenamefont {Hambric}\ \emph {et~al.}(2022)\citenamefont {Hambric}, \citenamefont {Li}, \citenamefont {Pelejo},\ and\ \citenamefont {Shi}}]{Hambric.etal2022}%
  \BibitemOpen
  \bibfield  {author} {\bibinfo {author} {\bibfnamefont {C.~L.}\ \bibnamefont {Hambric}}, \bibinfo {author} {\bibfnamefont {C.-K.}\ \bibnamefont {Li}}, \bibinfo {author} {\bibfnamefont {D.~C.}\ \bibnamefont {Pelejo}},\ and\ \bibinfo {author} {\bibfnamefont {J.}~\bibnamefont {Shi}},\ }\bibfield  {title} {\bibinfo {title} {Minimum number of non-zero-entries in a stable matrix exhibiting {{Turing}} instability},\ }\href {https://doi.org/10.3934/dcdss.2021128} {\bibfield  {journal} {\bibinfo  {journal} {Discrete and Continuous Dynamical Systems - S}\ }\textbf {\bibinfo {volume} {15}},\ \bibinfo {pages} {2497} (\bibinfo {year} {2022})}\BibitemShut {NoStop}%
\bibitem [{\citenamefont {Mincheva}\ and\ \citenamefont {Roussel}(2006)}]{Mincheva.Roussel2006}%
  \BibitemOpen
  \bibfield  {author} {\bibinfo {author} {\bibfnamefont {M.}~\bibnamefont {Mincheva}}\ and\ \bibinfo {author} {\bibfnamefont {M.~R.}\ \bibnamefont {Roussel}},\ }\bibfield  {title} {\bibinfo {title} {A graph-theoretic method for detecting potential {{Turing}} bifurcations},\ }\href {https://doi.org/10.1063/1.2397073} {\bibfield  {journal} {\bibinfo  {journal} {The Journal of Chemical Physics}\ }\textbf {\bibinfo {volume} {125}},\ \bibinfo {pages} {204102} (\bibinfo {year} {2006})}\BibitemShut {NoStop}%
\bibitem [{\citenamefont {Diego}\ \emph {et~al.}(2018)\citenamefont {Diego}, \citenamefont {Marcon}, \citenamefont {M{\"u}ller},\ and\ \citenamefont {Sharpe}}]{Diego.etal2018}%
  \BibitemOpen
  \bibfield  {author} {\bibinfo {author} {\bibfnamefont {X.}~\bibnamefont {Diego}}, \bibinfo {author} {\bibfnamefont {L.}~\bibnamefont {Marcon}}, \bibinfo {author} {\bibfnamefont {P.}~\bibnamefont {M{\"u}ller}},\ and\ \bibinfo {author} {\bibfnamefont {J.}~\bibnamefont {Sharpe}},\ }\bibfield  {title} {\bibinfo {title} {Key {{Features}} of {{Turing Systems}} are {{Determined Purely}} by {{Network Topology}}},\ }\href {https://doi.org/10.1103/PhysRevX.8.021071} {\bibfield  {journal} {\bibinfo  {journal} {Physical Review X}\ }\textbf {\bibinfo {volume} {8}},\ \bibinfo {pages} {021071} (\bibinfo {year} {2018})}\BibitemShut {NoStop}%
\bibitem [{\citenamefont {{Villar-Sep{\'u}lveda}}\ \emph {et~al.}(2025)\citenamefont {{Villar-Sep{\'u}lveda}}, \citenamefont {Champneys},\ and\ \citenamefont {Krause}}]{Villar-Sepulveda.etal2025}%
  \BibitemOpen
  \bibfield  {author} {\bibinfo {author} {\bibfnamefont {E.}~\bibnamefont {{Villar-Sep{\'u}lveda}}}, \bibinfo {author} {\bibfnamefont {A.~R.}\ \bibnamefont {Champneys}},\ and\ \bibinfo {author} {\bibfnamefont {A.~L.}\ \bibnamefont {Krause}},\ }\bibfield  {title} {\bibinfo {title} {Designing reaction-cross-diffusion systems with {{Turing}} and wave instabilities},\ }\href {https://doi.org/10.1007/s00285-025-02274-1} {\bibfield  {journal} {\bibinfo  {journal} {Journal of Mathematical Biology}\ }\textbf {\bibinfo {volume} {91}},\ \bibinfo {pages} {37} (\bibinfo {year} {2025})}\BibitemShut {NoStop}%
\bibitem [{\citenamefont {Gierer}\ and\ \citenamefont {Meinhardt}(1972)}]{Gierer.Meinhardt1972}%
  \BibitemOpen
  \bibfield  {author} {\bibinfo {author} {\bibfnamefont {A.}~\bibnamefont {Gierer}}\ and\ \bibinfo {author} {\bibfnamefont {H.}~\bibnamefont {Meinhardt}},\ }\bibfield  {title} {\bibinfo {title} {A theory of biological pattern formation},\ }\href {https://doi.org/10.1007/BF00289234} {\bibfield  {journal} {\bibinfo  {journal} {Kybernetik}\ }\textbf {\bibinfo {volume} {12}},\ \bibinfo {pages} {30} (\bibinfo {year} {1972})}\BibitemShut {NoStop}%
\bibitem [{\citenamefont {Murray}(2003)}]{Murray2003}%
  \BibitemOpen
  \bibinfo {editor} {\bibfnamefont {J.~D.}\ \bibnamefont {Murray}},\ ed.,\ \href {https://doi.org/10.1007/b98869} {\emph {\bibinfo {title} {Mathematical {{Biology}}: {{II}}: {{Spatial Models}} and {{Biomedical Applications}}}}},\ \bibinfo {series} {Interdisciplinary {{Applied Mathematics}}}, Vol.~\bibinfo {volume} {18}\ (\bibinfo  {publisher} {Springer New York},\ \bibinfo {address} {New York, NY},\ \bibinfo {year} {2003})\BibitemShut {NoStop}%
\bibitem [{\citenamefont {Landge}\ \emph {et~al.}(2020)\citenamefont {Landge}, \citenamefont {Jordan}, \citenamefont {Diego},\ and\ \citenamefont {M{\"u}ller}}]{Landge.etal2020}%
  \BibitemOpen
  \bibfield  {author} {\bibinfo {author} {\bibfnamefont {A.~N.}\ \bibnamefont {Landge}}, \bibinfo {author} {\bibfnamefont {B.~M.}\ \bibnamefont {Jordan}}, \bibinfo {author} {\bibfnamefont {X.}~\bibnamefont {Diego}},\ and\ \bibinfo {author} {\bibfnamefont {P.}~\bibnamefont {M{\"u}ller}},\ }\bibfield  {title} {\bibinfo {title} {Pattern formation mechanisms of self-organizing reaction-diffusion systems},\ }\href {https://doi.org/10.1016/j.ydbio.2019.10.031} {\bibfield  {journal} {\bibinfo  {journal} {Developmental Biology}\ }\textbf {\bibinfo {volume} {460}},\ \bibinfo {pages} {2} (\bibinfo {year} {2020})}\BibitemShut {NoStop}%
\bibitem [{\citenamefont {Smith}\ and\ \citenamefont {Dalchau}(2018)}]{Smith.Dalchau2018}%
  \BibitemOpen
  \bibfield  {author} {\bibinfo {author} {\bibfnamefont {S.}~\bibnamefont {Smith}}\ and\ \bibinfo {author} {\bibfnamefont {N.}~\bibnamefont {Dalchau}},\ }\bibfield  {title} {\bibinfo {title} {Model reduction enables {{Turing}} instability analysis of large reaction--diffusion models},\ }\href {https://doi.org/10.1098/rsif.2017.0805} {\bibfield  {journal} {\bibinfo  {journal} {Journal of The Royal Society Interface}\ }\textbf {\bibinfo {volume} {15}},\ \bibinfo {pages} {20170805} (\bibinfo {year} {2018})}\BibitemShut {NoStop}%
\bibitem [{\citenamefont {Marcon}\ \emph {et~al.}(2016)\citenamefont {Marcon}, \citenamefont {Diego}, \citenamefont {Sharpe},\ and\ \citenamefont {M{\"u}ller}}]{Marcon.etal2016}%
  \BibitemOpen
  \bibfield  {author} {\bibinfo {author} {\bibfnamefont {L.}~\bibnamefont {Marcon}}, \bibinfo {author} {\bibfnamefont {X.}~\bibnamefont {Diego}}, \bibinfo {author} {\bibfnamefont {J.}~\bibnamefont {Sharpe}},\ and\ \bibinfo {author} {\bibfnamefont {P.}~\bibnamefont {M{\"u}ller}},\ }\bibfield  {title} {\bibinfo {title} {High-throughput mathematical analysis identifies {{Turing}} networks for patterning with equally diffusing signals},\ }\href {https://doi.org/10.7554/eLife.14022} {\bibfield  {journal} {\bibinfo  {journal} {eLife}\ }\textbf {\bibinfo {volume} {5}},\ \bibinfo {pages} {e14022} (\bibinfo {year} {2016})}\BibitemShut {NoStop}%
\bibitem [{\citenamefont {Zheng}\ \emph {et~al.}(2016)\citenamefont {Zheng}, \citenamefont {Shao},\ and\ \citenamefont {Ouyang}}]{Zheng.etal2016}%
  \BibitemOpen
  \bibfield  {author} {\bibinfo {author} {\bibfnamefont {M.~M.}\ \bibnamefont {Zheng}}, \bibinfo {author} {\bibfnamefont {B.}~\bibnamefont {Shao}},\ and\ \bibinfo {author} {\bibfnamefont {Q.}~\bibnamefont {Ouyang}},\ }\bibfield  {title} {\bibinfo {title} {Identifying network topologies that can generate turing pattern},\ }\href {https://doi.org/10.1016/j.jtbi.2016.08.005} {\bibfield  {journal} {\bibinfo  {journal} {Journal of Theoretical Biology}\ }\textbf {\bibinfo {volume} {408}},\ \bibinfo {pages} {88} (\bibinfo {year} {2016})}\BibitemShut {NoStop}%
\bibitem [{\citenamefont {Scholes}\ \emph {et~al.}(2019)\citenamefont {Scholes}, \citenamefont {Schnoerr}, \citenamefont {Isalan},\ and\ \citenamefont {Stumpf}}]{Scholes.etal2019}%
  \BibitemOpen
  \bibfield  {author} {\bibinfo {author} {\bibfnamefont {N.~S.}\ \bibnamefont {Scholes}}, \bibinfo {author} {\bibfnamefont {D.}~\bibnamefont {Schnoerr}}, \bibinfo {author} {\bibfnamefont {M.}~\bibnamefont {Isalan}},\ and\ \bibinfo {author} {\bibfnamefont {M.~P.}\ \bibnamefont {Stumpf}},\ }\bibfield  {title} {\bibinfo {title} {A {{Comprehensive Network Atlas Reveals That Turing Patterns Are Common}} but {{Not Robust}}},\ }\href {https://doi.org/10.1016/j.cels.2019.07.007} {\bibfield  {journal} {\bibinfo  {journal} {Cell Systems}\ }\textbf {\bibinfo {volume} {9}},\ \bibinfo {pages} {243} (\bibinfo {year} {2019})}\BibitemShut {NoStop}%
\bibitem [{\citenamefont {Haas}\ and\ \citenamefont {Goldstein}(2021)}]{Haas.Goldstein2021}%
  \BibitemOpen
  \bibfield  {author} {\bibinfo {author} {\bibfnamefont {P.~A.}\ \bibnamefont {Haas}}\ and\ \bibinfo {author} {\bibfnamefont {R.~E.}\ \bibnamefont {Goldstein}},\ }\bibfield  {title} {\bibinfo {title} {Turing's {{Diffusive Threshold}} in {{Random Reaction-Diffusion Systems}}},\ }\href {https://doi.org/10.1103/PhysRevLett.126.238101} {\bibfield  {journal} {\bibinfo  {journal} {Physical Review Letters}\ }\textbf {\bibinfo {volume} {126}},\ \bibinfo {pages} {238101} (\bibinfo {year} {2021})}\BibitemShut {NoStop}%
\bibitem [{\citenamefont {Solomatina}\ \emph {et~al.}(2022)\citenamefont {Solomatina}, \citenamefont {Cezanne}, \citenamefont {Kalaidzidis}, \citenamefont {Zerial},\ and\ \citenamefont {Sbalzarini}}]{Solomatina.etal2022}%
  \BibitemOpen
  \bibfield  {author} {\bibinfo {author} {\bibfnamefont {A.}~\bibnamefont {Solomatina}}, \bibinfo {author} {\bibfnamefont {A.}~\bibnamefont {Cezanne}}, \bibinfo {author} {\bibfnamefont {Y.}~\bibnamefont {Kalaidzidis}}, \bibinfo {author} {\bibfnamefont {M.}~\bibnamefont {Zerial}},\ and\ \bibinfo {author} {\bibfnamefont {I.~F.}\ \bibnamefont {Sbalzarini}},\ }\bibfield  {title} {\bibinfo {title} {Design centering enables robustness screening of pattern formation models},\ }\href {https://doi.org/10.1093/bioinformatics/btac480} {\bibfield  {journal} {\bibinfo  {journal} {Bioinformatics}\ }\textbf {\bibinfo {volume} {38}},\ \bibinfo {pages} {ii134} (\bibinfo {year} {2022})}\BibitemShut {NoStop}%
\bibitem [{\citenamefont {Jilkine}\ and\ \citenamefont {{Edelstein-Keshet}}(2011)}]{Jilkine.Edelstein-Keshet2011}%
  \BibitemOpen
  \bibfield  {author} {\bibinfo {author} {\bibfnamefont {A.}~\bibnamefont {Jilkine}}\ and\ \bibinfo {author} {\bibfnamefont {L.}~\bibnamefont {{Edelstein-Keshet}}},\ }\bibfield  {title} {\bibinfo {title} {A {{Comparison}} of {{Mathematical Models}} for {{Polarization}} of {{Single Eukaryotic Cells}} in {{Response}} to {{Guided Cues}}},\ }\href {https://doi.org/10.1371/journal.pcbi.1001121} {\bibfield  {journal} {\bibinfo  {journal} {PLoS Computational Biology}\ }\textbf {\bibinfo {volume} {7}},\ \bibinfo {pages} {e1001121} (\bibinfo {year} {2011})}\BibitemShut {NoStop}%
\bibitem [{\citenamefont {Trong}\ \emph {et~al.}(2014)\citenamefont {Trong}, \citenamefont {Nicola}, \citenamefont {Goehring}, \citenamefont {Kumar},\ and\ \citenamefont {Grill}}]{Trong.etal2014}%
  \BibitemOpen
  \bibfield  {author} {\bibinfo {author} {\bibfnamefont {P.~K.}\ \bibnamefont {Trong}}, \bibinfo {author} {\bibfnamefont {E.~M.}\ \bibnamefont {Nicola}}, \bibinfo {author} {\bibfnamefont {N.~W.}\ \bibnamefont {Goehring}}, \bibinfo {author} {\bibfnamefont {K.~V.}\ \bibnamefont {Kumar}},\ and\ \bibinfo {author} {\bibfnamefont {S.~W.}\ \bibnamefont {Grill}},\ }\bibfield  {title} {\bibinfo {title} {Parameter-space topology of models for cell polarity},\ }\href {https://doi.org/10.1088/1367-2630/16/6/065009} {\bibfield  {journal} {\bibinfo  {journal} {New Journal of Physics}\ }\textbf {\bibinfo {volume} {16}},\ \bibinfo {pages} {065009} (\bibinfo {year} {2014})}\BibitemShut {NoStop}%
\bibitem [{\citenamefont {Halatek}\ \emph {et~al.}(2018)\citenamefont {Halatek}, \citenamefont {Brauns},\ and\ \citenamefont {Frey}}]{Halatek.etal2018}%
  \BibitemOpen
  \bibfield  {author} {\bibinfo {author} {\bibfnamefont {J.}~\bibnamefont {Halatek}}, \bibinfo {author} {\bibfnamefont {F.}~\bibnamefont {Brauns}},\ and\ \bibinfo {author} {\bibfnamefont {E.}~\bibnamefont {Frey}},\ }\bibfield  {title} {\bibinfo {title} {Self-organization principles of intracellular pattern formation},\ }\href {https://doi.org/10.1098/rstb.2017.0107} {\bibfield  {journal} {\bibinfo  {journal} {Philosophical Transactions of the Royal Society B: Biological Sciences}\ }\textbf {\bibinfo {volume} {373}},\ \bibinfo {pages} {20170107} (\bibinfo {year} {2018})}\BibitemShut {NoStop}%
\bibitem [{\citenamefont {Frey}\ and\ \citenamefont {Brauns}(2022)}]{Frey.Brauns2022}%
  \BibitemOpen
  \bibfield  {author} {\bibinfo {author} {\bibfnamefont {E.}~\bibnamefont {Frey}}\ and\ \bibinfo {author} {\bibfnamefont {F.}~\bibnamefont {Brauns}},\ }\bibfield  {title} {\bibinfo {title} {Self-organization of {{Protein Patterns}}},\ }in\ \href {https://doi.org/10.1093/oso/9780192858313.003.0011} {\emph {\bibinfo {booktitle} {Active {{Matter}} and {{Nonequilibrium Statistical Physics}}}}},\ \bibinfo {editor} {edited by\ \bibinfo {editor} {\bibfnamefont {J.}~\bibnamefont {Tailleur}}, \bibinfo {editor} {\bibfnamefont {G.}~\bibnamefont {Gompper}}, \bibinfo {editor} {\bibfnamefont {M.~C.}\ \bibnamefont {Marchetti}}, \bibinfo {editor} {\bibfnamefont {J.~M.}\ \bibnamefont {Yeomans}},\ and\ \bibinfo {editor} {\bibfnamefont {C.}~\bibnamefont {Salomon}}}\ (\bibinfo  {publisher} {Oxford University Press},\ \bibinfo {address} {Oxford, UK},\ \bibinfo {year} {2022})\ \bibinfo {edition} {1st}\ ed.,\ pp.\ \bibinfo {pages} {347--445}\BibitemShut {NoStop}%
\bibitem [{\citenamefont {Burkart}\ \emph {et~al.}(2022)\citenamefont {Burkart}, \citenamefont {Wigbers}, \citenamefont {W{\"u}rthner},\ and\ \citenamefont {Frey}}]{Burkart.etal2022a}%
  \BibitemOpen
  \bibfield  {author} {\bibinfo {author} {\bibfnamefont {T.}~\bibnamefont {Burkart}}, \bibinfo {author} {\bibfnamefont {M.~C.}\ \bibnamefont {Wigbers}}, \bibinfo {author} {\bibfnamefont {L.}~\bibnamefont {W{\"u}rthner}},\ and\ \bibinfo {author} {\bibfnamefont {E.}~\bibnamefont {Frey}},\ }\bibfield  {title} {\bibinfo {title} {Control of protein-based pattern formation via guiding cues},\ }\href {https://doi.org/10.1038/s42254-022-00461-3} {\bibfield  {journal} {\bibinfo  {journal} {Nature Reviews Physics}\ }\textbf {\bibinfo {volume} {4}},\ \bibinfo {pages} {511} (\bibinfo {year} {2022})}\BibitemShut {NoStop}%
\bibitem [{\citenamefont {Frey}\ and\ \citenamefont {Weyer}(2026)}]{Frey.Weyer2026}%
  \BibitemOpen
  \bibfield  {author} {\bibinfo {author} {\bibfnamefont {E.}~\bibnamefont {Frey}}\ and\ \bibinfo {author} {\bibfnamefont {H.}~\bibnamefont {Weyer}},\ }\bibfield  {title} {\bibinfo {title} {Pattern {{Formation Beyond Turing}}: {{Physical Principles}} of {{Mass-Conserving Reaction}}--{{Diffusion Systems}}},\ }\href {https://doi.org/10.1146/annurev-biophys-030822-031638} {\bibfield  {journal} {\bibinfo  {journal} {Annual Review of Biophysics}\ }\textbf {\bibinfo {volume} {55}},\ \bibinfo {pages} {493} (\bibinfo {year} {2026})}\BibitemShut {NoStop}%
\bibitem [{\citenamefont {Brauns}\ \emph {et~al.}(2020)\citenamefont {Brauns}, \citenamefont {Halatek},\ and\ \citenamefont {Frey}}]{Brauns.etal2020}%
  \BibitemOpen
  \bibfield  {author} {\bibinfo {author} {\bibfnamefont {F.}~\bibnamefont {Brauns}}, \bibinfo {author} {\bibfnamefont {J.}~\bibnamefont {Halatek}},\ and\ \bibinfo {author} {\bibfnamefont {E.}~\bibnamefont {Frey}},\ }\bibfield  {title} {\bibinfo {title} {Phase-{{Space Geometry}} of {{Mass-Conserving Reaction-Diffusion Dynamics}}},\ }\href {https://doi.org/10.1103/PhysRevX.10.041036} {\bibfield  {journal} {\bibinfo  {journal} {Physical Review X}\ }\textbf {\bibinfo {volume} {10}},\ \bibinfo {pages} {041036} (\bibinfo {year} {2020})}\BibitemShut {NoStop}%
\bibitem [{\citenamefont {Brauns}\ \emph {et~al.}(2021{\natexlab{a}})\citenamefont {Brauns}, \citenamefont {Halatek},\ and\ \citenamefont {Frey}}]{Brauns.etal2021}%
  \BibitemOpen
  \bibfield  {author} {\bibinfo {author} {\bibfnamefont {F.}~\bibnamefont {Brauns}}, \bibinfo {author} {\bibfnamefont {J.}~\bibnamefont {Halatek}},\ and\ \bibinfo {author} {\bibfnamefont {E.}~\bibnamefont {Frey}},\ }\bibfield  {title} {\bibinfo {title} {Diffusive coupling of two well-mixed compartments elucidates elementary principles of protein-based pattern formation},\ }\href {https://doi.org/10.1103/PhysRevResearch.3.013258} {\bibfield  {journal} {\bibinfo  {journal} {Physical Review Research}\ }\textbf {\bibinfo {volume} {3}},\ \bibinfo {pages} {013258} (\bibinfo {year} {2021}{\natexlab{a}})}\BibitemShut {NoStop}%
\bibitem [{\citenamefont {Huang}\ \emph {et~al.}(2003)\citenamefont {Huang}, \citenamefont {Meir},\ and\ \citenamefont {Wingreen}}]{Huang.etal2003}%
  \BibitemOpen
  \bibfield  {author} {\bibinfo {author} {\bibfnamefont {K.~C.}\ \bibnamefont {Huang}}, \bibinfo {author} {\bibfnamefont {Y.}~\bibnamefont {Meir}},\ and\ \bibinfo {author} {\bibfnamefont {N.~S.}\ \bibnamefont {Wingreen}},\ }\bibfield  {title} {\bibinfo {title} {Dynamic structures in {{Escherichia}} coli: {{Spontaneous}} formation of {{MinE}} rings and {{MinD}} polar zones},\ }\href {https://doi.org/10.1073/pnas.2135445100} {\bibfield  {journal} {\bibinfo  {journal} {Proceedings of the National Academy of Sciences}\ }\textbf {\bibinfo {volume} {100}},\ \bibinfo {pages} {12724} (\bibinfo {year} {2003})}\BibitemShut {NoStop}%
\bibitem [{\citenamefont {Fange}\ and\ \citenamefont {Elf}(2006)}]{Fange.Elf2006}%
  \BibitemOpen
  \bibfield  {author} {\bibinfo {author} {\bibfnamefont {D.}~\bibnamefont {Fange}}\ and\ \bibinfo {author} {\bibfnamefont {J.}~\bibnamefont {Elf}},\ }\bibfield  {title} {\bibinfo {title} {Noise-{{Induced Min Phenotypes}} in {{E}}. coli},\ }\href {https://doi.org/10.1371/journal.pcbi.0020080} {\bibfield  {journal} {\bibinfo  {journal} {PLoS Computational Biology}\ }\textbf {\bibinfo {volume} {2}},\ \bibinfo {pages} {e80} (\bibinfo {year} {2006})}\BibitemShut {NoStop}%
\bibitem [{\citenamefont {Halatek}\ and\ \citenamefont {Frey}(2012)}]{Halatek.Frey2012}%
  \BibitemOpen
  \bibfield  {author} {\bibinfo {author} {\bibfnamefont {J.}~\bibnamefont {Halatek}}\ and\ \bibinfo {author} {\bibfnamefont {E.}~\bibnamefont {Frey}},\ }\bibfield  {title} {\bibinfo {title} {Highly {{Canalized MinD Transfer}} and {{MinE Sequestration Explain}} the {{Origin}} of {{Robust MinCDE-Protein Dynamics}}},\ }\href {https://doi.org/10.1016/j.celrep.2012.04.005} {\bibfield  {journal} {\bibinfo  {journal} {Cell Reports}\ }\textbf {\bibinfo {volume} {1}},\ \bibinfo {pages} {741} (\bibinfo {year} {2012})}\BibitemShut {NoStop}%
\bibitem [{\citenamefont {Halatek}\ and\ \citenamefont {Frey}(2018)}]{Halatek.Frey2018}%
  \BibitemOpen
  \bibfield  {author} {\bibinfo {author} {\bibfnamefont {J.}~\bibnamefont {Halatek}}\ and\ \bibinfo {author} {\bibfnamefont {E.}~\bibnamefont {Frey}},\ }\bibfield  {title} {\bibinfo {title} {Rethinking pattern formation in reaction--diffusion systems},\ }\href {https://doi.org/10.1038/s41567-017-0040-5} {\bibfield  {journal} {\bibinfo  {journal} {Nature Physics}\ }\textbf {\bibinfo {volume} {14}},\ \bibinfo {pages} {507} (\bibinfo {year} {2018})}\BibitemShut {NoStop}%
\bibitem [{Note1()}]{Note1}%
  \BibitemOpen
  \bibinfo {note} {As an illustration, consider a system with two protein species A and B with component vector ${\protect \mathbf {u}=(m_\protect \text {AA}, \protect \, \protect \, m_\protect \text {AB}, \protect \, c_\protect \text {A}, \protect \, c_\protect \text {B})^{\protect \! \top }}$: a cytosolic monomer $c_\protect \text {A}$, a membrane dimer $m_{AA}$, a cytosolic monomer $c_\protect \text {B}$, and a heteromeric membrane complex $m_\protect \text {AB}$ containing one species A and one species B. The stoichiometric vectors are ${\protect \mathbf {s}_\protect \text {A}=(2,1,1,0)^{\protect \! \top }}$ and ${\protect \mathbf {s}_\protect \text {B}=(0,1,0,1)^{\protect \! \top }}$, giving the total species densities ${\rho _\protect \text {A}=2m_\protect \text {AA}+m_\protect \text {AB}+c_\protect \text {A}}$ and ${\rho _\protect \text {B}=m_\protect \text {AB}+c_\protect \text {B}}$. This example shows how $\protect \mathbf {s}_\alpha $ accounts for both higher-order states and heteromeric
  complexes.}\BibitemShut {Stop}%
\bibitem [{Note2()}]{Note2}%
  \BibitemOpen
  \bibinfo {note} {The species-specific block structure of $\protect \mathbf {c}$ introduced above can be expressed as an orthogonality condition on the cytosolic stoichiometric vectors: $\protect \mathbf {s}_c^\alpha \cdot \protect \mathbf {s}_c^\beta =0$ for all $\alpha \protect \neq \beta $.}\BibitemShut {Stop}%
\bibitem [{Note3()}]{Note3}%
  \BibitemOpen
  \bibinfo {note} {If membrane diffusion is finite, the mass-redistribution potentials include additional contributions from the membrane components, weighted by their respective diffusion coefficients; see Sec.~\ref {sec:approx-dispRel-oscillatory}.}\BibitemShut {Stop}%
\bibitem [{\citenamefont {Otsuji}\ \emph {et~al.}(2007)\citenamefont {Otsuji}, \citenamefont {Ishihara}, \citenamefont {Co}, \citenamefont {Kaibuchi}, \citenamefont {Mochizuki},\ and\ \citenamefont {Kuroda}}]{Otsuji.etal2007}%
  \BibitemOpen
  \bibfield  {author} {\bibinfo {author} {\bibfnamefont {M.}~\bibnamefont {Otsuji}}, \bibinfo {author} {\bibfnamefont {S.}~\bibnamefont {Ishihara}}, \bibinfo {author} {\bibfnamefont {C.}~\bibnamefont {Co}}, \bibinfo {author} {\bibfnamefont {K.}~\bibnamefont {Kaibuchi}}, \bibinfo {author} {\bibfnamefont {A.}~\bibnamefont {Mochizuki}},\ and\ \bibinfo {author} {\bibfnamefont {S.}~\bibnamefont {Kuroda}},\ }\bibfield  {title} {\bibinfo {title} {A {{Mass Conserved Reaction}}--{{Diffusion System Captures Properties}} of {{Cell Polarity}}},\ }\href {https://doi.org/10.1371/journal.pcbi.0030108} {\bibfield  {journal} {\bibinfo  {journal} {PLoS Computational Biology}\ }\textbf {\bibinfo {volume} {3}},\ \bibinfo {pages} {e108} (\bibinfo {year} {2007})}\BibitemShut {NoStop}%
\bibitem [{\citenamefont {Ishihara}\ \emph {et~al.}(2007)\citenamefont {Ishihara}, \citenamefont {Otsuji},\ and\ \citenamefont {Mochizuki}}]{Ishihara.etal2007}%
  \BibitemOpen
  \bibfield  {author} {\bibinfo {author} {\bibfnamefont {S.}~\bibnamefont {Ishihara}}, \bibinfo {author} {\bibfnamefont {M.}~\bibnamefont {Otsuji}},\ and\ \bibinfo {author} {\bibfnamefont {A.}~\bibnamefont {Mochizuki}},\ }\bibfield  {title} {\bibinfo {title} {Transient and steady state of mass-conserved reaction-diffusion systems},\ }\href {https://doi.org/10.1103/PhysRevE.75.015203} {\bibfield  {journal} {\bibinfo  {journal} {Phys. Rev. E}\ }\textbf {\bibinfo {volume} {75}},\ \bibinfo {pages} {015203} (\bibinfo {year} {2007})}\BibitemShut {NoStop}%
\bibitem [{\citenamefont {Brauns}\ \emph {et~al.}(2021{\natexlab{b}})\citenamefont {Brauns}, \citenamefont {Weyer}, \citenamefont {Halatek}, \citenamefont {Yoon},\ and\ \citenamefont {Frey}}]{Brauns.etal2021b}%
  \BibitemOpen
  \bibfield  {author} {\bibinfo {author} {\bibfnamefont {F.}~\bibnamefont {Brauns}}, \bibinfo {author} {\bibfnamefont {H.}~\bibnamefont {Weyer}}, \bibinfo {author} {\bibfnamefont {J.}~\bibnamefont {Halatek}}, \bibinfo {author} {\bibfnamefont {J.}~\bibnamefont {Yoon}},\ and\ \bibinfo {author} {\bibfnamefont {E.}~\bibnamefont {Frey}},\ }\bibfield  {title} {\bibinfo {title} {Wavelength {{Selection}} by {{Interrupted Coarsening}} in {{Reaction-Diffusion Systems}}},\ }\href {https://doi.org/10.1103/PhysRevLett.126.104101} {\bibfield  {journal} {\bibinfo  {journal} {Physical Review Letters}\ }\textbf {\bibinfo {volume} {126}},\ \bibinfo {pages} {104101} (\bibinfo {year} {2021}{\natexlab{b}})}\BibitemShut {NoStop}%
\bibitem [{Note4()}]{Note4}%
  \BibitemOpen
  \bibinfo {note} {We define the Jacobian matrix $\partial _{\protect \mathbf {y}}\protect \mathbf {w}$ of a vector-valued function $\protect \mathbf {w}$ with respect to the vector argument $\protect \mathbf {y}$ componentwise by ${\protect \big [ \partial _{\protect \mathbf {y}}\protect \mathbf {w} \protect \big ]_{\alpha \beta } \equiv \protect \frac {\partial w_\alpha }{\partial y_\beta } \equiv \partial _{y_\beta } w_\alpha }$.}\BibitemShut {Stop}%
\bibitem [{\citenamefont {Altschuler}\ \emph {et~al.}(2008)\citenamefont {Altschuler}, \citenamefont {Angenent}, \citenamefont {Wang},\ and\ \citenamefont {Wu}}]{Altschuler.etal2008}%
  \BibitemOpen
  \bibfield  {author} {\bibinfo {author} {\bibfnamefont {S.~J.}\ \bibnamefont {Altschuler}}, \bibinfo {author} {\bibfnamefont {S.~B.}\ \bibnamefont {Angenent}}, \bibinfo {author} {\bibfnamefont {Y.}~\bibnamefont {Wang}},\ and\ \bibinfo {author} {\bibfnamefont {L.~F.}\ \bibnamefont {Wu}},\ }\bibfield  {title} {\bibinfo {title} {On the spontaneous emergence of cell polarity},\ }\href {https://doi.org/10.1038/nature07119} {\bibfield  {journal} {\bibinfo  {journal} {Nature}\ }\textbf {\bibinfo {volume} {454}},\ \bibinfo {pages} {886} (\bibinfo {year} {2008})}\BibitemShut {NoStop}%
\bibitem [{\citenamefont {Goryachev}\ and\ \citenamefont {Pokhilko}(2008)}]{Goryachev.Pokhilko2008}%
  \BibitemOpen
  \bibfield  {author} {\bibinfo {author} {\bibfnamefont {A.~B.}\ \bibnamefont {Goryachev}}\ and\ \bibinfo {author} {\bibfnamefont {A.~V.}\ \bibnamefont {Pokhilko}},\ }\bibfield  {title} {\bibinfo {title} {Dynamics of {{Cdc42}} network embodies a {{Turing-type}} mechanism of yeast cell polarity},\ }\href {https://doi.org/10.1016/j.febslet.2008.03.029} {\bibfield  {journal} {\bibinfo  {journal} {FEBS Letters}\ }\textbf {\bibinfo {volume} {582}},\ \bibinfo {pages} {1437} (\bibinfo {year} {2008})}\BibitemShut {NoStop}%
\bibitem [{\citenamefont {Mori}\ \emph {et~al.}(2008)\citenamefont {Mori}, \citenamefont {Jilkine},\ and\ \citenamefont {{Edelstein-Keshet}}}]{Mori.etal2008}%
  \BibitemOpen
  \bibfield  {author} {\bibinfo {author} {\bibfnamefont {Y.}~\bibnamefont {Mori}}, \bibinfo {author} {\bibfnamefont {A.}~\bibnamefont {Jilkine}},\ and\ \bibinfo {author} {\bibfnamefont {L.}~\bibnamefont {{Edelstein-Keshet}}},\ }\bibfield  {title} {\bibinfo {title} {Wave-{{Pinning}} and {{Cell Polarity}} from a {{Bistable Reaction-Diffusion System}}},\ }\href {https://doi.org/10.1529/biophysj.107.120824} {\bibfield  {journal} {\bibinfo  {journal} {Biophysical Journal}\ }\textbf {\bibinfo {volume} {94}},\ \bibinfo {pages} {3684} (\bibinfo {year} {2008})}\BibitemShut {NoStop}%
\bibitem [{\citenamefont {{Edelstein-Keshet}}\ \emph {et~al.}(2013)\citenamefont {{Edelstein-Keshet}}, \citenamefont {Holmes}, \citenamefont {Zajac},\ and\ \citenamefont {Dutot}}]{Edelstein-Keshet.etal2013}%
  \BibitemOpen
  \bibfield  {author} {\bibinfo {author} {\bibfnamefont {L.}~\bibnamefont {{Edelstein-Keshet}}}, \bibinfo {author} {\bibfnamefont {W.~R.}\ \bibnamefont {Holmes}}, \bibinfo {author} {\bibfnamefont {M.}~\bibnamefont {Zajac}},\ and\ \bibinfo {author} {\bibfnamefont {M.}~\bibnamefont {Dutot}},\ }\bibfield  {title} {\bibinfo {title} {From simple to detailed models for cell polarization},\ }\href {https://doi.org/10.1098/rstb.2013.0003} {\bibfield  {journal} {\bibinfo  {journal} {Philosophical Transactions of the Royal Society B: Biological Sciences}\ }\textbf {\bibinfo {volume} {368}},\ \bibinfo {pages} {20130003} (\bibinfo {year} {2013})}\BibitemShut {NoStop}%
\bibitem [{\citenamefont {Cross}\ and\ \citenamefont {Hohenberg}(1993)}]{Cross.Hohenberg1993}%
  \BibitemOpen
  \bibfield  {author} {\bibinfo {author} {\bibfnamefont {M.~C.}\ \bibnamefont {Cross}}\ and\ \bibinfo {author} {\bibfnamefont {P.~C.}\ \bibnamefont {Hohenberg}},\ }\bibfield  {title} {\bibinfo {title} {Pattern formation outside of equilibrium},\ }\href {https://doi.org/10.1103/RevModPhys.65.851} {\bibfield  {journal} {\bibinfo  {journal} {Reviews of Modern Physics}\ }\textbf {\bibinfo {volume} {65}},\ \bibinfo {pages} {851} (\bibinfo {year} {1993})}\BibitemShut {NoStop}%
\bibitem [{Note5()}]{Note5}%
  \BibitemOpen
  \bibinfo {note} {Without loss of generality, we chose the normalization of the stoichiometric vectors by $s_{c,1}^\alpha =1$.}\BibitemShut {Stop}%
\bibitem [{Note6()}]{Note6}%
  \BibitemOpen
  \bibinfo {note} {Finite membrane diffusion ensures that the dispersion relation attains a negative real part at large wavenumbers $q$. In the approximation of zero membrane diffusion the dispersion relation remains positive in the limit ${q \to \infty }$ (see also Ref.~\cite {Villar-Sepulveda.etal2025}).}\BibitemShut {Stop}%
\bibitem [{Note7()}]{Note7}%
  \BibitemOpen
  \bibinfo {note} {They correspond to the so-called type-I instabilities in the classification of Cross and Hohenberg~\protect \citep {Cross.Hohenberg1993}. Because of the conservation law(s) ensuring zero eigenvalues at ${q=0}$, the short-wavelength instabilities arising in McRD systems are more specifically referred to as ``conserved Turing instabilities''~\protect \citep {Frohoff-Hulsmann.Thiele2023a}.}\BibitemShut {Stop}%
\bibitem [{\citenamefont {Gai}\ \emph {et~al.}(2020)\citenamefont {Gai}, \citenamefont {Iron},\ and\ \citenamefont {Kolokolnikov}}]{Gai.etal2020a}%
  \BibitemOpen
  \bibfield  {author} {\bibinfo {author} {\bibfnamefont {C.}~\bibnamefont {Gai}}, \bibinfo {author} {\bibfnamefont {D.}~\bibnamefont {Iron}},\ and\ \bibinfo {author} {\bibfnamefont {T.}~\bibnamefont {Kolokolnikov}},\ }\bibfield  {title} {\bibinfo {title} {Localized outbreaks in an {{S-I-R}} model with diffusion},\ }\bibfield  {journal} {\bibinfo  {journal} {Journal of Mathematical Biology}\ }\href {https://doi.org/10.1007/s00285-020-01466-1} {10.1007/s00285-020-01466-1} (\bibinfo {year} {2020})\BibitemShut {NoStop}%
\bibitem [{\citenamefont {Chiou}\ \emph {et~al.}(2021)\citenamefont {Chiou}, \citenamefont {Moran},\ and\ \citenamefont {Lew}}]{Chiou.etal2021}%
  \BibitemOpen
  \bibfield  {author} {\bibinfo {author} {\bibfnamefont {J.-g.}\ \bibnamefont {Chiou}}, \bibinfo {author} {\bibfnamefont {K.~D.}\ \bibnamefont {Moran}},\ and\ \bibinfo {author} {\bibfnamefont {D.~J.}\ \bibnamefont {Lew}},\ }\bibfield  {title} {\bibinfo {title} {How cells determine the number of polarity sites},\ }\href {https://doi.org/10.7554/eLife.58768} {\bibfield  {journal} {\bibinfo  {journal} {eLife}\ }\textbf {\bibinfo {volume} {10}},\ \bibinfo {pages} {e58768} (\bibinfo {year} {2021})}\BibitemShut {NoStop}%
\bibitem [{\citenamefont {Toffenetti}\ \emph {et~al.}(2026)\citenamefont {Toffenetti}, \citenamefont {Nettuno}, \citenamefont {Weyer},\ and\ \citenamefont {Frey}}]{Toffenetti.etal2026}%
  \BibitemOpen
  \bibfield  {author} {\bibinfo {author} {\bibfnamefont {D.}~\bibnamefont {Toffenetti}}, \bibinfo {author} {\bibfnamefont {B.}~\bibnamefont {Nettuno}}, \bibinfo {author} {\bibfnamefont {H.}~\bibnamefont {Weyer}},\ and\ \bibinfo {author} {\bibfnamefont {E.}~\bibnamefont {Frey}},\ }\href {https://doi.org/10.48550/arXiv.2605.15903} {\bibinfo {title} {Active {{Model B}}$^{-}$ from {{Mass-Conserving Reaction-Diffusion Systems}}}} (\bibinfo {year} {2026}),\ \Eprint {https://arxiv.org/abs/2605.15903} {arXiv:2605.15903 [cond-mat.soft]} \BibitemShut {NoStop}%
\bibitem [{\citenamefont {Mar{\'e}e}\ \emph {et~al.}(2006)\citenamefont {Mar{\'e}e}, \citenamefont {Jilkine}, \citenamefont {Dawes}, \citenamefont {Grieneisen},\ and\ \citenamefont {{Edelstein-Keshet}}}]{Maree.etal2006}%
  \BibitemOpen
  \bibfield  {author} {\bibinfo {author} {\bibfnamefont {A.~F.~M.}\ \bibnamefont {Mar{\'e}e}}, \bibinfo {author} {\bibfnamefont {A.}~\bibnamefont {Jilkine}}, \bibinfo {author} {\bibfnamefont {A.}~\bibnamefont {Dawes}}, \bibinfo {author} {\bibfnamefont {V.~A.}\ \bibnamefont {Grieneisen}},\ and\ \bibinfo {author} {\bibfnamefont {L.}~\bibnamefont {{Edelstein-Keshet}}},\ }\bibfield  {title} {\bibinfo {title} {Polarization and {{Movement}} of {{Keratocytes}}: {{A Multiscale Modelling Approach}}},\ }\href {https://doi.org/10.1007/s11538-006-9131-7} {\bibfield  {journal} {\bibinfo  {journal} {Bulletin of Mathematical Biology}\ }\textbf {\bibinfo {volume} {68}},\ \bibinfo {pages} {1169} (\bibinfo {year} {2006})}\BibitemShut {NoStop}%
\bibitem [{\citenamefont {Kl{\"u}nder}\ \emph {et~al.}(2013)\citenamefont {Kl{\"u}nder}, \citenamefont {Freisinger}, \citenamefont {{Wedlich-S{\"o}ldner}},\ and\ \citenamefont {Frey}}]{Klunder.etal2013}%
  \BibitemOpen
  \bibfield  {author} {\bibinfo {author} {\bibfnamefont {B.}~\bibnamefont {Kl{\"u}nder}}, \bibinfo {author} {\bibfnamefont {T.}~\bibnamefont {Freisinger}}, \bibinfo {author} {\bibfnamefont {R.}~\bibnamefont {{Wedlich-S{\"o}ldner}}},\ and\ \bibinfo {author} {\bibfnamefont {E.}~\bibnamefont {Frey}},\ }\bibfield  {title} {\bibinfo {title} {{{GDI-Mediated Cell Polarization}} in {{Yeast Provides Precise Spatial}} and {{Temporal Control}} of {{Cdc42 Signaling}}},\ }\href {https://doi.org/10.1371/journal.pcbi.1003396} {\bibfield  {journal} {\bibinfo  {journal} {PLOS Computational Biology}\ }\textbf {\bibinfo {volume} {9}},\ \bibinfo {pages} {e1003396} (\bibinfo {year} {2013})}\BibitemShut {NoStop}%
\bibitem [{Note8()}]{Note8}%
  \BibitemOpen
  \bibinfo {note} {The models for the Rab5 system have only single cytosolic components if the guanosine nucleotide dissociation inhibitor (GDI) is not modeled explicitly. If the GDI is modeled as done in Ref.~~\protect \citep {Solomatina.etal2022}, it has two cytosolic components (no membrane-bound states), of which one is the complex with Rab5-ADP. Because the complex with Rab5-ADP is the only cytosolic component of Rab5, the arguments from Sec.~\ref {sec:single-cyt-comp} apply to this cytosolic complex. Mass conservation of the GDI then implies that they also extend to its own individual cytosolic component. Therefore, also this extended model, although not strictly fulfilling the condition of one cytosolic component per species, does only show stationary long-wavelength instabilities.}\BibitemShut {Stop}%
\bibitem [{\citenamefont {Cezanne}\ \emph {et~al.}(2020)\citenamefont {Cezanne}, \citenamefont {Lauer}, \citenamefont {Solomatina}, \citenamefont {Sbalzarini},\ and\ \citenamefont {Zerial}}]{Cezanne.etal2020}%
  \BibitemOpen
  \bibfield  {author} {\bibinfo {author} {\bibfnamefont {A.}~\bibnamefont {Cezanne}}, \bibinfo {author} {\bibfnamefont {J.}~\bibnamefont {Lauer}}, \bibinfo {author} {\bibfnamefont {A.}~\bibnamefont {Solomatina}}, \bibinfo {author} {\bibfnamefont {I.~F.}\ \bibnamefont {Sbalzarini}},\ and\ \bibinfo {author} {\bibfnamefont {M.}~\bibnamefont {Zerial}},\ }\bibfield  {title} {\bibinfo {title} {A non-linear system patterns {{Rab5 GTPase}} on the membrane},\ }\href {https://doi.org/10.7554/eLife.54434} {\bibfield  {journal} {\bibinfo  {journal} {eLife}\ }\textbf {\bibinfo {volume} {9}},\ \bibinfo {pages} {e54434} (\bibinfo {year} {2020})}\BibitemShut {NoStop}%
\bibitem [{\citenamefont {Goehring}\ \emph {et~al.}(2011)\citenamefont {Goehring}, \citenamefont {Trong}, \citenamefont {Bois}, \citenamefont {Chowdhury}, \citenamefont {Nicola}, \citenamefont {Hyman},\ and\ \citenamefont {Grill}}]{Goehring.etal2011}%
  \BibitemOpen
  \bibfield  {author} {\bibinfo {author} {\bibfnamefont {N.~W.}\ \bibnamefont {Goehring}}, \bibinfo {author} {\bibfnamefont {P.~K.}\ \bibnamefont {Trong}}, \bibinfo {author} {\bibfnamefont {J.~S.}\ \bibnamefont {Bois}}, \bibinfo {author} {\bibfnamefont {D.}~\bibnamefont {Chowdhury}}, \bibinfo {author} {\bibfnamefont {E.~M.}\ \bibnamefont {Nicola}}, \bibinfo {author} {\bibfnamefont {A.~A.}\ \bibnamefont {Hyman}},\ and\ \bibinfo {author} {\bibfnamefont {S.~W.}\ \bibnamefont {Grill}},\ }\bibfield  {title} {\bibinfo {title} {Polarization of {{PAR Proteins}} by {{Advective Triggering}} of a {{Pattern-Forming System}}},\ }\href {https://doi.org/10.1126/science.1208619} {\bibfield  {journal} {\bibinfo  {journal} {Science}\ }\textbf {\bibinfo {volume} {334}},\ \bibinfo {pages} {1137} (\bibinfo {year} {2011})}\BibitemShut {NoStop}%
\bibitem [{\citenamefont {Hansen}\ \emph {et~al.}(2019)\citenamefont {Hansen}, \citenamefont {Huang}, \citenamefont {Lee}, \citenamefont {Bieling}, \citenamefont {Christensen},\ and\ \citenamefont {Groves}}]{Hansen.etal2019}%
  \BibitemOpen
  \bibfield  {author} {\bibinfo {author} {\bibfnamefont {S.~D.}\ \bibnamefont {Hansen}}, \bibinfo {author} {\bibfnamefont {W.~Y.~C.}\ \bibnamefont {Huang}}, \bibinfo {author} {\bibfnamefont {Y.~K.}\ \bibnamefont {Lee}}, \bibinfo {author} {\bibfnamefont {P.}~\bibnamefont {Bieling}}, \bibinfo {author} {\bibfnamefont {S.~M.}\ \bibnamefont {Christensen}},\ and\ \bibinfo {author} {\bibfnamefont {J.~T.}\ \bibnamefont {Groves}},\ }\bibfield  {title} {\bibinfo {title} {Stochastic geometry sensing and polarization in a lipid kinase--phosphatase competitive reaction},\ }\href {https://doi.org/10.1073/pnas.1901744116} {\bibfield  {journal} {\bibinfo  {journal} {Proceedings of the National Academy of Sciences}\ }\textbf {\bibinfo {volume} {116}},\ \bibinfo {pages} {15013} (\bibinfo {year} {2019})}\BibitemShut {NoStop}%
\bibitem [{\citenamefont {Ge{\ss}ele}\ \emph {et~al.}(2020)\citenamefont {Ge{\ss}ele}, \citenamefont {Halatek}, \citenamefont {W{\"u}rthner},\ and\ \citenamefont {Frey}}]{Gessele.etal2020}%
  \BibitemOpen
  \bibfield  {author} {\bibinfo {author} {\bibfnamefont {R.}~\bibnamefont {Ge{\ss}ele}}, \bibinfo {author} {\bibfnamefont {J.}~\bibnamefont {Halatek}}, \bibinfo {author} {\bibfnamefont {L.}~\bibnamefont {W{\"u}rthner}},\ and\ \bibinfo {author} {\bibfnamefont {E.}~\bibnamefont {Frey}},\ }\bibfield  {title} {\bibinfo {title} {Geometric cues stabilise long-axis polarisation of {{PAR}} protein patterns in {{C}}. elegans},\ }\href {https://doi.org/10.1038/s41467-020-14317-w} {\bibfield  {journal} {\bibinfo  {journal} {Nature Communications}\ }\textbf {\bibinfo {volume} {11}},\ \bibinfo {pages} {539} (\bibinfo {year} {2020})}\BibitemShut {NoStop}%
\bibitem [{\citenamefont {Denk}\ \emph {et~al.}(2018)\citenamefont {Denk}, \citenamefont {Kretschmer}, \citenamefont {Halatek}, \citenamefont {Hartl}, \citenamefont {Schwille},\ and\ \citenamefont {Frey}}]{Denk.etal2018}%
  \BibitemOpen
  \bibfield  {author} {\bibinfo {author} {\bibfnamefont {J.}~\bibnamefont {Denk}}, \bibinfo {author} {\bibfnamefont {S.}~\bibnamefont {Kretschmer}}, \bibinfo {author} {\bibfnamefont {J.}~\bibnamefont {Halatek}}, \bibinfo {author} {\bibfnamefont {C.}~\bibnamefont {Hartl}}, \bibinfo {author} {\bibfnamefont {P.}~\bibnamefont {Schwille}},\ and\ \bibinfo {author} {\bibfnamefont {E.}~\bibnamefont {Frey}},\ }\bibfield  {title} {\bibinfo {title} {{{MinE}} conformational switching confers robustness on self-organized {{Min}} protein patterns},\ }\href {https://doi.org/10.1073/pnas.1719801115} {\bibfield  {journal} {\bibinfo  {journal} {Proceedings of the National Academy of Sciences}\ }\textbf {\bibinfo {volume} {115}},\ \bibinfo {pages} {4553} (\bibinfo {year} {2018})}\BibitemShut {NoStop}%
\bibitem [{\citenamefont {Ren}\ \emph {et~al.}(2025)\citenamefont {Ren}, \citenamefont {Weyer}, \citenamefont {Sandler}, \citenamefont {W{\"u}rthner}, \citenamefont {Fu}, \citenamefont {Tangtartharakul}, \citenamefont {Li}, \citenamefont {Sou}, \citenamefont {Villarreal}, \citenamefont {Kim}, \citenamefont {Frey},\ and\ \citenamefont {Jun}}]{Ren.etal2025}%
  \BibitemOpen
  \bibfield  {author} {\bibinfo {author} {\bibfnamefont {Z.}~\bibnamefont {Ren}}, \bibinfo {author} {\bibfnamefont {H.}~\bibnamefont {Weyer}}, \bibinfo {author} {\bibfnamefont {M.}~\bibnamefont {Sandler}}, \bibinfo {author} {\bibfnamefont {L.}~\bibnamefont {W{\"u}rthner}}, \bibinfo {author} {\bibfnamefont {H.}~\bibnamefont {Fu}}, \bibinfo {author} {\bibfnamefont {C.~B.}\ \bibnamefont {Tangtartharakul}}, \bibinfo {author} {\bibfnamefont {D.}~\bibnamefont {Li}}, \bibinfo {author} {\bibfnamefont {C.}~\bibnamefont {Sou}}, \bibinfo {author} {\bibfnamefont {D.}~\bibnamefont {Villarreal}}, \bibinfo {author} {\bibfnamefont {J.~E.}\ \bibnamefont {Kim}}, \bibinfo {author} {\bibfnamefont {E.}~\bibnamefont {Frey}},\ and\ \bibinfo {author} {\bibfnamefont {S.}~\bibnamefont {Jun}},\ }\bibfield  {title} {\bibinfo {title} {Robust and resource-optimal dynamic pattern formation of {{Min}} proteins in vivo},\ }\bibfield  {journal} {\bibinfo  {journal} {Nature Physics}\ }\href {https://doi.org/10.1038/s41567-025-02878-w}
  {10.1038/s41567-025-02878-w} (\bibinfo {year} {2025})\BibitemShut {NoStop}%
\bibitem [{\citenamefont {Meindlhumer}\ \emph {et~al.}(2023)\citenamefont {Meindlhumer}, \citenamefont {Brauns}, \citenamefont {Fin{\v z}gar}, \citenamefont {Kerssemakers}, \citenamefont {Dekker},\ and\ \citenamefont {Frey}}]{Meindlhumer.etal2023}%
  \BibitemOpen
  \bibfield  {author} {\bibinfo {author} {\bibfnamefont {S.}~\bibnamefont {Meindlhumer}}, \bibinfo {author} {\bibfnamefont {F.}~\bibnamefont {Brauns}}, \bibinfo {author} {\bibfnamefont {J.~R.}\ \bibnamefont {Fin{\v z}gar}}, \bibinfo {author} {\bibfnamefont {J.}~\bibnamefont {Kerssemakers}}, \bibinfo {author} {\bibfnamefont {C.}~\bibnamefont {Dekker}},\ and\ \bibinfo {author} {\bibfnamefont {E.}~\bibnamefont {Frey}},\ }\bibfield  {title} {\bibinfo {title} {Directing {{Min}} protein patterns with advective bulk flow},\ }\href {https://doi.org/10.1038/s41467-023-35997-0} {\bibfield  {journal} {\bibinfo  {journal} {Nature Communications}\ }\textbf {\bibinfo {volume} {14}},\ \bibinfo {pages} {450} (\bibinfo {year} {2023})}\BibitemShut {NoStop}%
\bibitem [{\citenamefont {Szeto}\ \emph {et~al.}(2002)\citenamefont {Szeto}, \citenamefont {Rowland}, \citenamefont {Rothfield},\ and\ \citenamefont {King}}]{Szeto.etal2002}%
  \BibitemOpen
  \bibfield  {author} {\bibinfo {author} {\bibfnamefont {T.~H.}\ \bibnamefont {Szeto}}, \bibinfo {author} {\bibfnamefont {S.~L.}\ \bibnamefont {Rowland}}, \bibinfo {author} {\bibfnamefont {L.~I.}\ \bibnamefont {Rothfield}},\ and\ \bibinfo {author} {\bibfnamefont {G.~F.}\ \bibnamefont {King}},\ }\bibfield  {title} {\bibinfo {title} {Membrane localization of {{MinD}} is mediated by a {{C-terminal}} motif that is conserved across eubacteria, archaea, and chloroplasts},\ }\href {https://doi.org/10.1073/pnas.232590599} {\bibfield  {journal} {\bibinfo  {journal} {Proceedings of the National Academy of Sciences}\ }\textbf {\bibinfo {volume} {99}},\ \bibinfo {pages} {15693} (\bibinfo {year} {2002})}\BibitemShut {NoStop}%
\bibitem [{\citenamefont {Hu}\ and\ \citenamefont {Lutkenhaus}(2003)}]{Hu.Lutkenhaus2003}%
  \BibitemOpen
  \bibfield  {author} {\bibinfo {author} {\bibfnamefont {Z.}~\bibnamefont {Hu}}\ and\ \bibinfo {author} {\bibfnamefont {J.}~\bibnamefont {Lutkenhaus}},\ }\bibfield  {title} {\bibinfo {title} {A conserved sequence at the {{C-terminus}} of {{MinD}} is required for binding to the membrane and targeting {{MinC}} to the septum: {{Role}} of {{C-terminus}} of {{MinD}}},\ }\href {https://doi.org/10.1046/j.1365-2958.2003.03321.x} {\bibfield  {journal} {\bibinfo  {journal} {Molecular Microbiology}\ }\textbf {\bibinfo {volume} {47}},\ \bibinfo {pages} {345} (\bibinfo {year} {2003})}\BibitemShut {NoStop}%
\bibitem [{\citenamefont {Szeto}\ \emph {et~al.}(2003)\citenamefont {Szeto}, \citenamefont {Rowland}, \citenamefont {Habrukowich},\ and\ \citenamefont {King}}]{Szeto.etal2003}%
  \BibitemOpen
  \bibfield  {author} {\bibinfo {author} {\bibfnamefont {T.~H.}\ \bibnamefont {Szeto}}, \bibinfo {author} {\bibfnamefont {S.~L.}\ \bibnamefont {Rowland}}, \bibinfo {author} {\bibfnamefont {C.~L.}\ \bibnamefont {Habrukowich}},\ and\ \bibinfo {author} {\bibfnamefont {G.~F.}\ \bibnamefont {King}},\ }\bibfield  {title} {\bibinfo {title} {The {{MinD Membrane Targeting Sequence Is}} a {{Transplantable Lipid-binding Helix}}},\ }\href {https://doi.org/10.1074/jbc.M306876200} {\bibfield  {journal} {\bibinfo  {journal} {Journal of Biological Chemistry}\ }\textbf {\bibinfo {volume} {278}},\ \bibinfo {pages} {40050} (\bibinfo {year} {2003})}\BibitemShut {NoStop}%
\bibitem [{\citenamefont {Mileykovskaya}\ \emph {et~al.}(2003)\citenamefont {Mileykovskaya}, \citenamefont {Fishov}, \citenamefont {Fu}, \citenamefont {Corbin}, \citenamefont {Margolin},\ and\ \citenamefont {Dowhan}}]{Mileykovskaya.etal2003}%
  \BibitemOpen
  \bibfield  {author} {\bibinfo {author} {\bibfnamefont {E.}~\bibnamefont {Mileykovskaya}}, \bibinfo {author} {\bibfnamefont {I.}~\bibnamefont {Fishov}}, \bibinfo {author} {\bibfnamefont {X.}~\bibnamefont {Fu}}, \bibinfo {author} {\bibfnamefont {B.~D.}\ \bibnamefont {Corbin}}, \bibinfo {author} {\bibfnamefont {W.}~\bibnamefont {Margolin}},\ and\ \bibinfo {author} {\bibfnamefont {W.}~\bibnamefont {Dowhan}},\ }\bibfield  {title} {\bibinfo {title} {Effects of {{Phospholipid Composition}} on {{MinD-Membrane Interactions}} {\emph{in }}{{{\emph{Vitro}}}} and {\emph{in }}{{{\emph{Vivo}}}}},\ }\href {https://doi.org/10.1074/jbc.M302603200} {\bibfield  {journal} {\bibinfo  {journal} {Journal of Biological Chemistry}\ }\textbf {\bibinfo {volume} {278}},\ \bibinfo {pages} {22193} (\bibinfo {year} {2003})}\BibitemShut {NoStop}%
\bibitem [{\citenamefont {Taghbalout}\ \emph {et~al.}(2006)\citenamefont {Taghbalout}, \citenamefont {Ma},\ and\ \citenamefont {Rothfield}}]{Taghbalout.etal2006}%
  \BibitemOpen
  \bibfield  {author} {\bibinfo {author} {\bibfnamefont {A.}~\bibnamefont {Taghbalout}}, \bibinfo {author} {\bibfnamefont {L.}~\bibnamefont {Ma}},\ and\ \bibinfo {author} {\bibfnamefont {L.}~\bibnamefont {Rothfield}},\ }\bibfield  {title} {\bibinfo {title} {Role {{Of MinD-Membrane Association}} in {{Min Protein Interactions}}},\ }\href {https://doi.org/10.1128/JB.188.8.2993-3001.2006} {\bibfield  {journal} {\bibinfo  {journal} {Journal of Bacteriology}\ }\textbf {\bibinfo {volume} {188}},\ \bibinfo {pages} {2993} (\bibinfo {year} {2006})}\BibitemShut {NoStop}%
\bibitem [{\citenamefont {Ramm}\ \emph {et~al.}(2019)\citenamefont {Ramm}, \citenamefont {Heermann},\ and\ \citenamefont {Schwille}}]{Ramm.etal2019}%
  \BibitemOpen
  \bibfield  {author} {\bibinfo {author} {\bibfnamefont {B.}~\bibnamefont {Ramm}}, \bibinfo {author} {\bibfnamefont {T.}~\bibnamefont {Heermann}},\ and\ \bibinfo {author} {\bibfnamefont {P.}~\bibnamefont {Schwille}},\ }\bibfield  {title} {\bibinfo {title} {The {{E}}. coli {{MinCDE}} system in the regulation of protein patterns and gradients},\ }\href {https://doi.org/10.1007/s00018-019-03218-x} {\bibfield  {journal} {\bibinfo  {journal} {Cellular and Molecular Life Sciences}\ }\textbf {\bibinfo {volume} {76}},\ \bibinfo {pages} {4245} (\bibinfo {year} {2019})}\BibitemShut {NoStop}%
\bibitem [{\citenamefont {Wiggins}(2003)}]{Wiggins2003}%
  \BibitemOpen
  \bibfield  {author} {\bibinfo {author} {\bibfnamefont {S.}~\bibnamefont {Wiggins}},\ }\href {https://doi.org/10.1007/b97481} {\emph {\bibinfo {title} {Introduction to {{Applied Nonlinear Dynamical Systems}} and {{Chaos}}}}},\ \bibinfo {series} {Texts in {{Applied Mathematics}}}, Vol.~\bibinfo {volume} {2}\ (\bibinfo  {publisher} {Springer-Verlag},\ \bibinfo {address} {New York},\ \bibinfo {year} {2003})\BibitemShut {NoStop}%
\bibitem [{Note9()}]{Note9}%
  \BibitemOpen
  \bibinfo {note} {In systems with an extended bulk (bulk--boundary coupling), we compare the changes in membrane density to the changes in the cytosolic densities integrated perpendicular to the membrane over the height of the cytosolic bulk.}\BibitemShut {Stop}%
\bibitem [{Note10()}]{Note10}%
  \BibitemOpen
  \bibinfo {note} {To treat the switch model within our framework, we describe the reactive MinE state within a quasi-steady-state approximation, justified by its fast deactivation~\protect \citep {Meindlhumer.etal2023}.}\BibitemShut {Stop}%
\bibitem [{\citenamefont {Schl{\"o}gl}(1972)}]{Schlogl1972}%
  \BibitemOpen
  \bibfield  {author} {\bibinfo {author} {\bibfnamefont {F.}~\bibnamefont {Schl{\"o}gl}},\ }\bibfield  {title} {\bibinfo {title} {Chemical reaction models for non-equilibrium phase transitions},\ }\href {https://doi.org/10.1007/BF01379769} {\bibfield  {journal} {\bibinfo  {journal} {Zeitschrift f{\"u}r Physik}\ }\textbf {\bibinfo {volume} {253}},\ \bibinfo {pages} {147} (\bibinfo {year} {1972})}\BibitemShut {NoStop}%
\bibitem [{\citenamefont {Tateno}\ and\ \citenamefont {Ishihara}(2021)}]{Tateno.Ishihara2021}%
  \BibitemOpen
  \bibfield  {author} {\bibinfo {author} {\bibfnamefont {M.}~\bibnamefont {Tateno}}\ and\ \bibinfo {author} {\bibfnamefont {S.}~\bibnamefont {Ishihara}},\ }\bibfield  {title} {\bibinfo {title} {Interfacial-curvature-driven coarsening in mass-conserved reaction-diffusion systems},\ }\href {https://doi.org/10.1103/PhysRevResearch.3.023198} {\bibfield  {journal} {\bibinfo  {journal} {Physical Review Research}\ }\textbf {\bibinfo {volume} {3}},\ \bibinfo {pages} {023198} (\bibinfo {year} {2021})}\BibitemShut {NoStop}%
\bibitem [{\citenamefont {Weyer}\ \emph {et~al.}(2023)\citenamefont {Weyer}, \citenamefont {Brauns},\ and\ \citenamefont {Frey}}]{Weyer.etal2023}%
  \BibitemOpen
  \bibfield  {author} {\bibinfo {author} {\bibfnamefont {H.}~\bibnamefont {Weyer}}, \bibinfo {author} {\bibfnamefont {F.}~\bibnamefont {Brauns}},\ and\ \bibinfo {author} {\bibfnamefont {E.}~\bibnamefont {Frey}},\ }\bibfield  {title} {\bibinfo {title} {Coarsening and wavelength selection far from equilibrium: {{A}} unifying framework based on singular perturbation theory},\ }\href {https://doi.org/10.1103/PhysRevE.108.064202} {\bibfield  {journal} {\bibinfo  {journal} {Physical Review E}\ }\textbf {\bibinfo {volume} {108}},\ \bibinfo {pages} {064202} (\bibinfo {year} {2023})}\BibitemShut {NoStop}%
\bibitem [{\citenamefont {Miller}\ \emph {et~al.}(2023)\citenamefont {Miller}, \citenamefont {Fortunato}, \citenamefont {Novaga}, \citenamefont {Shvartsman},\ and\ \citenamefont {Muratov}}]{Miller.etal2023}%
  \BibitemOpen
  \bibfield  {author} {\bibinfo {author} {\bibfnamefont {P.~W.}\ \bibnamefont {Miller}}, \bibinfo {author} {\bibfnamefont {D.}~\bibnamefont {Fortunato}}, \bibinfo {author} {\bibfnamefont {M.}~\bibnamefont {Novaga}}, \bibinfo {author} {\bibfnamefont {S.~Y.}\ \bibnamefont {Shvartsman}},\ and\ \bibinfo {author} {\bibfnamefont {C.~B.}\ \bibnamefont {Muratov}},\ }\bibfield  {title} {\bibinfo {title} {Generation and {{Motion}} of {{Interfaces}} in a {{Mass-Conserving Reaction-Diffusion System}}},\ }\href {https://doi.org/10.1137/22M152548X} {\bibfield  {journal} {\bibinfo  {journal} {SIAM Journal on Applied Dynamical Systems}\ }\textbf {\bibinfo {volume} {22}},\ \bibinfo {pages} {2408} (\bibinfo {year} {2023})}\BibitemShut {NoStop}%
\bibitem [{\citenamefont {Weyer}\ \emph {et~al.}(2025)\citenamefont {Weyer}, \citenamefont {Roth},\ and\ \citenamefont {Frey}}]{Weyer.etal2025a}%
  \BibitemOpen
  \bibfield  {author} {\bibinfo {author} {\bibfnamefont {H.}~\bibnamefont {Weyer}}, \bibinfo {author} {\bibfnamefont {T.~A.}\ \bibnamefont {Roth}},\ and\ \bibinfo {author} {\bibfnamefont {E.}~\bibnamefont {Frey}},\ }\bibfield  {title} {\bibinfo {title} {Protein pattern morphology and dynamics emerging from effective interfacial tension},\ }\bibfield  {journal} {\bibinfo  {journal} {Nature Physics}\ }\href {https://doi.org/10.1038/s41567-025-03101-6} {10.1038/s41567-025-03101-6} (\bibinfo {year} {2025})\BibitemShut {NoStop}%
\bibitem [{\citenamefont {Zhou}\ and\ \citenamefont {Frey}(2026)}]{Zhou.Frey2026}%
  \BibitemOpen
  \bibfield  {author} {\bibinfo {author} {\bibfnamefont {D.}~\bibnamefont {Zhou}}\ and\ \bibinfo {author} {\bibfnamefont {E.}~\bibnamefont {Frey}},\ }\href {https://doi.org/10.48550/arXiv.2605.15158} {\bibinfo {title} {Duality {{Between Chemical Potential Dynamics}} and {{Reaction-Diffusion Systems}}}} (\bibinfo {year} {2026}),\ \Eprint {https://arxiv.org/abs/2605.15158} {arXiv:2605.15158 [cond-mat.soft]} \BibitemShut {NoStop}%
\bibitem [{\citenamefont {Weyer}\ \emph {et~al.}(2026)\citenamefont {Weyer}, \citenamefont {Leung},\ and\ \citenamefont {Frey}}]{weyer2026github}%
  \BibitemOpen
  \bibfield  {author} {\bibinfo {author} {\bibfnamefont {H.}~\bibnamefont {Weyer}}, \bibinfo {author} {\bibfnamefont {C.~Y.}\ \bibnamefont {Leung}},\ and\ \bibinfo {author} {\bibfnamefont {E.}~\bibnamefont {Frey}},\ }\href@noop {} {\bibinfo {title} {{Multicomponent-slope-criteria: Code for "Classification of Intracellular Protein Patterns from Reactive Equilibria"}}},\ \bibinfo {howpublished} {\url{https://github.com/henrikweyer/Multicomponent-slope-criteria}} (\bibinfo {year} {2026}),\ \bibinfo {note} {gitHub repository}\BibitemShut {NoStop}%
\bibitem [{\citenamefont {Raskin}\ and\ \citenamefont {{de Boer}}(1999)}]{Raskin.DeBoer1999}%
  \BibitemOpen
  \bibfield  {author} {\bibinfo {author} {\bibfnamefont {D.~M.}\ \bibnamefont {Raskin}}\ and\ \bibinfo {author} {\bibfnamefont {P.~A.~J.}\ \bibnamefont {{de Boer}}},\ }\bibfield  {title} {\bibinfo {title} {Rapid pole-to-pole oscillation of a protein required for directing division to the middle of {{Escherichia}} coli},\ }\href {https://doi.org/10.1073/pnas.96.9.4971} {\bibfield  {journal} {\bibinfo  {journal} {Proceedings of the National Academy of Sciences}\ }\textbf {\bibinfo {volume} {96}},\ \bibinfo {pages} {4971} (\bibinfo {year} {1999})}\BibitemShut {NoStop}%
\bibitem [{\citenamefont {Hu}\ and\ \citenamefont {Lutkenhaus}(1999)}]{Hu.Lutkenhaus1999}%
  \BibitemOpen
  \bibfield  {author} {\bibinfo {author} {\bibfnamefont {Z.}~\bibnamefont {Hu}}\ and\ \bibinfo {author} {\bibfnamefont {J.}~\bibnamefont {Lutkenhaus}},\ }\bibfield  {title} {\bibinfo {title} {Topological regulation of cell division in {{Escherichia}} coli involves rapid pole to pole oscillation of the division inhibitor {{MinC}} under the control of {{MinD}} and {{MinE}}},\ }\href {https://doi.org/10.1046/j.1365-2958.1999.01575.x} {\bibfield  {journal} {\bibinfo  {journal} {Molecular Microbiology}\ }\textbf {\bibinfo {volume} {34}},\ \bibinfo {pages} {82} (\bibinfo {year} {1999})}\BibitemShut {NoStop}%
\bibitem [{\citenamefont {Hu}\ and\ \citenamefont {Lutkenhaus}(2001)}]{Hu.Lutkenhaus2001}%
  \BibitemOpen
  \bibfield  {author} {\bibinfo {author} {\bibfnamefont {Z.}~\bibnamefont {Hu}}\ and\ \bibinfo {author} {\bibfnamefont {J.}~\bibnamefont {Lutkenhaus}},\ }\bibfield  {title} {\bibinfo {title} {Topological {{Regulation}} of {{Cell Division}} in {{E}}. coli},\ }\href {https://doi.org/10.1016/S1097-2765(01)00273-8} {\bibfield  {journal} {\bibinfo  {journal} {Molecular Cell}\ }\textbf {\bibinfo {volume} {7}},\ \bibinfo {pages} {1337} (\bibinfo {year} {2001})}\BibitemShut {NoStop}%
\bibitem [{\citenamefont {Lutkenhaus}(2007)}]{Lutkenhaus2007}%
  \BibitemOpen
  \bibfield  {author} {\bibinfo {author} {\bibfnamefont {J.}~\bibnamefont {Lutkenhaus}},\ }\bibfield  {title} {\bibinfo {title} {Assembly {{Dynamics}} of the {{Bacterial MinCDE System}} and {{Spatial Regulation}} of the {{Z Ring}}},\ }\href {https://doi.org/10.1146/annurev.biochem.75.103004.142652} {\bibfield  {journal} {\bibinfo  {journal} {Annual Review of Biochemistry}\ }\textbf {\bibinfo {volume} {76}},\ \bibinfo {pages} {539} (\bibinfo {year} {2007})}\BibitemShut {NoStop}%
\bibitem [{\citenamefont {Heermann}\ \emph {et~al.}(2021)\citenamefont {Heermann}, \citenamefont {Steiert}, \citenamefont {Ramm}, \citenamefont {Hundt},\ and\ \citenamefont {Schwille}}]{Heermann.etal2021}%
  \BibitemOpen
  \bibfield  {author} {\bibinfo {author} {\bibfnamefont {T.}~\bibnamefont {Heermann}}, \bibinfo {author} {\bibfnamefont {F.}~\bibnamefont {Steiert}}, \bibinfo {author} {\bibfnamefont {B.}~\bibnamefont {Ramm}}, \bibinfo {author} {\bibfnamefont {N.}~\bibnamefont {Hundt}},\ and\ \bibinfo {author} {\bibfnamefont {P.}~\bibnamefont {Schwille}},\ }\bibfield  {title} {\bibinfo {title} {Mass-sensitive particle tracking to elucidate the membrane-associated {{MinDE}} reaction cycle},\ }\href {https://doi.org/10.1038/s41592-021-01260-x} {\bibfield  {journal} {\bibinfo  {journal} {Nature Methods}\ }\textbf {\bibinfo {volume} {18}},\ \bibinfo {pages} {1239} (\bibinfo {year} {2021})}\BibitemShut {NoStop}%
\bibitem [{\citenamefont {Tostevin}\ and\ \citenamefont {Howard}(2008)}]{Tostevin.Howard2008}%
  \BibitemOpen
  \bibfield  {author} {\bibinfo {author} {\bibfnamefont {F.}~\bibnamefont {Tostevin}}\ and\ \bibinfo {author} {\bibfnamefont {M.}~\bibnamefont {Howard}},\ }\bibfield  {title} {\bibinfo {title} {Modeling the establishment of par protein polarity in the one-cell {{C.~elegans}} embryo},\ }\href {https://doi.org/10.1529/biophysj.108.135152} {\bibfield  {journal} {\bibinfo  {journal} {Biophysical Journal}\ }\textbf {\bibinfo {volume} {95}},\ \bibinfo {pages} {4512} (\bibinfo {year} {2008})}\BibitemShut {NoStop}%
\bibitem [{Note11()}]{Note11}%
  \BibitemOpen
  \bibinfo {note} {A detachment term into the binding-competent species could be added and would not change the following derivation because such a term would only change the mathematical form of ${\protect \mathbf R}_m(\protect \mathbf {m},\protect \mathbf {c}_1)$ but the membrane-reaction term would still only depend on the binding-competent cytosolic components.}\BibitemShut {Stop}%
\bibitem [{Note12()}]{Note12}%
  \BibitemOpen
  \bibinfo {note} {For example, such a density dependence may emerge for cytosolic dimerization of MinD upon nucleotide exchange (cf.\ discussion in Appendix~\ref {app:cytSeries}).}\BibitemShut {Stop}%
\bibitem [{Note13()}]{Note13}%
  \BibitemOpen
  \bibinfo {note} {For any matrix with full row rank, the dimension of its null space equals the number of columns minus the number of rows.}\BibitemShut {Stop}%
\bibitem [{Note14()}]{Note14}%
  \BibitemOpen
  \bibinfo {note} {Due to the assumption of a single binding-competent cytosolic component per species, the Jacobian $[\partial _{\protect \bm {\rho }}\protect \mathbf {c}_1^*]_{\protect \mathrm {hss}}$ is a square matrix and generically invertible; it becomes singular only on a codimension-one subset where ${\det \left [\partial _{\protect \bm {\rho }} \protect \mathbf {c}_1^{*}\right ]_{\protect \mathrm {hss}}=0}$.}\BibitemShut {Stop}%
\bibitem [{\citenamefont {Hu}\ \emph {et~al.}(2002)\citenamefont {Hu}, \citenamefont {Gogol},\ and\ \citenamefont {Lutkenhaus}}]{Hu.etal2002}%
  \BibitemOpen
  \bibfield  {author} {\bibinfo {author} {\bibfnamefont {Z.}~\bibnamefont {Hu}}, \bibinfo {author} {\bibfnamefont {E.~P.}\ \bibnamefont {Gogol}},\ and\ \bibinfo {author} {\bibfnamefont {J.}~\bibnamefont {Lutkenhaus}},\ }\bibfield  {title} {\bibinfo {title} {Dynamic assembly of {{MinD}} on phospholipid vesicles regulated by {{ATP}} and {{MinE}}},\ }\href {https://doi.org/10.1073/pnas.102059099} {\bibfield  {journal} {\bibinfo  {journal} {Proceedings of the National Academy of Sciences}\ }\textbf {\bibinfo {volume} {99}},\ \bibinfo {pages} {6761} (\bibinfo {year} {2002})}\BibitemShut {NoStop}%
\bibitem [{\citenamefont {Lackner}\ \emph {et~al.}(2003)\citenamefont {Lackner}, \citenamefont {Raskin},\ and\ \citenamefont {{de Boer}}}]{Lackner.etal2003}%
  \BibitemOpen
  \bibfield  {author} {\bibinfo {author} {\bibfnamefont {L.~L.}\ \bibnamefont {Lackner}}, \bibinfo {author} {\bibfnamefont {D.~M.}\ \bibnamefont {Raskin}},\ and\ \bibinfo {author} {\bibfnamefont {P.~A.~J.}\ \bibnamefont {{de Boer}}},\ }\bibfield  {title} {\bibinfo {title} {{{ATP-Dependent Interactions}} between {{{\emph{Escherichia}}}}{\emph{ coli}} {{Min Proteins}} and the {{Phospholipid Membrane In Vitro}}},\ }\href {https://doi.org/10.1128/JB.185.3.735-749.2003} {\bibfield  {journal} {\bibinfo  {journal} {Journal of Bacteriology}\ }\textbf {\bibinfo {volume} {185}},\ \bibinfo {pages} {735} (\bibinfo {year} {2003})}\BibitemShut {NoStop}%
\bibitem [{\citenamefont {Hu}\ \emph {et~al.}(2003)\citenamefont {Hu}, \citenamefont {Saez},\ and\ \citenamefont {Lutkenhaus}}]{Hu.etal2003}%
  \BibitemOpen
  \bibfield  {author} {\bibinfo {author} {\bibfnamefont {Z.}~\bibnamefont {Hu}}, \bibinfo {author} {\bibfnamefont {C.}~\bibnamefont {Saez}},\ and\ \bibinfo {author} {\bibfnamefont {J.}~\bibnamefont {Lutkenhaus}},\ }\bibfield  {title} {\bibinfo {title} {Recruitment of {{MinC}}, an {{Inhibitor}} of {{Z-Ring Formation}}, to the {{Membrane}} in {{{\emph{Escherichia}}}}{\emph{ coli}} : {{Role}} of {{MinD}} and {{MinE}}},\ }\href {https://doi.org/10.1128/JB.185.1.196-203.2003} {\bibfield  {journal} {\bibinfo  {journal} {Journal of Bacteriology}\ }\textbf {\bibinfo {volume} {185}},\ \bibinfo {pages} {196} (\bibinfo {year} {2003})}\BibitemShut {NoStop}%
\bibitem [{Note15()}]{Note15}%
  \BibitemOpen
  \bibinfo {note} {Note that the filtering factor also depends on the species index $\alpha $, $P_i^{k,\alpha }(q)$, but we suppress this index in order to simplify the notation.}\BibitemShut {Stop}%
\bibitem [{\citenamefont {Thalmeier}\ \emph {et~al.}(2016)\citenamefont {Thalmeier}, \citenamefont {Halatek},\ and\ \citenamefont {Frey}}]{Thalmeier.etal2016}%
  \BibitemOpen
  \bibfield  {author} {\bibinfo {author} {\bibfnamefont {D.}~\bibnamefont {Thalmeier}}, \bibinfo {author} {\bibfnamefont {J.}~\bibnamefont {Halatek}},\ and\ \bibinfo {author} {\bibfnamefont {E.}~\bibnamefont {Frey}},\ }\bibfield  {title} {\bibinfo {title} {Geometry-induced protein pattern formation},\ }\href {https://doi.org/10.1073/pnas.1515191113} {\bibfield  {journal} {\bibinfo  {journal} {Proceedings of the National Academy of Sciences}\ }\textbf {\bibinfo {volume} {113}},\ \bibinfo {pages} {548} (\bibinfo {year} {2016})}\BibitemShut {NoStop}%
\bibitem [{\citenamefont {W{\"u}rthner}\ \emph {et~al.}(2022)\citenamefont {W{\"u}rthner}, \citenamefont {Brauns}, \citenamefont {Pawlik}, \citenamefont {Halatek}, \citenamefont {Kerssemakers}, \citenamefont {Dekker},\ and\ \citenamefont {Frey}}]{Wurthner.etal2022}%
  \BibitemOpen
  \bibfield  {author} {\bibinfo {author} {\bibfnamefont {L.}~\bibnamefont {W{\"u}rthner}}, \bibinfo {author} {\bibfnamefont {F.}~\bibnamefont {Brauns}}, \bibinfo {author} {\bibfnamefont {G.}~\bibnamefont {Pawlik}}, \bibinfo {author} {\bibfnamefont {J.}~\bibnamefont {Halatek}}, \bibinfo {author} {\bibfnamefont {J.}~\bibnamefont {Kerssemakers}}, \bibinfo {author} {\bibfnamefont {C.}~\bibnamefont {Dekker}},\ and\ \bibinfo {author} {\bibfnamefont {E.}~\bibnamefont {Frey}},\ }\bibfield  {title} {\bibinfo {title} {Bridging scales in a multiscale pattern-forming system},\ }\href {https://doi.org/10.1073/pnas.2206888119} {\bibfield  {journal} {\bibinfo  {journal} {Proceedings of the National Academy of Sciences}\ }\textbf {\bibinfo {volume} {119}},\ \bibinfo {pages} {e2206888119} (\bibinfo {year} {2022})}\BibitemShut {NoStop}%
\bibitem [{\citenamefont {Burkart}\ \emph {et~al.}(2024)\citenamefont {Burkart}, \citenamefont {M{\"u}ller},\ and\ \citenamefont {Frey}}]{Burkart.Mueller.Frey2024}%
  \BibitemOpen
  \bibfield  {author} {\bibinfo {author} {\bibfnamefont {T.}~\bibnamefont {Burkart}}, \bibinfo {author} {\bibfnamefont {B.~J.}\ \bibnamefont {M{\"u}ller}},\ and\ \bibinfo {author} {\bibfnamefont {E.}~\bibnamefont {Frey}},\ }\bibfield  {title} {\bibinfo {title} {Dimensionality reduction in bulk--boundary reaction--diffusion systems},\ }\href {https://doi.org/10.1103/PhysRevE.110.034412} {\bibfield  {journal} {\bibinfo  {journal} {Phys. Rev. E}\ }\textbf {\bibinfo {volume} {110}},\ \bibinfo {pages} {034412} (\bibinfo {year} {2024})}\BibitemShut {NoStop}%
\bibitem [{\citenamefont {Levine}\ and\ \citenamefont {Rappel}(2005)}]{Levine.Rappel2005}%
  \BibitemOpen
  \bibfield  {author} {\bibinfo {author} {\bibfnamefont {H.}~\bibnamefont {Levine}}\ and\ \bibinfo {author} {\bibfnamefont {W.-J.}\ \bibnamefont {Rappel}},\ }\bibfield  {title} {\bibinfo {title} {Membrane-bound {{Turing}} patterns},\ }\href {https://doi.org/10.1103/PhysRevE.72.061912} {\bibfield  {journal} {\bibinfo  {journal} {Physical Review E}\ }\textbf {\bibinfo {volume} {72}},\ \bibinfo {pages} {061912} (\bibinfo {year} {2005})}\BibitemShut {NoStop}%
\bibitem [{\citenamefont {{Frohoff-H{\"u}lsmann}}\ and\ \citenamefont {Thiele}(2023)}]{Frohoff-Hulsmann.Thiele2023a}%
  \BibitemOpen
  \bibfield  {author} {\bibinfo {author} {\bibfnamefont {T.}~\bibnamefont {{Frohoff-H{\"u}lsmann}}}\ and\ \bibinfo {author} {\bibfnamefont {U.}~\bibnamefont {Thiele}},\ }\bibfield  {title} {\bibinfo {title} {Nonreciprocal {{Cahn-Hilliard Model Emerges}} as a {{Universal Amplitude Equation}}},\ }\href {https://doi.org/10.1103/PhysRevLett.131.107201} {\bibfield  {journal} {\bibinfo  {journal} {Physical Review Letters}\ }\textbf {\bibinfo {volume} {131}},\ \bibinfo {pages} {107201} (\bibinfo {year} {2023})}\BibitemShut {NoStop}%
\end{thebibliography}
%

\end{document}